\documentclass{ar-1col}
\usepackage[comma]{natbib}
\usepackage{url}
\usepackage{amssymb}
\usepackage{hyperref}
\usepackage{colortbl}
\usepackage{graphics}
\usepackage{graphicx}
\jname{Xxxx. Xxx. Xxx. Xxx.}
\jvol{AA}
\jyear{YYYY}
\doi{10.1146/((please add article doi))}

\newcommand{\apj} {Astrophys. J.}
\newcommand{\apjs}{Astrophys. J. Suppl.}
\newcommand{\apjl}{Astrophys. J. Lett.}
\newcommand{\aj}{Astronom. J.}
\newcommand{\aap} {Astron. Astrophys.}
\newcommand{\aaps} {Astron. Astrophys. Suppl.}
\newcommand{\araa}{Ann. Rev. Astron. Astrophys.}
\newcommand{\mnras}{MNRAS}
\newcommand{\nat}{Nature}
\newcommand{\bain}{Bull. Astron. Inst. Netherlands}
\newcommand{\rmxaa}{Revista Mexicana de Astronom\'ia y 
         Astrof\'isica}
\newcommand{\pasa}{Pub. Astro. Soc. Australia}        
\newcommand{\pasp}{Pub. Astro. Soc. Pac.}
\newcommand{\pasj}{Pub. Astro. Soc. Japan}
\newcommand{\jgr}{J. Geophys. Res.}
\newcommand{\physrep}{Physics Reports}
\newcommand{\jcp}{J. Chem. Physics}
\newcommand{\prl}{Phys. Rev. Lett.}

\newcommand{\hi} {\mbox{\rm H\,{\small\footnotesize I}}}
\newcommand{\hii} {\mbox{\rm  H\,{\small\footnotesize II}}}
\newcommand{\cii} {\mbox{\rm C\,{\small\footnotesize II}}}
\newcommand{\ci} {\mbox{\rm C\,{\small\footnotesize I}}}
\newcommand{\oi} {\mbox{\rm O\,{\small\footnotesize I}}}
\newcommand{\oiii} {\mbox{\rm O\,{\small\footnotesize III}}}
\newcommand{\siii} {\mbox{\rm Si\,{\small\footnotesize II}}}
\newcommand{\suiii} {\mbox{\rm S\,{\small\footnotesize III}}}
\newcommand{\nii} {\mbox{\rm N\,{\small\footnotesize II}}}
\newcommand{\niii} {\mbox{\rm N\,{\small\footnotesize III}}}
\newcommand{\feii} {\mbox{\rm Fe\,{\small\footnotesize II}}}
\newcommand{\si} {\mbox{\rm S\,{\small\footnotesize I}}}
\newcommand{\sii} {\mbox{\rm S\,{\small\footnotesize II}}}
\newcommand{\Hxn}{$H_{\rm X}/n$}
\newcommand{\HH}{H$_{2}$}
\newcommand{\Av}{$A_{\rm V}$}

\newcommand{\herschel}{\textit{Herschel}}

\newcommand{\spitzer}{\textit{Spitzer}}

\newcommand{\Lsun}{L$_{\odot}$}
\newcommand{\Msun}{M$_{\odot}$}
\newcommand{\Zsun}{Z$_{\odot}$}
\newcommand{\mic}{$\mu$m}
\newcommand{\cm}{${\rm cm^{-3}}$}

\definecolor{table-yellow}{rgb}{1,1,0.8}
\definecolor{table-blue}{rgb}{0.7, 0.8, 0.9}
\definecolor{table-green}{rgb}{0.82, 0.94, 0.75}
\definecolor{table-orange}{rgb}{0.96, 0.76, 0.76}
\definecolor{table-text-red}{rgb}{0.9, 0.17, 0.31}
\definecolor{table-text-green}{rgb}{0.05, 0.5, 0.06}
\begin{document}

\markboth{Wolfire, Vallini, Chevance}{PDRs and XDRs}

\title{Photodissociation and X-Ray Dominated Regions}

\author{Mark G.\ Wolfire,$^1$ Livia Vallini,$^2$ and M\'elanie Chevance$^3$
\affil{$^1$Department of Astronomy, University of Maryland, College Park, MD, 20742, USA; email: mwolfire@gmail.com}
\affil{$^2$Scuola Normale Superiore, Piazza dei Cavalieri 7, I-56126, Pisa, Italy}
\affil{$^3$Astronomisches Rechen-Institut, Zentrum f\"ur Astronomie der Universit\"at Heidelberg, M\"onchhofstrasse 12-14, D-69120 Heidelberg, Germany}}

\begin{abstract}
The radiation from stars and active galactic nuclei (AGN) creates photodissociation regions (PDRs) and X-ray dominated regions (XDRs), where the chemistry
or heating are dominated by far-ultraviolet (FUV) radiation
or X-ray radiation, respectively. PDRs include a wide range of environments from the diffuse interstellar medium to dense star-forming regions.
XDRs are found 
in the center of galaxies hosting AGN, in protostellar disks, and in the vicinity of X-ray binaries. 
 In this review, we describe the dominant thermal, chemical, and radiation transfer processes in PDRs and XDRs, as well as a brief description of models and their use to analyze observations. We then present recent results from Milky Way, nearby extragalactic, and high-redshift observations.

Several important results are:

\begin{itemize}

\item 
Velocity resolved PDR lines reveal the kinematics
of the neutral\\
atomic gas and
provide constraints on the
stellar feedback process.\\ 
Their 
interpretation is, however, in dispute as
observations suggest a\\ prominent
role for
stellar winds while they are much less important \\in 
theoretical models.

\item A significant fraction of molecular mass
resides in CO-dark gas\\
especially in low-metallicity/highly irradiated environments.

\item
 The CO ladder 
 and 
 {\mbox [{\rm C\,{\small\footnotesize I}}]}/[\cii] ratios can determine if FUV or X-rays\\ dominate the ISM heating of extragalactic sources. 

\item With ALMA, PDR and XDR tracers are now routinely detected\\
on galactic scales over cosmic time. This makes it possible to link \\
the star formation history of the Universe to the evolution of the \\
physical and chemical properties of the gas.

\end{itemize}

\end{abstract}

\begin{keywords}
photodissociation regions, 
x-ray dominated regions, 
interstellar medium, active galactic nuclei, infrared emission, star forming regions
\end{keywords}
\maketitle

\tableofcontents

\section{INTRODUCTION}
Massive stars produce a prodigious amount of radiative energy that interacts with the interstellar medium (ISM). The radiation drives chemical processing and heats the gas and dust which cool in bright line and continuum emission. 
Regions where the
heating or chemistry are dominated by  far-ultraviolet  
(FUV; $6\,{\rm eV} < h\nu < 13.6\,{\rm eV}$) 
radiation
are called photodissociation regions \citep[or PDRs;][hereafter TH85]{TielensHollenbach1985}.\footnote{
Sometime referred to as Photon-Dominated Regions (also PDRs). This term was
first coined by Alex Dalgarno and Amiel Sternberg to contrast PDRs with optically thick
gas in which the chemistry is driven by cosmic-ray ionization, rather than by
FUV photons that stimulate molecule production via ionization and
heating in addition to destroying molecules via photodissociation (A. Sternberg,
private communication). In this review we will adopt the original and more widely used term 
Photodissociation Regions.}
 PDRs can have a range of incident FUV flux, 
densities, and column densities, depending on
the distance of the stars that produce the FUV radiation and the gas environment where the
FUV is absorbed.
The ``classic" PDR discussed in TH85 is one in which FUV radiation
is emitted by an O or early B star and passes through an \hii\ region to be absorbed in an adjoining molecular cloud. In general, these are high density and high FUV field PDRs
\citep[e.g.,][]{TielensHollenbach1985b, Meixner1992, Tielens1993}
and is  where feedback from radiative and mechanical energy of massive stars acts on the neutral gas.
The FUV radiation from the OB association may also illuminate more distant parts of the molecular cloud and produce PDRs with low radiation fields but greater
area than the high intensity PDRs \citep{Stacey1993}.
In addition, the OB stars contribute FUV radiation  to the 
interstellar radiation field 
that illuminates molecular clouds but also illuminates the diffuse ISM where the field
strength, density and column densities are lower than the classic PDRs \citep{Wolfire1995,Wolfire2003}. PDRs can also be found
 in reflection nebula 
\citep{Chokshi1988,Steiman-Cameron1997},
in planetary nebula \citep{Graham1993, Latter2000}, 
on the surfaces of pillars and globules \citep{Mookerjea2019,Goicoechea2020,Schneider2021}, in embedded
protostars \citep{vanKempen2010, Visser2012}, and 
protostellar and protoplanetary disks \citep{Aikawa2002,Gorti2004, Woitke2010, Kamp2010, Oberg2021}.

PDRs include all regions
of the neutral ISM where FUV radiation plays a role in the physics and/or chemistry. This includes the atomic gas, but also the deeper molecular layers where FUV radiation
plays a role in the organic inventory. Most of the molecular gas in giant molecular clouds resides in PDRs and thus almost all of the atomic gas and molecular gas
in the Galaxy is in PDRs. This conclusion holds  for other galaxies as well 
so that most of the non-stellar baryons within galaxies are in PDRs.

Much of the infrared line and continuum radiation from galaxies arises
in PDRs \citep{Crawford1985,Stacey2010}. The FUV radiation is mainly absorbed by grains and is radiated away 
as infrared continuum, but a fraction of the FUV radiation heats the gas via the photoelectric effect
on small grains and large molecules. The PDR line emission is the dominant coolant
in the neutral gas and can be as bright
as $0.1-1$\% of the infrared continuum. 
The line and continuum emission 
can be used to determine the gas physical conditions in the PDR gas, and thus in most
of the mass of the ISM. 

The observation and modeling  of PDRs is also important for 
star formation. The line emission reveals the local environment
that gives rise to star formation and the feedback processes that
might inhibit it. PDRs measure the star formation rate by probing
the radiative energy produced by embedded massive stars.
In addition, PDRs play an important role in the global
processing of material between thermal phases, which results in cold dense
gas from which molecules and stars can form \citep{Ostriker2010}. The ionization produced
by FUV radiation in molecular clouds 
provides coupling to magnetic fields, helping to
regulate star formation \citep{McKee1989}.

In addition to FUV radiation, X-ray radiation 
can dominate the heating, ionization and chemical composition in  X-ray dominated regions \citep[XDRs;][]{Maloney1996}\footnote{The term X-ray Dissociation Regions - in analogy to the Photodissociation Regions - was first coined by \citet{Maloney1996} to refer to the portion of a gas cloud illuminated by an X-ray source where the molecule dissociation, but also the ionization, heating, and chemical composition are driven by the X-rays. For this reason the term XDR is now widely used with the more general meaning of X-ray Dominated Region. In this review we will adopt this definition.}. The X-rays can be produced by a variety of sources and processes 
such as young stellar objects, gas accretion onto a super massive black hole, and X-ray binaries.
Early seminal works by \citet{Krolik1983} and \citet{Lepp1983} concentrated on the effects of X-rays from embedded stellar sources within molecular clouds, and on the impact of X-rays on the molecular gas within the obscuring torus of Active Galactic Nuclei \citep[AGN; e.g.,][]{Krolik1989}. The first comprehensive analysis of the chemistry, ionization, and thermal balance of XDRs was that of \citet{Maloney1996}. Since then, the physics and chemistry of XDRs has been examined  from several different perspectives concentrating on protoplanetary disks \citep{Glassgold1997, Igea1999, Stauber2005, Ercolano2008, Owen2011, Meijerink2012, Aresu2012}, galactic centers of AGN host sources \citep[e.g.,][]{Meijerink2007, Meijerink2011, vanderWerf2010, Garcia-Burillo2010, Harada2013, Mingozzi2018}, and X-ray binaries \citep{Moser2017,Lebouteiller2017}. 

Several previous reviews have been published
on both PDRs and XDRs \citep[e.g.,][]{Hollenbach1997,Hollenbach1999RvMP,Sternberg2005,Snow2006,Bolatto2013}. 
In this review we provide the basic micro-physics 
required to understand the dominant thermal, chemical, and radiative processes  in PDRs and XDRs while striving to 
summarize  results
from theoretical, observational, and laboratory work
since the previous reviews. In order to limit the scope of such a vast
field of research  we will only lightly touch upon topics 
that are covered in
recent reviews such as the emission from Polycyclic Aromatic Hydrocarbons
\citep[PAHs;][]{Tielens2008}, chemistry in diffuse gas \citep{Gerin2016}, and protoplanetary disks \cite[e.g.,][]{Bergin2007}.

\section{Fundamental Processes in  PDRs and XDRs}
\label{sec:physicsPDRsXDRs}

\begin{figure}[ht]
\includegraphics[width=5in]{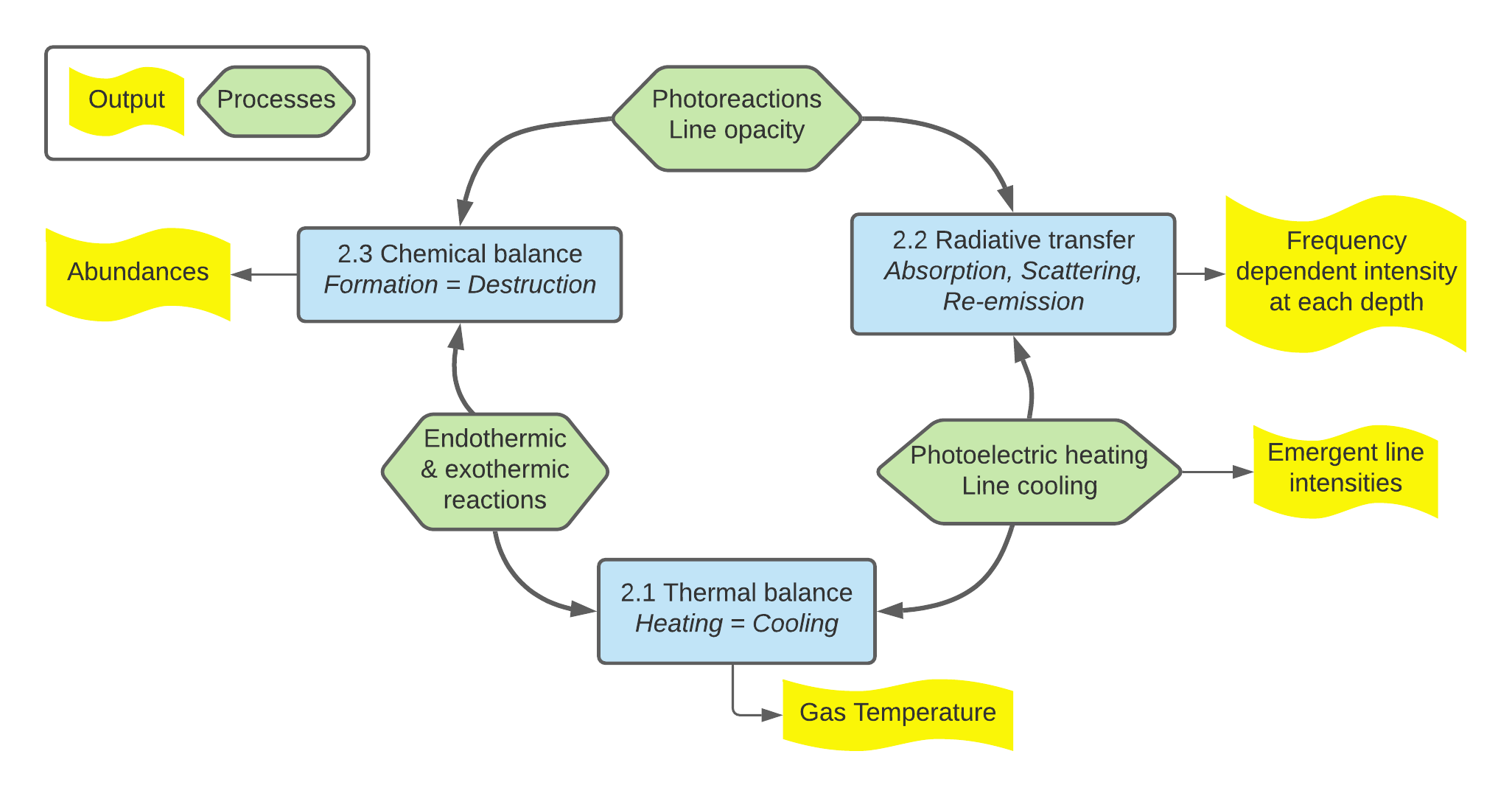}
\caption{Schematics of the iterative calculation required between chemical balance, thermal balance and radiative transfer (blue boxes) in PDR models. Numbers indicate subsection in text. Some relevant processes are indicated in green.
The main outputs of the calculation are indicated in yellow.} 
\label{fig:Iteration}
\end{figure}

All PDR and XDR  models must self-consistently
solve for the gas temperature,
the radiation field, and the abundances of atomic and molecular species, as a function of depth into
the cloud. 
These components are coupled 
and thus require an iterative
scheme to solve (see {\bf Figure \ref{fig:Iteration}}). Some simplification can be made by assuming a plane-parallel geometry,
thermal balance in which
the heating rate equals 
the cooling rate, chemical balance
in which the formation rate of each
species is balanced by the destruction rate, and parameterized fits to
the depth dependence of photo processes. In this section we briefly 
discuss the three main components of a model under
these assumptions and discuss 
models that lift them in later sections.
Note that while we will mostly refer to a prototypical PDR model, the same considerations apply to XDR models, modulo their peculiar heating mechanisms that are addressed in Section \ref{sec:xdr_heating}.

A common iterative scheme is to start at the outer
edge of a cloud, with an initial guess of temperature,
abundances, and radiation field and solve for 
abundances in 
chemical balance. Using these abundances, the temperature
is adjusted lower (higher) if the cooling rate is 
higher (lower) than the heating rate and the chemistry
is recalculated. Both the chemistry and temperature are iterated
until both chemical balance and thermal equilibrium are
achieved. This cycle is repeated at the next step into the cloud where fitted functions are used for the  dependence of
radiative processes with depth.

\subsection{Thermal Balance}
The heating rates are generally of the form 
$n\Gamma$ erg ${\rm cm^{-3}}$ ${\rm s^{-1}}$ where
$n$ is the hydrogen nucleus density and $\Gamma$ is the
heating rate per hydrogen in units erg ${\rm s^{-1}}$. 
For example, an important heating process in PDRs is the photoelectric effect on small grains and PAHs
which is given by
\begin{equation}
    n\Gamma = 10^{-24} n G \epsilon\,\, {\rm erg}
    \, {\rm cm^{-3}}\, {\rm s^{-1}}\, , 
\end{equation}
where $n$ is the hydrogen nucleus density, $G$ is a measure of the integrated FUV radiation 
field, and $\epsilon$ is the photoelectric
heating efficiency. TH85 defined $G_0$ to be the
incident field strength in units of the \cite{Habing1968} interstellar radiation field, while $G$ is the local (attenuated) field strength within the cloud. The Habing field has an integrated FUV flux of $1.6\times 10^{-3}$ ${\rm erg}\, {\rm cm^{-2}}\, {\rm s^{-1}}$.
Since 
$\epsilon$ depends
on $G$, the gas temperature $T$, and the electron abundance
$n_{e}$, the photoelectric heating rate is not fixed until $G$,
$T$, and $n_{e}$ are converged.

The cooling rate
is generally of the form $n^2\Lambda$ erg ${\rm cm^{-3}}$ ${\rm s^{-1}}$ where $\Lambda$ is
the cooling rate coefficient
in erg ${\rm cm^3}$ ${\rm s^{-1}}$. 
To illustrate the cooling process,  we consider
collisional excitation followed by radiative de-excitation of a two-level system of species $i$, with  a density $n_{\rm i}$, 
and collisions with atomic hydrogen of density $n$.
In the absence of background radiation 
the cooling rate is 
\begin{equation}
n^2 \Lambda = n x_{\rm i} \frac{g_{\rm u}/g_{\rm l} {\rm e}^{-E_{\rm ul}/kT}A_{\rm ul}E_{\rm ul}\beta_{\rm esc}(\tau_{\rm lu})}{(g_{\rm u}/g_{\rm l}   {\rm e}^{-E_{\rm ul}/kT} + 1)+A_{\rm ul}\beta_{\rm esc}(\tau_{\rm lu })/n\gamma_{\rm ul}} \, , 
\label{eq:coolingrate}
\end{equation}
where $x_{\rm i} = n_{\rm i}/n$ is the fractional abundance
of species $i$, $g_{\rm u}$ and $g_{\rm l}$ are
the statistical weights of the upper and
lower states, $E_{\rm ul}$ is
the energy of the transition, $k$ is the Boltzmann constant, $T$ is the
gas temperature, $A_{\rm ul}$ is the Einstein A coefficient, 
$\beta_{\rm esc}(\tau_{\rm lu})$ is the escape 
probability,  $\tau_{\rm lu}$ is the optical
depth in the line, and 
$\gamma_{\rm ul}$ is the rate coefficient for collisional de-excitation. $\gamma_{ul}$ is 
often fit with a form $\gamma_{\rm ul}=CT^\alpha$.
The derivation of equation \ref{eq:coolingrate} can
be found, for example, in \cite{Tielens2005}. 

\begin{marginnote}
\entry{$\beta_{\rm esc}(\tau_{lu})$}{The 
probability that a photon  escapes the cloud
given a line optical depth $\tau_{\rm lu}$ to the surface.}
\end{marginnote}

A useful characterization of a line transition is given by the critical density, $n_{\rm cr}$, which is the density at which
the rate of collisional de-excitations  is equal to the rate of spontaneous radiative de-excitations,
$n_{\rm cr} = A_{\rm ul}/\gamma_{\rm ul}$\footnote{For a 
multi-level system, $n_{\rm cr}$ includes the sum of all
radiative and collisional rates out of 
level u to lower levels.}.
 Through $\gamma_{\rm ul}$,
$n_{\rm cr}$ has a weak dependence on temperature.
For densities, $n\ll n_{\rm cr}$ the cooling rate per volume is 
\begin{equation}
n^2 \Lambda = n^2 x_i \gamma_{\rm ul}
g_{\rm u}/g_{\rm l} {\rm e}^{-E_{\rm ul}/kT} E_{\rm ul}\, ,
\end{equation}
which is proportional to 
$n^2$ and each collisional excitation results in a photon.
For densities  $n\gg n_{\rm cr}$  
the levels are thermalized (they depend only
on gas temperature and are populated according to the Boltzmann distribution)  and the cooling rate is 
proportional to $n$ (see equation \ref{eq:coolingrate}).  Note that even though the rate
of collisional de-excitation exceeds the rate of radiative
de-excitation, the upper level is still 
radiatively de-excited at the Einstein $A$ rate.  

\begin{marginnote}[]
\entry{$n_{\rm cr}$}{Critical density at which
the rate of collisional de-excitations  is equal to the rate of spontaneous radiative de-excitations.}.
\end{marginnote}

The emergent line intensity is given by the integral
of the cooling rate from the cloud surface into the cloud,
\begin{equation}
I = \frac{1}{2\pi}\int_0^z dz\, n^2  \Lambda\, ,
\end{equation}
where the factor $1/2\pi$ is appropriate for a
semi-infinite slab where photons escape through only the front surface, and $n^2\Lambda$ is given
by equation \ref{eq:coolingrate} and accounts
for optical depth effects in the line.
The line optical depth, $\tau_{\rm lu}$, and hence $\beta_{\rm esc}$ depends on the
densities of the species in the upper and lower levels and
the Doppler line width, $\delta v_{\rm D}$, which includes
both  thermal and turbulent broadening\footnote{The 
Doppler width is related to the
full width at half maximum width by 
$\delta v_{\rm FWHM} = 1.665 \delta v_{\rm D}$.} (see Section \ref{sec:PDR_cooling}).
The effects of background radiation, multiple collision partners, and multilevel systems  can be included in
the  excitation, cooling,
and line emission following 
 \cite{Tielens2005} and \cite{Draine2011}.  Thermal balance is achieved when $n^2\Lambda = n\Gamma$. 
Heating rates for PDRs are proportional to $G_0n\epsilon$
or $\propto n^{1.7}G_0^{0.3}$ for high charge parameter
(see equation \ref{eq:peheatingeps}) ($\propto nH_{\rm X}$
for XDRs; see Section
\ref{sec:xdr_heating}) and cooling
rates are proportional to $n^2$, and thus the thermal structure
is a function of $(G_0/n)^{0.3}$ (or \Hxn\ for XDRs, see Section \ref{sec:xdr_structure}).
 The $[\cii]$ ($^2P_{3/2} - \,^2\!P_{1/2}$) 158\,\mic\ 
line is an example of a two-level, fine-structure transition,
with $E_{\rm ul}/k = 92$ K, 
$A_{\rm ul} = 2.4\times 10^{-6}$ ${\rm s^{-1}}$, and at $T=100$ K, critical densities
of 9, 3000, and 6100 ${\rm cm^{-3}}$ for
collisions with $e^-$, H, and  ${\rm H_2}$ respectively. 
The statistical weights are given by $2J+1$ so that
$g_{\rm u} = 4$ and $g_{\rm l} = 2$. \cite{Goldsmith2012} give
analytic solutions for the excitation of ${\rm C^+}$ and the $[\cii]$ line intensity in various limits
of the critical density, optical depth, and background radiation fields. 
A similar analysis for the $[\oi]$ lines is
given in \cite{Goldsmith2019}.

\subsection{Radiation Transfer}
The equation of transfer for the FUV continuum radiation field (or X-ray continuum, see Section~\ref{subsec:PDRXDR_inputparams}),
including non-isotropic grain scattering, has been solved
using a number of methods including  spherical harmonics \citep{Flannery1980,LePetit2006, Goicoechea2007}, and ray tracing \citep{Rollig2013,yorke1980}.
Photo rates at each depth into the cloud are given by $4\pi \int J(\lambda)(\lambda/hc)
\sigma(\lambda)d\lambda$ where $\sigma(\lambda)$ is
the cross section of the process, and $J(\lambda)$ is the local mean intensity 
defined as the angle average over the specific
intensity $I(\lambda)$,  $J(\lambda)=1/(4\pi) \int_{\rm 4\pi} I(\lambda) d\Omega$, where $d\Omega$ is the differential element of solid angle.  Instead of explicitly solving
for $J(\lambda)$, a simplification can be made
by using pre-calculated fits to the photo rates which take into account the depth dependence of the radiation field. These are calculated for a specific
angle of incidence and 
spectral energy distribution, grain type, and geometry  \citep[e.g.,][]{Heays2017}.
For a radiation field that is normally incident on a plane parallel layer, the photo rates generally take
the form 
\begin{equation}
k_i = \alpha G_0 {\rm e}^{-\gamma_{\rm exp} A_{\rm V}}\,\, {\rm s^{-1}} 
\label{eq:photorate}
\end{equation}
where $\alpha$ is the 
unattenuated rate in the local interstellar radiation
field, $G_0$ is the incident field and
and $\gamma_{\rm exp}$ gives the (exponential) depth
dependence of the dust opacity relative to the
visual extinction, $A_{\rm V}$, into the cloud. The effects of
scattering are usually included in $\gamma_{\rm exp}$ for a specific grain
model.
The hydrogen nucleus column density,
$N = N_{\rm H} + 2N_{\rm H_2}$, is related to $A_{\rm V}$ by $N=1.9\times 10^{21} A_{\rm V}$ $\rm cm^{-2}$
for the local Galaxy.  Several different 
radiation fields are in use besides the Habing field, including
the \cite{Draine1978} field, $\chi$, and the \cite{Mathis1983} field,
$U$\footnote{The symbol $U$ is also commonly used for the ionization parameter in \hii\ regions.}. The integrated field strengths are related to $G_0$ by $\chi{\sim} G_0/1.7$ and $U{\sim}G_0/1.1$.\footnote{Note that $\alpha$ calculated for a Draine field will be approximately 1.7 times higher than that for a Habing field, but this scaling does not account for differences in the spectral energy distribution.} See also Section  \ref{subsec:PDRXDR_inputparams} for conversion between different radiation fields.

In addition to the dust opacity, gas opacity can significantly modify the radiation field, mainly due to 
absorption by H, ${\rm H_2}$, C, and CO, and in
protostellar disks OH and ${\rm H_2O}$.
The gas opacity decreases the photorates faster with depth than by 
dust opacity alone \citep{vanDishoeck1988,Visser2009}.

\subsection{Chemical Balance}
In chemical balance a kinetic approach is
used to find the abundances. 
For each species the destruction and formation rates
are calculated and the abundances found for which
the formation rates and destruction rates are equal. 

Two-body reaction rates between species $i$ and $j$  are given by 
$n_in_j k_{ij}$ 
${\rm cm^{-3}}$ ${\rm s^{-1}}$ where $n_i$ and $n_j$ are the volume
densities of $i$ and $j$, and $k_{ij}$ is the reaction rate coefficient in
${\rm cm^3}$ ${\rm s^{-1}}$. The densities of PDRs considered in this review are sufficiently low so that three-body reactions are not important but they can be important at
higher densities or lower grain abundances found, for example, in protostellar disks.

Fits to $k_{ij}$ are often given
as 
\begin{equation}
    k_{\rm ij} = a \left( \frac{T}{300\, {\rm K}} \right)^b 
        {\rm e}^{-c/T}
\end{equation}
where $c$ is 
the activation energy for the reaction to proceed in temperature units.  
For example, the reaction ${\rm C^+ + H_2\rightarrow CH^+ + H}$ has a rate coefficient $k = 1.5\times10^{-10} \exp(-4640/T)$ ${\rm cm^{3}}$ ${\rm s^{-1}}$ (with $b=0$).
The photodestruction rates and cosmic-ray 
destruction rates of a species $i$ are 
proportional to $G_0n_i$ and $\zeta n_i$
respectively where $\zeta$ is the cosmic-ray ionization rate.
 For two-body formation rates 
$(\propto n^2)$
balanced by photo- or cosmic-ray destruction
($\propto nG_0$, $\propto n\zeta$)
the chemical structure is a function 
of $G_0/n$ and $\zeta/n$.

The previous subsections serve as illustrations for the basic processes in PDRs and the interdependence of chemistry, thermal processes, and radiation transfer. The same apply to XDRs, whose peculiar physics and chemistry are discussed in Section~\ref{sec:xdr}. In the next two sections we discuss PDRs and XDRs in more detail. 

\section{The Physics of PDRs}
\label{sec:pdr}
\subsection{1-D Structure}
\label{sec:PDR_structure}

\begin{figure}
\includegraphics[width=4.5in]{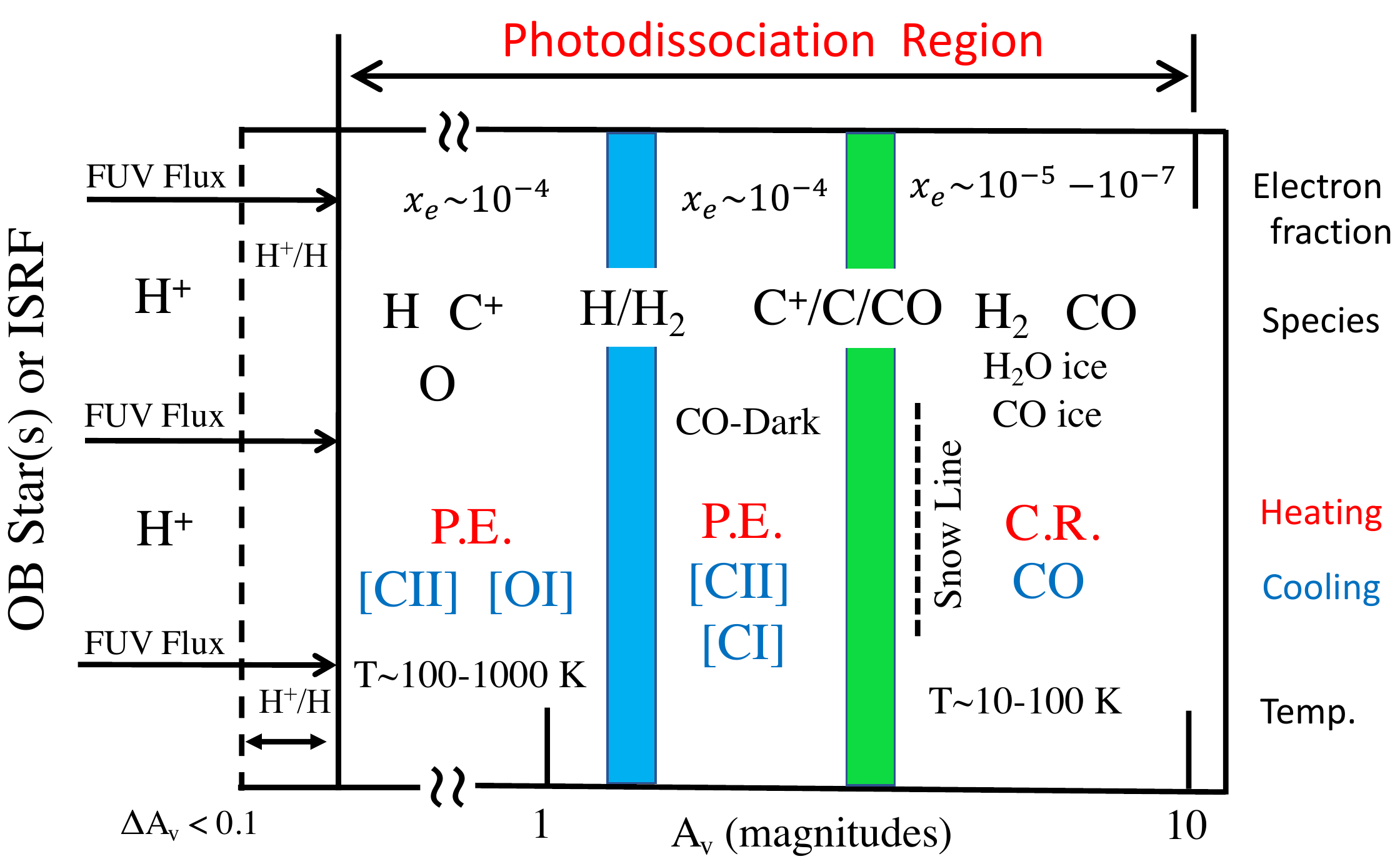}
\caption{Schematic of a PDR as a function of visual
extinction, $A_{\rm V}$, into the cloud where $A_{\rm V}$ is related to the 
hydrogen nucleus column density by $N=1.9\times 10^{21}$ ${\rm cm^{-2}}A_{\rm V}$. 
Typically, in the outer layers, the gas consists of neutral H, ${\rm He}$, O, and singly
ionized states of metals (e.g., ${\rm C^+}$, ${\rm Si^+}$, 
${\rm Fe^+}$). The electron fraction is $x_e{\sim}10^{-4}$ and is provided by the photoionization of C to ${\rm C^+}$. The gas temperature can be $T{\sim}100-1000$\,K, heated  by the photoelectric effect (P.E.)
on small grains and PAHs,
and cooled by fine-structure lines of [\cii] and
[\oi]. At $A_{\rm V}{\sim}2$, H is converted
to \HH,
and at $A_{\rm V}{\sim}2-4$, ${\rm C^+}$ recombines with electrons, and C and CO are formed. CO-dark molecular gas lies between the \HH\ and CO layers where [\cii] and [\ci] dominate the cooling.
Atoms and molecules freeze out on grains beyond the snow line.  At greater depths the gas temperatures are $T{\sim}10-100$\,K, dominated by cosmic-ray (C.R.)  and
gas-grain heating, and CO rotational line cooling. The electron fraction drops to 
$x_e{\sim} 10^{-7}$. 
Lower $G_0/n$ moves the transitions to lower $A_{\rm V}$. 
Figure adapted with permission from TH85, \copyright AAS.}
\label{fig:PDRSchematic}
\end{figure}
We first discuss the basic structure of PDRs
in terms of a one-dimensional (1-D) layer. 
{\bf Figure \ref{fig:PDRSchematic}} shows the 1-D structure of
a  classic PDR adjacent to an \hii\ region as a function of
$A_{\rm V}$ into the cloud (the exact value of $A_{\rm V}$
depends on $G_0/n$). The stellar
extreme-ultraviolet (EUV; $h\nu > 13.6$ eV) photons, emitted by the OB stars, ionize the
surrounding gas and produce an \hii\ region. 
The  FUV
radiation field that emerges from the \hii\ region
illuminates the (hydrogen) neutral gas beyond and produces the PDR.   Molecules are photodissociated and metals with ionization potentials less than 13.6 eV are singly ionized (e.g., ${\rm C^+}$, ${\rm Si^+}$, ${\rm S^+}$).  Atomic hydrogen, helium, and oxygen are neutral
since these have ionization potentials greater than 13.6 eV. Here, the gas is
heated mainly by ejection of electrons from small  grains and PAHs and cooled by
fine-structure line emission and  attains a temperature typically $T{\sim} 100-1000$\,K. The gas temperature exceeds the grain temperature since the gas cools by line radiation while grains cool by continuum radiation. The absorption of
FUV radiation 
excites emission from
PAHs (e.g., \citealt{Allamandola1989,Peeters2002,Peeters2017}, see also reviews by \citealt{Puget1989,Tielens2008, Tielens2021}) 
while the FUV and optical radiation heats larger
grains that emit an IR continuum. 
Deeper into the cloud
the FUV radiation diminishes due to dust opacity and
at $A_{\rm V}{\sim} 2$, H is converted to ${\rm H_2}$.
 Deeper into the cloud ($A_{\rm V}{\sim} 2-4$), ${\rm C^+}$ recombines
with electrons to produce atomic C, and CO can form through a series of ion-neutral reactions \citep{vanDishoeck1988,Sternberg1995}. 
The region between ${\rm H_2}$ formation and CO formation 
is known as the CO-poor \citep{Lada1988, vanDishoeck1990} or CO-dark molecular gas \citep{Grenier2005, Wolfire2010}
and is \HH\
molecular gas that is not associated with CO millimeter line emission. The \HH\ and CO photodissociate via line absorption and can therefore  self-shield and thus appear closer to the surface than ${\rm H_2O}$ or ${\rm O_2}$,  or other molecules which dissociate via FUV continuum.
In the deepest layers the gas 
 attains
temperatures $T{\sim} 10-30$ K and is heated mainly
by cosmic-ray ionization and, in dense regions, from gas collisions
with grains heated by the IR radiation from the surface. 
The gas is cooled
by rotational transitions of CO. 
Depending on the
physical conditions, atoms and molecules may freeze out on grain surfaces at $A_{\rm V}\gtrsim 3$, greatly affecting
the gas-phase abundances and altering the gas cooling and
chemistry.  Although we have presented the
PDR as a static layer sandwiched between  the \hii\ region and the  molecular cloud,
a complete picture involves the evolution of the \hii\
region, the PDR internal dynamics, and disruption of the cloud. We will touch upon these topics in Section \ref{sec:multiD_PDR}.
The PDR structure changes depending on the type and evolution of the illuminating source. For example, 
while O and early B stars produce \hii\ regions, later B stars
produce reflection nebulae with FUV radiation but little EUV radiation and hence no 
\hii\ region.  Lower $G_0/n$, or illumination by cooler stars, moves the transitions 
between H/\HH\ and ${\rm C^+}$/C/CO to lower $A_{\rm V}$.

\subsection{Chemistry}
\label{sec:PDR_chemistry}

The basic chemical pathways \citep[e.g.,][]{vanDishoeck1986, Sternberg1995}
have not changed significantly since
the last reviews  although many of the rates have been updated
through laboratory and theoretical work
and several processes are now found to be more significant. For example, the dissociative recombination 
of ${\rm H_3^+}$ measured by \cite{McCall2004} was determined to be higher than previous estimates and led to 
an order of magnitude increase in the cosmic-ray ionization rate
required to match observations of molecular ions 
\citep{Indriolo2007,Hollenbach2012, Neufeld2017}. Freeze-out of O and CO onto grains, and 
reactions occurring on  grain surfaces  were not accounted for but are now found 
to dramatically change gas phase abundances, mainly for $A_{\rm V} \gtrsim 3$ \citep[e.g.,][]{Hollenbach2009}, 
and are essential for the production of complex molecules.  
\subsubsection{Photoreactions}
\label{sec:PDR_chemistry_photoreactions}
The photo rates have been reviewed extensively by \cite{Heays2017} for ionization and dissociation of a number
of atomic and molecular species. Rates are provided in
free space for a \cite{Draine1978} interstellar radiation field.
The depth dependence of the dust opacity, including scattering, for an isotropically
illuminated layer is fit both as $\exp(-\gamma_{\rm exp} A_{\rm V})$, and as
$E_2(\gamma_{\rm E2} A_{\rm V})$ where $E_2$ is the 2nd exponential
integral function and results from the formal solution to
the transfer equation. 
Isotropic radiation falls off faster than
normally incident radiation at the 
cloud edge. 
 The dependence  for a normally incident field can be taken from Heays et al.\ to
be $\propto \exp(-\gamma_{\rm E2}A_{\rm V})$.
Only the normally
incident rays penetrate to large depth and 
the depth dependence of both isotropic and normally incident fields become the same. 
A source of FUV radiation can
be produced in cloud interiors when cosmic rays excite 
\HH\ to electronic levels that are then radiatively 
de-excited \citep{Prasad1983}. The rates for these photon interactions are
also found in \cite{Heays2017}.

\subsubsection{Photodissociation of \HH}
\label{sec:PDR_chemistry_photodissociationofH2}
The photodissociation of \HH\ proceeds by line absorption of
photons at
$912 < \lambda < 1108$ \AA\ within the Lyman and Werner 
bands \citep{Black1987,Sternberg1989, Abgrall1992,Sternberg2014}. The molecule is pumped from 
the ground electronic state $X^1\Sigma_g^+$ to an excited $B^1\Sigma_u$ or $C^1\Pi_u$ electronic
state. For ${\sim} 90$\% of the pumps 
a radiative de-excitation (emitting a UV photon) 
lands in a bound vibrational state of the electronic ground state $X$, where a cascade
through vibration and rotation levels ensues producing an
IR line spectrum \citep[see e.g.,][for calculations of
FUV pumped and collisional \HH\ line spectra]{Shaw2005,Roueff2019,Zhang2021}. 
However, at high gas density, the FUV pumped  $v,J$ levels are collisionally de-excited and the energy
goes into heat rather than a radiative cascade.
The remaining ${\sim} 10$\% of the pumps
land in the vibrational continuum thereby dissociating the
\HH.  When the FUV absorption lines become optically thick, the FUV pumping and dissociation rates 
rapidly drop with increasing \HH\ column.  
This optical depth effect is known as \HH\ self-shielding and is 
an essential part of \HH\ chemistry. Fits for \HH\ self-shielding as a 
function of the \HH\ column density  
$f_{\rm H_2, shield}(N_{\rm H_2})$ are given in \cite{DraineBertoldi1996}, yielding a photodissociation rate
per \HH\ of $D=5.8\times 10^{-11}\chi 
f_{\rm H_2, shield}\exp(-3.1A_{\rm V})$ ${\rm s^{-1}}$ 
\citep[see][for a comprehensive analysis of \HH\ photodissociation]{Sternberg2014}\footnote{Here we have written the photo rate using $\chi$, the \citet{Draine1978} field, since it has been calculated explicitly for that energy distribution.}.
Similar to \HH, CO also photodissociates in lines but the lines are 
nearly 100\% predissociated, i.e.,  absorption leads directly to dissociation. The absorption from \HH\ can also shield  CO from photodissociation
deeper into the cloud. Shielding functions for CO including \HH\ and
CO self-shielding are given in \cite{Visser2009}. Atomic C is photoionized by continuum radiation at $\lambda <1102$ \AA\ and can also shield molecules at larger depth \citep{Rollins2012}.

 \begin{marginnote}[]
\entry{Adsorption}{Adsorption is when an atom
or molecule attaches to a grain surface 
upon collision.}
\end{marginnote}

\begin{marginnote}[]
\entry{Physisorb/chemisorb}
{Physisorb is  attachment
by weak long-range van der Waals forces. 
Chemisorb is attachment by strong  chemical bonds.}
\end{marginnote}

\begin{marginnote}[]
\entry{Desorption}{Desorption is when an atom or 
molecule becomes detached from a grain surface.}
\end{marginnote}
\subsubsection{Formation of \HH}
\label{sec:PDR_chemistry_formationofH2}
The formation of \HH\ in the ISM occurs on grain surfaces (e.g.,  \citealt{Cazaux2004, Cazaux2010}, see also review by
\citealt{Wakelam2017}). In diffuse gas,  \HH\ is formed by
the Langmuir-Hinshelwood (LH) mechanism where
H atoms are physisorbed to the surface by   
weak long-range van der Waals forces. 
The H atoms migrate around the surface until they
meet and react to form \HH.
A rate coefficient of $k_{\rm H_2}=3\times 10^{-17}$ 
${\rm cm^3}$ ${\rm s^{-1}}$ was fit to observations of 
\HH\ columns in diffuse gas and has been known as the ``standard" or diffuse gas value \citep{Jura1974}.  The rate per volume 
 is given by 
$R_{\rm H_2}=nn_{\rm H}k_{\rm H_2}$ ${\rm cm^{-3}}$ ${\rm s^{-1}}$
where $n$  is the hydrogen nucleus density and $n_{\rm H}$ is the density of atomic hydrogen.  
Theoretical modeling  \citep{Hollenbach1979}
using observationally constrained parameters reproduce 
this rate and provide a functional form for the rate coefficient
\begin{equation}
k_{\rm H_2}=3\times 10^{-17} \left( \frac{T}{100\, {\rm K}} \right)^{1/2}  S(T,T_{\rm gr})\eta(T_{\rm gr}) Z'_d\, {\rm cm^3}
\, {\rm s^{-1}}\, ,
\label{eq:H2formation}
\end{equation}
where $S(T, T_{\rm gr})$ is the sticking coefficient for H atoms striking grains of temperature
$T_{\rm gr}$,  $\eta(T_{\rm gr})$ is the fraction of atoms that react 
to form \HH\ before
being thermally desorbed, and $Z'_d$ is a scale factor for the dust abundance with
$Z'_d=1$ the local Galactic abundance. 
Usually in PDRs, $T > T_{\rm gr}$.
On grains with regular surfaces and only physisorbed sites, H will
thermally desorb before it can react with another H to
form \HH\ at a critical grain temperature of $T_{\rm cr}{\sim}20$ K.
Surface defects or an amorphous surface  could increase the adsorption
energies, and 
$T_{\rm cr}$ is then the grain temperature for which a H is desorbed before finding a defect site. 
\cite{Hollenbach1971} find $T_{\rm cr}{\sim} 65$ K for a binding energy of ${\sim} 600$ K which allows for a sufficiently
high grain temperature for \HH\ formation in diffuse gas.\footnote{We note, however, that a more stringent constraint
comes from comparing thermal desorption times with
the time for another H to be adsorbed. For
$T_{\rm cr}{\sim} 50$ K a binding energy of 1500 K is required. Whether such sites exist on an amorphous silicate surface is unclear.}
For cold gas and grains, $\eta$
and $S$ are ${\sim}1$ and equation \ref{eq:H2formation} with
these values are often used in both diffuse and dense PDRs.

However, for the higher $T$  and $T_{\rm  gr}$ found in PDRs, both $S$ and $\eta$ decrease since H will bounce rather
than stick and adsorbed atoms will be rapidly desorbed, thus
driving down the ${\rm H_2}$ formation rate by the LH mechanism.  But the analysis of the \HH\ rotational and vibrational line emission in 
both moderate and low $G_0$ PDRs
indicate the required rate coefficient is
1-5 times higher than that in diffuse gas
\citep{Habart2004}. 
The higher formation rates draw the ${\rm H/H_2}$ transition closer to the warm surface where collisional excitation enhances the rotational line emission. 
What could maintain or even increase \HH\ formation? 
 The effects of higher velocity collisions and
reduced sticking coefficient at higher $T$ may approximately cancel \citep{Kaufman1999} leaving a constant rate, but this does
not account for the thermal desorption that would occur for higher
grain temperatures.
Another mechanism that might
be at work is the
Eley-Rideal (ER) process 
\citep[see e.g.,][]{Cuppen2008, Cuppen2010, LeBourlot2012, Wakelam2017}.
For ER
the H atoms
are chemisorbed to the surface by strong 
chemical bonds that hold the H atom in place.
A hot atom from the gas phase may strike the bound
atom and react to form \HH. 
The resulting formation rate is a function of 
the energy barrier required to 
attach the H atom to the surface,
surface coverage of chemisorbed H so that a
second H will strike it,
the total surface area of grains per hydrogen nucleus, and the gas temperature. For a graphitic grain the barrier is measured to be ${\sim} 2000$ K \citep{Sha2002} for attachment to the surface (basal plane)
and high gas temperatures are required to overcome the barrier.
After the first attached H, 
the barrier
becomes minor for additional H in neighboring sites and high coverage in clusters can be achieved \citep{Hornekaer2006}.  
On silicates
there is no  barrier for a H atom adsorption. 
\cite{LeBourlot2012} and \cite{Bron2014} adopt a barrier
of ${\sim 300}$ K for carbon particles suggesting that defects might lower the barrier,
although direct interaction will be limited by the surface coverage
 of defect sites. With a barrier of 300 K,  and
 sufficiently high  surface coverage of defects 
and grain surface area, 
the rate coefficient at $T{\sim} 2\times 10^3$ K is  3-4 times the diffuse rate and provides a
good fit to \HH\ line emission in the Orion Bar and NGC\,7023 
\citep{Bron2014, Joblin2018}.  Since the diffuse rate is essentially a
maximum  in which every two H collisions result in an \HH, the inferred
high \HH\ formation rate in PDRs likely requires activation of new surface routes \citep[see also discussion in][]{Tielens2021}.

\cite{Bron2014}  considered the effects of temperature fluctuations in very small grains and PAHs  
caused by single photon absorption. Between photon 
absorptions, an \HH\ may form by the LH mechanism before
another photon arrives and desorbs the H. The process has
a low efficiency for $G_0/n{\sim 1}$ but can be important
for $G_0 < 200$ and $n>10^3$ ${\rm cm^{-3}}$.
\begin{marginnote}
\entry{Endothermic reaction}{Chemical reaction where there is a net
energy input between energy state of reactants and
energy state of products}
\end{marginnote}
\begin{marginnote}
\entry{Exothermic reaction}{Chemical reaction where there is a net  energy release between energy state of reactants and energy state of products}
\end{marginnote}
\begin{marginnote}
\entry{Activation energy}{Energy required for a reaction
to occur. Both endothermic and exothermic reactions may require an activation energy.}
\end{marginnote}

Another possibility, that takes advantage of the increased
surface area of PAHs, are H abstraction reactions 
on  PAHs that have adsorbed a large number of H
\citep[a superhydrogenated PAH;][]{Thrower2012}. An incident H from the gas phase reacts with H on the PAH to form \HH. 
The barrier for attachment is minor $T{\sim 100}$ K and easily achieved at PDR temperatures.  
However, the additional H is easily photodesorbed \citep{Andrews2016} and
if the abstraction cross section is as low as that measured
for the PAH coronene \citep{Mannella2012} then this process will be not be efficient.  Another process is by way of photodesorption
\citep{Castellanos2018}. An absorbed UV photon will excite the PAH,  allow H to roam, and the energy is dissipated in 
a loss of an H or \HH. Whether H or \HH\ is 
produced is quite sensitive to PAH
structure, and might work in only a narrow band of PAH types.
For optimum conditions, the \HH\ formation rate coefficient could be as high as ${\sim}2\times 10^{-17} (T/100\,{\rm K})^{1/2}$ ${\rm cm^{3}}$ ${\rm s^{-1}}$ (Tielens 2021, private communication).

Model fits to observations of the mid-infrared \HH\ line emission
 and the CO ladder can help to constrain the
\HH\ formation rate (e.g., \citealt{Habart2004,Sheffer2011,Stock2015,Joblin2018,Wu2018}, and future observations of \HH\ with JWST). As already mentioned, higher \HH\ formation rates
draws the \HH\ to warmer surface layers and increases the \HH\ line emission but can also produce a small fraction of warm CO through reactions with excited \HH\ (Sections \ref{sec:PDR_chemistry_reactionswithH2*} and 
\ref{subsec:PDRsHighFUV}).

Although there has been significant advances in our understanding from theoretical and
laboratory work \citep[see review by][]{Wakelam2017}, the \HH\ formation process 
is still uncertain, especially as a function of the gas and grain temperatures. Observations suggest
$k_{\rm H_2}{\sim} 3\times 10^{-17}$  ${\rm cm^{-3}}$ ${\rm s^{-1}}$ or larger, even
in regions of fairly high gas temperature ($T{\sim} 1000$ K) and warm ($T_{\rm gr}{\sim} 75$ K) grains, as in the Orion
Bar PDR.
Further experimental and quantum chemical studies as well as confirmation by
comparison to astronomical observations are warranted.

\subsubsection{Reactions with ${\rm H_2^*}$}
\label{sec:PDR_chemistry_reactionswithH2*}
The internal energy of  vibrationally or rotationally excited  \HH, ${\rm H_2^*}$, 
can be used to increase the rate 
of endothermic chemical reactions \citep[TH85,][]{Sternberg1995, Agundez2010}. 
The reactions 
${\rm C^+}\, + {\rm H_2^*}\rightarrow {\rm CH^+} + {\rm H}$ and
${\rm S^+}\, + {\rm H_2^*}\rightarrow {\rm SH^+} + {\rm H}$
are especially important for the formation of ${\rm CH^+}$ and ${\rm SH^+}$. The 
activation energies of $\Delta E/k = 4640$ K and $\Delta E/k = 9860$ K respectively are  
too high to be achieved by gas temperatures in much of the PDR. However, these endothermic reactions can
be driven near the surface of PDRs where the ${\rm H_2}$ is FUV pumped or where gas
temperatures are sufficiently high to drive the reactions (with rates $\propto
\exp[-\Delta E/kT]$). Such enhanced rates are required
to match observations of ${\rm SH^+}$ and ${\rm CH^+}$ \citep{Nagy2013, Zanchet2019, Goicoechea2019}, and can lead to
the production of ${\rm HCO^+}$ and CO. 

 \begin{marginnote}
\entry{$\zeta_{\rm p}$} {Primary  cosmic-ray ionization
rate per hydrogen atom with units ${\rm s^{-1}}$.}
\end{marginnote}  
\subsubsection{Cosmic-ray reactions}
\label{sec:PDR_chemistry_comicrayreactions}
  Since there are no stellar EUV  photons beyond the \hii\ region boundary,  
H and He can not be ionized by the radiation field but 
can be ionized
by cosmic rays that penetrate the PDR.\footnote{\cite{Wolfire2003, Wolfire2010} included a thermal soft X-ray component produced by
 local Galactic white dwarfs and super novae remnants to partially ionize the diffuse WNM and molecular cloud surfaces but these photons do not penetrate more than  $N{\sim} 10^{19}$ ${\rm cm^{-2}}$. Although they are important for low column density
 diffuse gas, they are not important for dense PDRs.}
It is this ionization by cosmic rays that drives much of the ion-neutral
 chemistry in PDRs (an exception are reactions with ${{\rm H}_2^*}$). Typical ion-neutral reaction rate coefficients are ${\sim} 10^{-9}$
 ${\rm cm^{-3}}$ ${\rm s^{-1}}$ with no temperature dependence nor activation barrier, and thus ion-neutral chemistry
 proceeds rapidly compared to neutral-neutral reactions  which have typical rate  coefficients of ${\sim}10^{-11}$ ${\rm cm^{-3}}$ ${\rm s^{-1}}$.  Typical ionization rates in diffuse gas are
 inferred from observations of ${\rm ArH^+}$, ${\rm OH^+}$, 
 ${\rm H_2O^+}$, and ${\rm H_3^+}$ to be $\zeta_{\rm p}\sim 2\times 10^{-16}$ ${\rm s^{-1}}$ per H atom (\citealt{Hollenbach2012,Indriolo2015, Neufeld2017}, see also \citealt{Shaw2021} for the effects of
variable PAH abundance).
 Each ionization of H, \HH\ or He, by a cosmic-ray electron of energy $\lesssim 10$ MeV, produces a secondary electron of energy ${\sim}35$ eV which can further ionize
 another H or \HH\ (the fast cosmic ray being considered as the primary particle). 
 The total rate including secondary ionizations in H and \HH\ gas is $\zeta_{\rm H}{\sim}1.5\zeta_{\rm p}$ and $\zeta_{\rm H_2}{\sim}2.3 \zeta_{\rm p}$
 respectively, and depends in detail on the molecular and
 electron fractions \citep{Cravens1978,Glassgold1974,Dalgarno1999,Glassgold2012}. Higher electron fractions result in more 
 energy going into heat rather than secondary
 ionizations.  The dependence of the ionization rate with column
 density into the PDR is not well known and depends on the incident
 cosmic-ray spectrum and interaction with turbulence and the magnetic field in the
 cloud
\citep{Padovani2009,Silsbee2019}. The low energy cosmic
rays are absorbed  at increasing column density and
decrease the total ionization rate compared to the rate at the surface.
 Observations indicate $\zeta_{\rm p}\sim 1.1\times 10^{-17}$ ${\rm s^{-1}}$ in protostellar envelopes
 \citep{vanderTak2000}
 and  marginal evidence for a $1/N$ dependence in clouds \citep{Padovani2009,Neufeld2017}. 
 Higher ionization rates  initially increase the abundance of molecular
 ions \citep{Hollenbach2012,LePetit2016}, however,  
 for $\zeta_{\rm p}/n_2 \gtrsim 10^{-14}$ ${\rm s^{-1}}$ with
 $n_2 = n/(100\,{\rm cm^{-3}})$,  the abundance of molecular ions decreases due
 to   the destruction of \HH\ by cosmic rays and due to an increasing electron abundance from ionizations leading  to faster ${\rm H_3}^+$ dissociative 
 recombination \citep[e.g.,][]{LePetit2016}.  Increased ionization rates destroy CO by reactions with ${\rm He^+}$, requiring higher densities for formation and pushing CO deeper into the cloud \citep[e.g.,][]{Gong2017}. Similarly,
  internal sources of
 cosmic rays from e.g., protostars can change the chemical structure by warming the gas in the interior and increasing the ${\rm He^+}$ density  which destroys CO and increases C
 \citep{Gaches2019}.
 
 \subsubsection{Grain assisted recombination}
\label{sec:PDR_chemistry_grainrecombination}
 The effects of an increased cosmic-ray ionization rate are partly mitigated by grain assisted
 recombination where mainly light ions (${\rm H^+}$, ${\rm He^+}$ and ${\rm C^+}$)
 recombine with electrons on ${\rm PAH^-}$ or charge exchange
 with ${\rm PAH^0}$.  Since the ${\rm PAH^-}$ and ${\rm PAH^0}$ species 
 are the result of electron attachment to ${\rm PAH^0}$ and ${\rm PAH^+}$, the result is to 
 reduce the abundance of free electrons and increase the abundance of neutral species. If not for this additional recombination, the production of CO would be inhibited for rates as low as $\zeta_{\rm p}=10^{-17}/n_2$
 ${\rm s^{-1}}$ 
 and most of the carbon would remain as C. The effects of grain assisted recombination have long been known from absorption line studies in diffuse clouds \citep[e.g.,][]{Lepp1988, Weingartner2001c}. Many modelers adopt the \cite{Draine1987}
 formalism to obtain the reaction rates as a function of
 the number of carbon atoms in a PAH. The rate 
must be either integrated over  the distribution in  PAH size and abundance or
multiplied by a typical size and abundance.
 \cite{Wolfire2008,Wolfire2003} calibrated
the theoretical rates to match the observed  ${\rm C^+/C}$ ratio in
diffuse gas and suggested a correction factor $\Phi_{\rm PAH}=0.5$
is needed, although \cite{Liszt2011} suggests rates should be recalibrated with the generally lower C columns from \cite{Burgh2010}. Fits for the \cite{Draine1987} rates  near $T=300$ K can be found in \cite{Hollenbach2012}. 
 We emphasize that the rates and required integrals are quite uncertain and urge that further work is needed.

 The photoionization of metals, in particular
 S, produces free electrons and affects the ion-neutral chemistry at intermediate depths $A_{\rm V}{\sim}1-5$. Similar to the effects
 of cosmic rays, a high electron abundance from metal ions can
 suppress the ion-neutral chemistry. The 
 effects are mitigated by recombination on
 PAHs or freeze-out of metals \citep[e.g.,][]{Hollenbach2012}.
 
\subsubsection{Grain surface reactions}
\label{sec:PDR_chemistry_grainsurfaces} 
Atoms and molecules collide with grains, and 
if the grains are 
 sufficiently cold,  they will be adsorbed on grain surfaces
thereby  depleting the species from the gas phase and freeze out
as ice on grains \citep[e.g.,][]{Hollenbach2009,Esplugues2016,Esplugues2017, Tielens2021}. Freeze out affects the gas-phase chemistry and
cooling, and induces grain-surface chemistry which can produce
simple molecules (e.g., ${\rm H_2O}$) and complex organic molecules \citep[COMs; e.g., methanol ${\rm CH_3OH}$;][]{Garrod2008}, which can then be desorbed from the grain back into the gas. Reviews of 
surface chemistry in PDRs and dark cores can be found 
in \cite{Cuppen2017} and in pre- and protostellar environments in \cite{Oberg2021}.

Similar to H,
 adsorbed species may diffuse across the surface, although heavier
 atoms or molecules diffuse slower, 
 and undergo
 chemical reactions when meeting (the LH mechanism) or
 may be hit by a gas-phase species and chemically react (the 
 ER mechanism). 
 The atoms and molecules on the grain surface 
 can be desorbed if they can overcome their
 binding energy. Potential desorption processes important
 near the cloud surface include photodesorption 
\citep[e.g.,][]{Oberg2009}, thermal desorption
\citep{Tielens1987}, and thermal fluctuations \citep{Bron2014}, 
while deeper into the cloud, cosmic-ray desorption \citep[e.g.,][]{Hasegawa1993}, cosmic-ray induced FUV photodesorption,
and chemical desorption \citep{Dulieu2013,Garrod2007} dominate.
Photon and cosmic-ray desorptions are individual
photon or cosmic-ray events, thermal 
desorption depends on the equilibrium (large) grain temperature, thermal fluctuations 
are the temperature spikes of small grains and PAHs after absorbing
a photon, and chemical desorption  relies
on exothermic chemical reactions on the grain surface. The cosmic-ray rates, and chemical desorption in particular, are not
well constrained.
In addition, thermal desorption and chemical reaction rates are 
sensitive to variations in grain temperature \citep{Esplugues2019}. Additional photoprocessing can occur 
on the grain surface and species can be photodissociated during
photodesorption. The formation of molecules is then a competition
between the rates of adsorption and surface reactions versus the desorption rates. See e.g.,  \cite{Hollenbach2009}, \cite{vanDishoeck2021}, and \cite{Tielens2021}
for  ${\rm H_2O}$ ice formation.

The ice is observed to be mainly composed of  
CO, ${\rm H_2O}$,  ${\rm CO_2}$,  
${\rm NH_3}$ (ammonia), ${\rm H_2CO}$ (formaldehyde), ${\rm CH_4}$ (methane), and ${\rm CH_3OH}$ (methanol)  \citep[e.g.,][]{Gibb2004} 
but can vary considerably
depending on grain temperature, radiation field strength, gas
density, and depth into the cloud. For example, ${\rm H_2O}$ ice can be thermally desorbed at 
grain temperatures $T_{\rm gr} \gtrsim 100$ K while CO is
desorbed at 
grain temperatures $T_{\rm gr} \gtrsim 20$ K. The depth for water ice formation is proportional to $\ln(G_0/n)$ and is typically \Av\ ${\sim} 1.6$ for moderate density ($n {\sim}10^3\, {\rm cm^{-3}}$) and low
FUV field ($G_0\sim 5$) sources \citep[see review and papers therein by][]{Boogert2015}.\footnote{In previous PDR reviews the ${\rm O/O_2}$ transition was depicted at
$A_{\rm V}\gtrsim 10$. However in \cite{Hollenbach2009} it was shown that the abundance of ${\rm O_2}$ peaks at \Av $\sim 5$. At lower \Av\
the photodesorption of OH and ${\rm H_2O}$ followed by the gas-phase reaction ${\rm OH\,+O\rightarrow O_2 + H}$ increases  ${\rm O_2}$
while at larger \Av\ freeze-out of O in ${\rm H_2O}$ ice diminishes all O bearing gas-phase species.}
However, grain surface chemistry  can produce 
minimum ${\rm H_2O}$, OH, and NH  gas-phase abundances even in diffuse
gas \citep{Crutcher1976, Sonnentrucker2015} although the fraction in ice is
insignificant. 
The time scale for several processes can become  longer
than cloud lifetimes and these require a time dependent approach (see  Section~\ref{subsec:1Dtimedependent}).

PDR chemical networks have been extended and updated to include e.g., N chemistry \citep[e.g.,][]{YoungOwl2002,Boger2005,Li2013,Persson2014}, S chemistry  
\citep{Sternberg1995, Goicoechea2021}, and H, C, and N isotopes \citep{LePetit2002,Heays2014,Szucs2014,Roueff2015,Rollig2013b, Visser2018}. Photo rates and 
cross sections can be found on-line at 
\url{https://home.strw.leidenuniv.nl/~ewine/photo/}
and dielectronic and radiative recombination rates
at \url{http://amdpp.phys.strath.ac.uk/tamoc}.
Data bases for gas-phase reactions are KIDA  \citep{Wakelam2012} and UMIST \citep{McElroy2013} although the latter is becoming somewhat dated.

\subsection{Gas Heating}
\label{sec:PDR_heating}

Several processes may contribute to the gas heating
depending on the incident FUV field strength, cosmic-ray ionization rate, gas density, and depth into the cloud. At moderate depths ($A_V\lesssim 5$), the dominant gas heating process  is from photoelectric ejection of electrons from 
small grains and PAHs \citep{Bakes1994,Weingartner2001}. The fraction of FUV photon energy 
that goes into gas heating is the heating efficiency, $\epsilon$, and is a function of 
the grain charge. 
The charge parameter $\gamma = GT^{1/2}/n_{e}$ ${\rm K^{1/2}}$ 
${\rm cm^3}$,  
is  proportional to the
ratio of the ionization rate of grains to the electron recombination rate where
$G$ is the local (attenuated) field strength,
$T$ is the gas
temperature, and $n_{e}$ is the electron density. For 
$\gamma \lesssim 10^2$ ${\rm K^{1/2}}$ 
${\rm cm^3}$
grains are neutral 
and $\epsilon$ is at a maximum of about
${\sim} 5$\%, while for higher $\gamma$,
grains become charged and
$\epsilon$ drops due to the 
electron kinetic
energy loss in escaping the Coulomb potential and because
fewer photons can ionize a more highly charged species.
Since the ionization to recombination rate is  proportional to the grain size, 
small grains
and PAHs have on average lower charge than larger species.
In addition, the larger grains
contribute less because the electron escape length is smaller than the photon absorption depth; i.e, the yield goes down.
Thus, small grains and PAHs dominate the heating with 
half
coming from grains smaller than 
${\sim} 15$\,\AA. Note that the electrons contributed
from PAHs are generally an insignificant fraction of
the free electron abundance since the PAH fractional abundance 
$(n_{\rm PAH}/n\sim 10^{-7})$ is much less than that of ${\rm C^+}$ $(n_{\rm C^+}/n\sim 10^{-4})$. 

The  heating rate per unit volume is given by $n\Gamma_{\rm PE}=10^{-24}Gn\epsilon$
erg $\rm cm^{-3}$ ${\rm s^{-1}}$ with $\epsilon$ from \cite{Bakes1994} (hereafter BT94), given by
\begin{equation}
\epsilon = \frac{4.87\times 10^{-2}}{1.0+4\times 10^{-3}\gamma^{0.73}}+ \frac{3.65\times 10^{-2}(T/10^4)^{0.7}}{1+2\times 10^{-2}\gamma}\, .
\label{eq:peheatingeps}
\end{equation}
The second term is only important at high temperatures where
a higher recombination rate of electrons with ionized small
grains and PAHs results in lower positive
charge
and higher $\epsilon$. In addition, 
for gas temperatures $T \gtrsim 3\times 10^3$ K and $\gamma \gtrsim 10^4$, recombination 
may become a significant gas cooling process. The photoelectric
heating rate is often reported as the net heating minus the cooling.
\cite{Wolfire2008} introduced a parameter  
$\Phi_{\rm PAH}{\sim} 0.5$, 
that modifies the charge parameter as $\gamma' = \gamma/\Phi_{\rm PAH}$
to account for a reduced recombination rate on grains,
needed to explain the observed column densities of C in the
diffuse ISM. 
The theoretical efficiency produces good agreement with
gas temperatures, thermal pressures, and [\cii] cooling rates in diffuse gas \citep{Wolfire1995,Wolfire2003, Jenkins2011, Gerin2015} where $T{\sim} 100$ K, $n_{e}{\sim} 0.01$, $G_0{\sim} 1.7$, and $\gamma{\sim} 1.7\times 10^3$ 
${\rm K^{1/2}}$ ${\rm cm^3}$.  The agreement with observed fine-structure line intensities and line ratios in dense PDRs is usually
quite good \citep[e.g.,][]{Stacey1991,Hollenbach1999RvMP}. 
The analytic fit for $\epsilon$ given by \cite{Weingartner2001}  (hereafter WD01) for a B0 star radiation field  is a factor of ${\sim} 1.7$ lower than BT94
at $\gamma{\sim} 10^5$ ${\rm K^{1/2}}$ ${\rm cm^3}$ and ${\sim} 4.1$ lower at extreme values of $\gamma {\sim} 10^6$ ${\rm K^{1/2}}$ ${\rm cm^3}$.
At low values of  $\gamma {\sim} 10^2$ ${\rm K^{1/2}}$ ${\rm cm^3}$,
WD01 is a factor 2.6 higher than BT94.
The $\epsilon$ given by the WD01 fit is nearly the
same as that 
derived from the Meudon PDR code (Section~\ref{sec:Overview}) at the cloud surface for an incident 
\cite{Draine1978} field. Although the shape of various heating
efficiencies that are in use are quite similar, there are differences by factors of a few especially for high $\gamma$. Observations such as those discussed next might help to constrain the efficiency.

\begin{figure}[ht]
\includegraphics[width=4.5in]{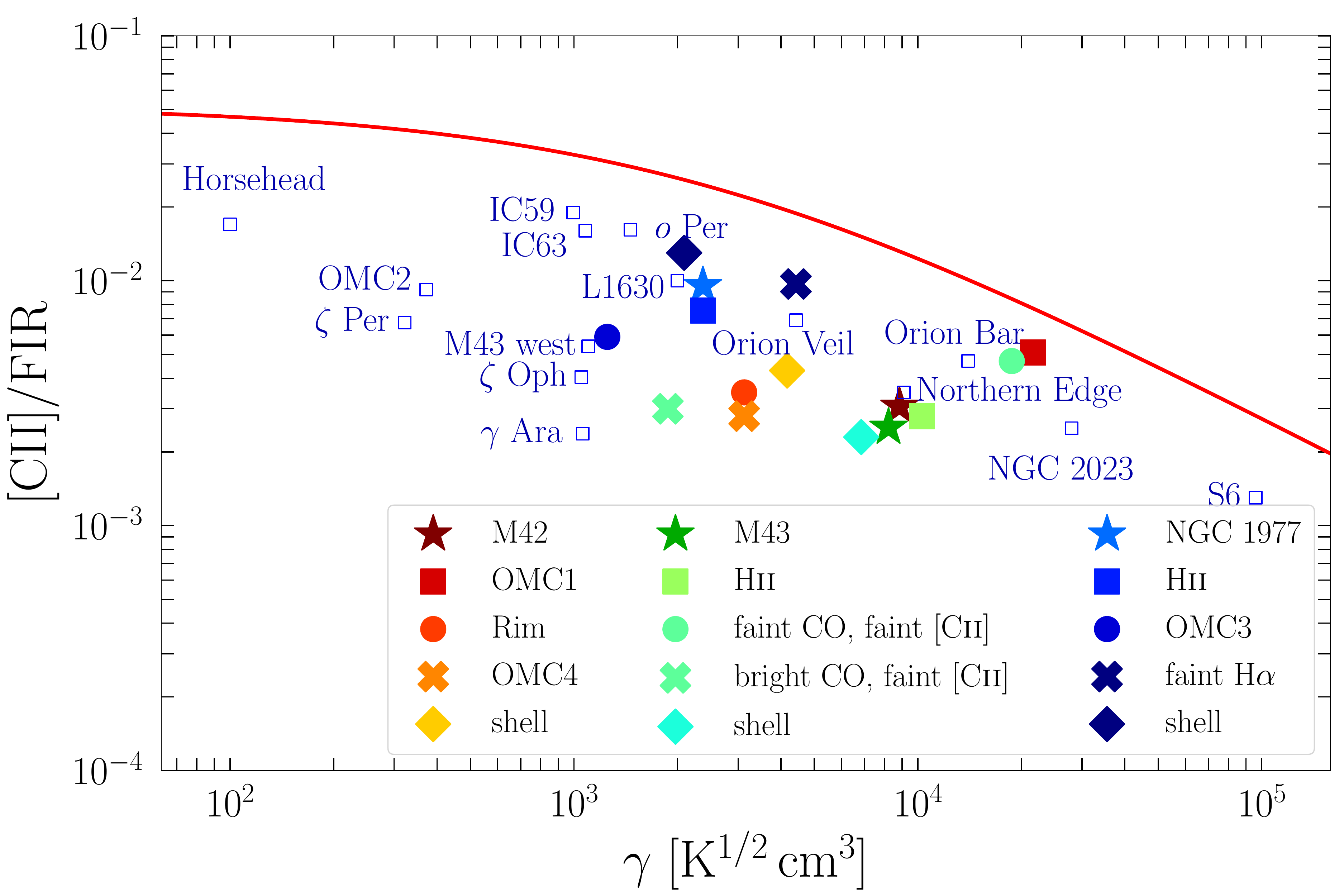}
\caption{Observations of the [\cii] intensity to infrared dust
continuum intensity, [\cii]/FIR,   
versus charge parameter $\gamma=G_0 T^{1/2}/n_{e}$ $({\rm K^{1/2}\, cm^{3}})$.  The curve overlaid on the observations is  
the heating efficiency $\epsilon$ from  \cite{Bakes1994}. Points show
a mix of dense PDRs and diffuse
gas lines-of-sight. OMC1 includes [\oi]. 
See \cite{Salas2019} and \cite{Pabst2021a, Pabst2022} for source descriptions.
Figure adapted with permission from \cite{Pabst2022} \copyright ESO.}
\label{fig:Pabsteps}
\end{figure}

If [\cii] is the dominant coolant and the integrated
far-infrared (FIR) dust continuum is a measure of the incident radiative energy,
then the ratio of [\cii] intensity to FIR intensity 
([\cii]/FIR) provides an observational check of the theoretical 
$\epsilon$. 
However, 
observations of several sources, including dense PDRs and diffuse gas, of [\cii]/FIR 
({\bf Figure~\ref{fig:Pabsteps}}) present a puzzling result \citep{Salas2019,Pabst2022}\footnote{Here FIR is defined as the integrated 40\,\mic-500\,\mic\  continuum. Another measure in use is the total infrared (TIR) 3\,\mic-1100\,\mic\ continuum \citep{Dale2002}. ${\rm TIR}{\approx}2\times{\rm FIR}$ depending on grain temperature.}.  When compared to the BT94 efficiency, observational points fall below the theoretical curve. 
This may be the
result of optical photons contributing to
grain heating that do not heat the gas, a
substantial contribution from [\oi] cooling,
or lower heating rates in dense PDRs where
the PAH distribution is uncertain. 

The photoelectric heating
diminishes with depth into the cloud
mainly due to dust opacity which  
reduces the integrated FUV radiation field.
Reddening of  the spectral energy distribution also reduces $\epsilon$ by ${\sim} 25$\% at $A_{\rm V}=1$ and $\gamma = 10^4$.
For a normally incident
B0 star spectrum and $R_V=3.1$ grains\footnote{The total-to-selective extinction  $R_{\rm V}=A_{\rm V}/(E_{\rm B-V})$ has a typical value of  3.1 in the diffuse ISM while molecular clouds can have larger $R_{\rm V}{\sim}5.5$.}, the FUV field falls 
approximately as $\propto \exp(-1.8 A_V)$.
A detailed treatment of the  FUV radiation transfer \citep{LePetit2006, Rollig2013} and of the
grain charge on a  distribution of grains 
allows for different 
spectral energy distributions, angle of incidence,  and different grain size distributions,
than those used for the analytic expressions
\citep{vanHoof2004}.
A cooler incident spectrum \citep{Spaans1994},
shifts the grain charge to more neutral grains but ultimately produces
less efficient heating since there are fewer high energy photons. \cite{Spaans1994} provide a simple formula that ``corrects"  equation \ref{eq:peheatingeps} for the cooler radiation field and also quantitatively assesses the effect on photodissociation and photoionization.
Increasing $R_{\rm V}$  shifts the grain size distribution to larger grains, which have lower heating efficiency due to grain charging, and lower yield, 
and allows for increased penetration of FUV radiation into the cloud \citep{Weingartner2001, Abel2008}.

 Additional sources of heat come from \HH\ formation,  photodissociation, and collisional de-excitation of excited
 rovibrational levels 
 \citep{Sternberg1989}.  
 The formation of ${\rm H_2}$ on grains releases $4.48$ eV of binding energy. 
The distribution of the energy is uncertain and depends on the formation process. 
For the
 the Eley-Rideal (ER) formation mechanism (see { Section \ref{sec:PDR_chemistry}})
${\sim} 0.6$\,eV  
 goes into kinetic energy of the molecule which can heat the gas,  2.7\,eV goes into internal
 rotational and vibrational excitation, 
 and most of 
 the remainder goes into grain heating
  \citep{sizun2010}. 
 For the Langmuir-Hinshelwood (LH) mechanism
  equipartition is generally assumed with ${\sim} 1.5$\,eV going to the grain, kinetic energy, and 
  excitation. These  ``formation pumped" excited $(v,J)$  levels  can be collisionally de-excited 
  at densities above the critical density
   $n\gtrsim n_{\rm cr}\approx 1.1\times 10^5/\sqrt{T_3}$ ${\rm cm^{-3}}$  where 
   $T_3=T/(1000\,{\rm K})$ \citep{Rollig2006}, and contribute to  gas heating.
  The total formation heating is the sum of the translational heating and de-excitation  of excited levels.  
  The photodissociated \HH\ 
  attains ${\sim}0.5$\,eV in kinetic energy that also adds to gas heating  \citep{Stephens1973, Abgrall2000}.

The ${\rm H_2}$ rovibrational  levels can also be excited by FUV pumping, and contribute to gas heating, called vibrational heating, through collisional  de-excitation of the excited levels (\citealt{Sternberg1989, Burton1990} see also \citealt{Rollig2006} for
  an analytic fit  to the \citealt{Sternberg1995} ${\rm H_2}$ model). 
At the cloud edge,  the formation  and vibration heating 
 can dominate that of photoelectric heating
  \citep{Burton1990,LeBourlot2012,Rollig2013}.
 There, 
 the gas temperature and the FUV field are high and the destruction and formation cycle is rapid
 thereby increasing the formation heating. In addition,
 rates of formation go as $R_{\rm H_2} \propto nn_{\rm H}$ which is always higher at the surface.
  When the density is greater than the critical density for de-excitation, since there are 9 pumps for
every formation, the vibrational heating by FUV pumping tends to dominate.
    At lower densities, the formation (translational) heating can be important.

Cosmic-ray ionization of H, ${\rm H_2}$, and He creates secondary electrons that heat the gas
by similar processes discussed for
X-ray photoelectrons (Section \ref{sec:xdr_heating}). 
Typical
heating rates are $Q {\sim} 10$ eV per ion pair in
diffuse molecular gas and $Q {\sim} 13$ eV 
per ion pair in 
molecular clouds with the heating rate 
per unit volume given by $n\Gamma = \zeta_{\rm p} Q n$
\citep{Glassgold2012}.
For cosmic-ray ionization rates typical of
those found  
in the Galactic disk, 
cosmic-ray heating is less important than photoelectric heating at the cloud surface even for radiation fields as
low as the interstellar field, but can be important in cloud
interiors that are shielded from FUV radiation. 
For  higher cosmic-ray ionization
rates as found in the central molecular zone near the Galactic center 
\citep[${\sim}1-10\times 10^{-14}$ ${\rm s^{-1}}$;][]{LePetit2016}
or supernovae remnants \citep[${\sim}10^{-10}$ ${\rm s^{-1}}$;][]{Priestley2017}, the cosmic-ray heating can dominate
that of photoelectric heating even close to the cloud
surface \citep{Bayet2011}. 

Gas collisions with grains can be
either a heating or cooling process depending on if the gas temperature is lower or higher
than the grain temperature. Grain temperatures are determined from radiative
equilibrium using a full radiation transfer treatment with a distribution of grain sizes \citep[e.g.,][]{Rollig2013}, or with a simple fitted formula
\citep{Hollenbach1991,Hocuk2017}.
The gas is usually warmer than grains on PDR surfaces to $A_{\rm V}{\sim} 3$ but can be cooler than the IR heated grains at greater depths.
Collisional de-excitation of fine-structure levels that are pumped by the infrared continuum radiation can also lead to
gas heating. This is
most likely to occur for the 
[\oi] 63\,\mic\ transition which can become
optically thick to the continuum radiation and is at a wavelength where the dust continuum is strong.

\subsection{Gas Cooling}
\label{sec:PDR_cooling}

In the atomic gas, cooling proceeds mainly through 
collisional excitation and radiative de-excitation of atomic fine-structure levels. The dominant coolants
are [\cii] 158\,\mic\ and 
[\oi] 63\,\mic, and 145\,\mic\ 
 line emission. As ${\rm C^+}$ recombines to C the fine-structure lines of [\ci] 370, 609 $\mu$m become important.
In molecular gas rotational transitions of CO dominate
the cooling \citep[e.g.,][]{Neufeld1993}.
For high density and
high FUV field PDRs, 
[\siii] 35\,\mic\ and ro-vibrational transitions of ${\rm H_2}$ can contribute to gas cooling \citep{Abel2005,Kaufman2006}.
The 
calculation for the level populations are carried out
in statistical equilibrium using non-LTE rate equations. Radiative
pumping, self-absorption, and line
transfer is generally handled using an escape probability formalism. The escape probability depends on $\tau_{\rm lu}$, the cloud geometry, and whether a microturbulent or large velocity gradient (LVG) approach is used. 
The
 line optical depth  integrated from the cloud surface $z=0$ to depth $z$ is given by:
\begin{equation}
    \tau_{\rm lu} = \int_0^z dz 
    \frac{A_{\rm ul} c^3}{8 \pi \nu_{\rm ul}^3}n_{\rm u}\left[ \frac{n_{\rm l}g_{\rm u}}{n_{\rm u}g_{\rm l}}-1\right]
    \frac{1}{\delta v_{\rm D}}\, ,
\end{equation}
where $n_{\rm u}$ and $n_{\rm l}$ are the densities
of the species in the upper and lower levels, and
$\delta v_{\rm D}$ is
the Doppler line width which includes both thermal 
and turbulent broadening.

In a  microturbulent model the velocity gradient 
is small compared to $\delta v_{\rm D}$
and the full column of the species along a line of sight 
contributes to the optical depth.
In a large velocity gradient (LVG)
model (e.g., RADEX, \citealt{vanderTak2007}) the velocity gradient, $dv/dz$, is sufficiently large so that the integration is limited to the 
velocity coherent length $\delta v_{\rm D}/(dv/dz)$. Most PDR models 
adopt a microturbulent approach because the physical size of the emitting
region is narrow and the velocity gradient is small compared 
to the line width.
The $\beta_{\rm esc}$ for  a microturbulent, semi-infinite, plane-parallel layer can be found in TH85. See also 
\cite{Draine2011} and \cite{Tielens2021} for 
a derivation of escape probability formalism used 
in line radiation transfer.
At the cloud surface $\beta_{\rm esc} = 0.5$ 
and decreases as  $1/\tau_{\rm lu}$ at large $\tau_{\rm lu}$.
In spherical geometry, the escape probability in microturbulent gas  is given in  \cite{Stoerzer1996}  and in the LVG limit
in \cite{Goldsmith2012}. 

It is important to use up-to-date collisional excitation 
rates in order to find the gas cooling and emitted
line intensities. The LAMBDA  
 \citep{vanderTak2020} 
 and BASECOL \citep{ba2020} databases are on-going efforts to provide
rates on-line in a standard format. 
The dominant collision partners are usually 
atomic hydrogen near the PDR surface and 
molecular hydrogen at larger depth. Electron impact excitation may also be important for molecular ions \citep{Hamilton2018}.
At intermediate
depth both atomic and molecular hydrogen can contribute to the collisional
excitation rates. 

Several excitation rates have been calculated separately
for collisions with Ortho and Para ${\rm H_2}$.
Collision rates involving Ortho ${\rm H_2}$ are 
generally faster than those for Para ${\rm H_2}$
since the quadrupole interaction averages to zero for Para, 
although the difference is much greater for rotational transitions than for fine-structure transitions. 
Ortho to Para conversion of \HH\ should be accounted for in 
the chemical network \citep[e.g.,][]{Sternberg1999, Bron2016}. \cite{Wiesenfeld2014} note that the collisional rates for \HH\ with ${\rm C^+}$ are
quite close to those for H with ${\rm C^+}$, contrary to the factor of 2 difference expected previously. Impacts of electrons with cations
(e.g., ${\rm C^+}$)
may also be important for sufficiently high electron fraction.
For $T{\sim}300$ K, excitation of ${\rm C^+}$ 
by $e^{{\rm-}}$ and ${\rm H}$  are 
comparable at $x_{e}=\gamma^{\rm H}/\gamma^{e}{\sim}5\times 10^{-3}$. See \cite{Lique2018}
for a recent derivation of the de-excitation coefficients 
of \oi\ by $\rm H_2$, H, and He, but contact the authors
for a corrected data set and extended to 8000 K.

\section{The Physics of XDRs}
\label{sec:xdr}
\subsection{1-D Structure}
The 1-D structure of X-ray dominated regions (XDRs) differs from that of PDRs in two important ways: XDRs are characterized by much larger column densities ($N\approx 10^{22}\rm cm^{-2}$) of warm gas ($T\approx 100-500$ K), and they have a peculiar enhanced abundance of molecular ions. The physical mechanisms causing the different 1-D structure of XDRs vs PDRs are linked to the deep penetration of X-rays and to the production of photoelectrons following primary ionizations. By analogy with PDRs, where the thermal and chemical conditions are determined by $G_0/n$, the XDR structure can be parameterized in terms of $H_{\rm X}/n$, namely the ratio between the energy deposition rate per particle:
\begin{equation}
H_{\rm X} = \int_{E_{\rm min}}^{E_{\rm max}} \sigma_{\rm pe}(E) F_{\rm X}(E) dE,
\label{eq:XDR_en_dep_rate}
\end{equation}
and the gas density $n$. This follows from equating the heating and molecular destruction rates induced by X-ray photons (which are proportional to $n H_{\rm X}$), to the cooling and  molecular formation rates (which are proportional to $n^2$). At equilibrium, the thermal and chemical conditions of XDRs are thus governed by $H_{\rm X}/n$. In equation \ref{eq:XDR_en_dep_rate}, $F_{\rm X}(E)$ is the local photon energy flux per unit energy interval, that is generally assumed to be a power-law, $F_{\rm X}(E)=F_0 (E/ {\rm 1 \,keV})^{-\alpha}$ (see Section~\ref{subsec:PDRXDR_inputparams} for a discussion regarding this and other functional forms), and $\sigma_{\rm pe}(E)$ is the photoelectric absorption cross section per hydrogen nucleus \citep[][]{Morrison1983, Wilms2000}.
Given that $\sigma_{\rm pe}(E)$ goes roughly as $E^{-3}$ \citep[][]{Maloney1996}, the lowest energy photons are attenuated more than higher
energy photons with increasing column density.
The gas is optically thick to X-rays with $E\leq E_0$ where the energy $E_0$ for which $\tau(E)= 1$ depends on the column density. For molecular clouds with  $N\approx 10^{22}\rm \, cm^{-2}$ nearly all photons with $E\lesssim1$ keV are absorbed before the cloud center is reached. Considering a flux $F_{\rm X}$ from $E_{\rm min}=1\,\rm keV$ to $E_{\rm max}=100 \,\rm keV$ the energy deposition rate can be written as $H_{\rm X} = 4.8 \times 10^{-24} F_X/(N/10^{22} {\rm cm^{-2}})^{0.9}$ \citep[see][for the complete derivation]{Maloney1996}. Note its slow decline  with column density 
as opposed to the FUV induced photoelectric heating in PDRs (see Section~\ref{sec:PDR_heating}) that is instead confined in a much thinner layer 
by the exponential attenuation of the FUV flux due to dust.
In XDRs the X-ray  energy that is converted to heat and then re-radiated as line cooling -- i.e., the heating efficiency -- is larger \citep[up to $\epsilon\approx 1$,][]{Maloney1996} than in PDRs \citep[$\epsilon\approx 10^{-2}-10^{-3}$,][and discussion in Section~\ref{sec:PDR_heating}]{Hollenbach1999RvMP}. The chemistry is also affected  up to high column densities (see Section~\ref{sec:xdr_chemistry}) by ion-molecule reactions initiated by X-rays and this results in different emergent line emission and ratios that can be then used to distinguish XDRs from PDRs (see Section~\ref{subsec:XDR_diagnostics}).

\begin{figure}[h]
    \centering
    \includegraphics[scale=0.51]{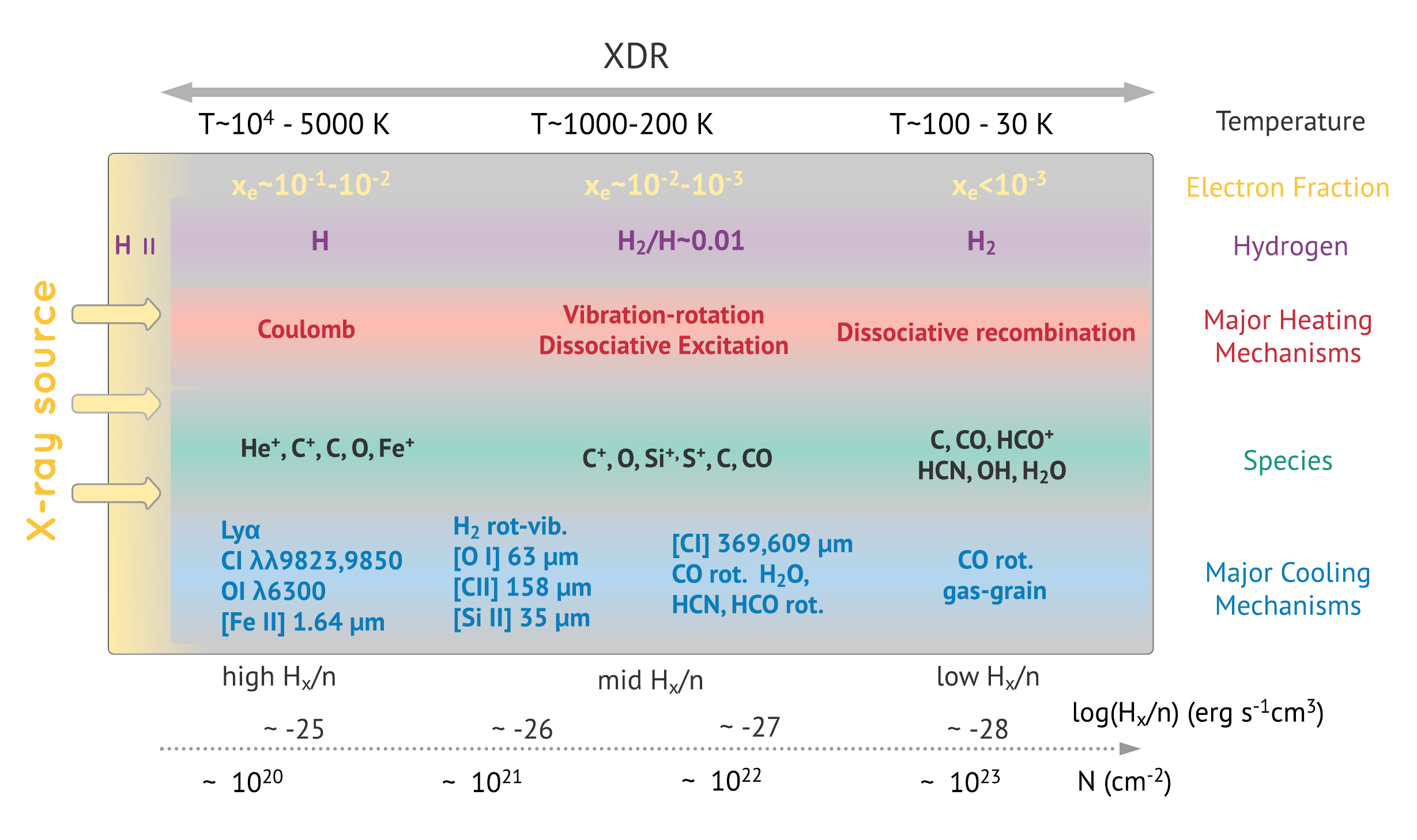}
    \caption{Schematic 1-D structure of an X-ray dominated region (XDR) as a function of the column density $N$. The ratio of the local X-ray energy deposition rate per particle to the total density, \Hxn, decreases with increasing column density~(see Section~\ref{sec:xdr_structure}). We assume no incident FUV photons, a 1-100 keV flux $F_X\approx100 \rm\, erg\,s^{-1}\, cm^{-2}$, and gas density $n=10^5\rm \,cm^{-3}$. The schematic highlights the major heating/cooling processes and the approximate temperature, electron fraction, and chemical composition as a function of depth.}
    \label{fig:XDR_1D}
\end{figure}
In general, a 1-D gas slab ({\bf Figure~\ref{fig:XDR_1D}}) illuminated by a nearby X-ray source with no attenuating foreground gas or dust has an external ionized layer. {\bf Figure \ref{fig:XDR_1D}} spans the $10^{-29}\leq{\rm H_X}/n < 10^{-25}\rm {erg\, s^{-1} \, cm^{3}}$ range extending from substantially ionized gas to cold molecular gas. The corresponding effective ionization parameter ($\xi_{eff}$), which is related to the energy deposition rate as ${\rm H_X}/n = 3.8 \times 10^{-25} \xi_{eff}\rm \, erg\,s^{-1}\, cm^{3}$, ranges from $10^{-4}\leq \xi_{eff} \leq 1$. If the slab is shielded from the direct light of the X-ray source by e.g., a circumnuclear torus or a shadowing gas cloud, then the external \hii~region will not be present. 
The external \hii~region is followed by an FUV-produced PDR only if the ratio of the X-ray flux to the gas density is low and the X-rays do not dominate over FUV heating at low column densities. Note that FUV photons can be produced internally following the various degradation paths of the initial X-ray photon and this is  discussed in detail in  Section \ref{sec:xdr_heating}.
At high column densities, and for high X ray incident flux (so as to dominate cosmic rays) 
regardless of the presence or not of an outer \hii/PDR layer, X-rays completely dominate the heating and the chemical composition, and this is the actual XDR.
A peculiar feature of XDRs is the much less abrupt $\rm C^+/C/CO$  transition as compared to PDRs. This is produced by 
the slow decrease of \Hxn, and by the internally generated FUV photons resulting from collisions with secondary electrons which maintain fairly constant $\rm C^+$ and C abundance until the CO dominates at large column densities $N\approx 10^{22} - 10^{23}\rm \, cm^{-2}$  \citep{Meijerink2005}.

An important consideration regarding the X-ray photoelectric cross-section, $\sigma_{\rm pe}$, is that despite being much less abundant than H and He, heavy elements (C, O, Mg, etc) dominate $\sigma_{\rm pe}$ above $0.5$ keV \citep{Wilms2000}, and hence they are the major sources of photoelectrons. Note that these elements  are readily incorporated into dust grains, but there  remains a significant fraction of C and O (mostly in the form of CO), and noble elements, in the gas phase. As such, the deposition of X-ray energy occurs through absorption by both gas and dust. 
\citet{Bethell2011} provide a simple polynomial fit to the X-ray photoelectric cross-section for a mixture of gas and dust, with specific focus on protoplanetary disks. For energies below 1 keV the gas is the main opacity source while at energies $E>1$ keV the metals in dust grains dominate the total opacity. \citet[][]{Rab2018} also include the contribution of PAHs, concluding that they play a negligible role in the X-ray radiative transfer.
\label{sec:xdr_structure}
\begin{figure}
    \centering
    \includegraphics[scale=0.4]{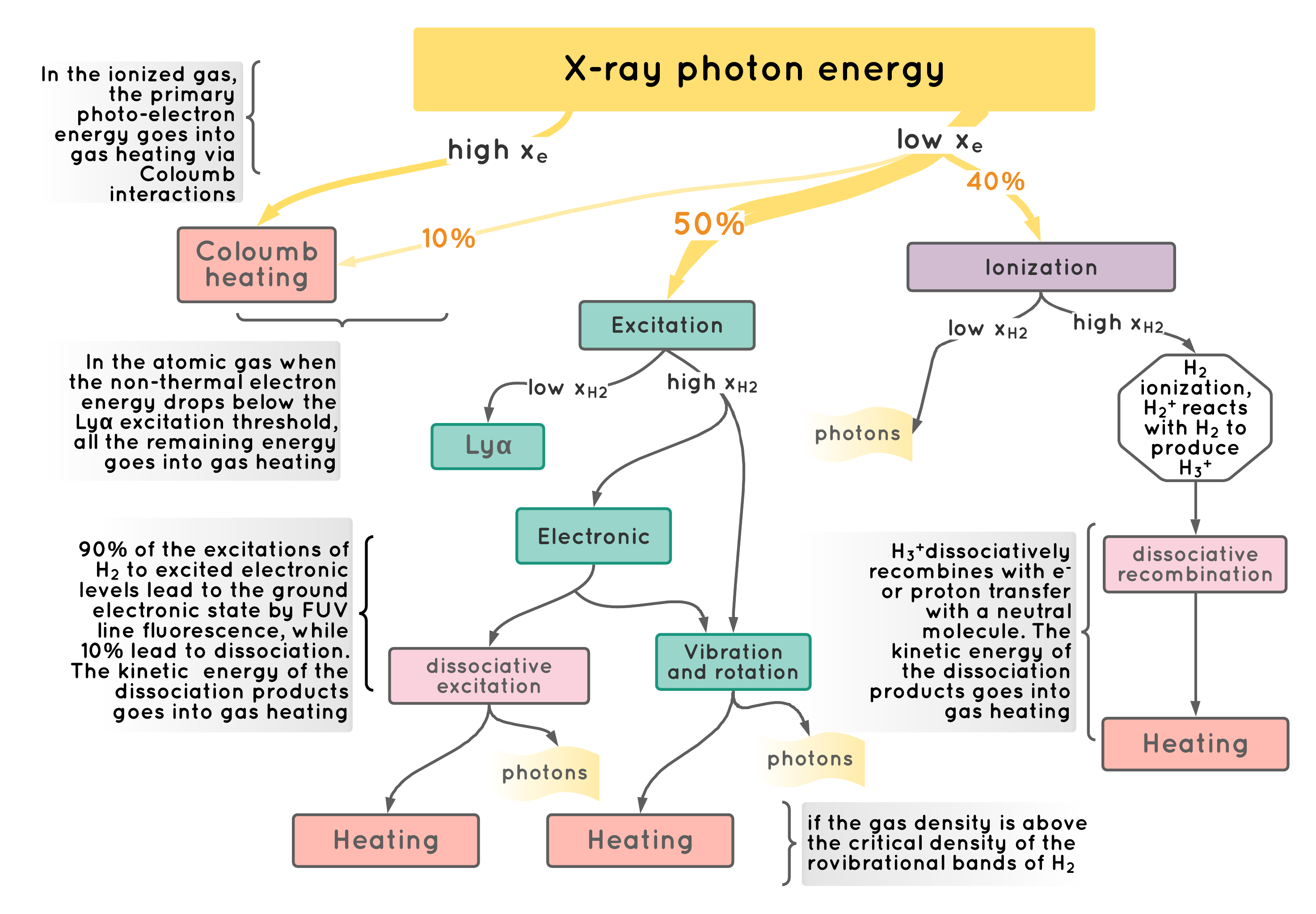}
        \caption{Flow chart of the energy deposition and loss routes in XDRs for the ionized, atomic, and molecular gas. The different mechanisms driving the energy deposition depend on the ionized fraction ($x_{e}$) and \HH~fraction ($x_{\rm H_2}$) of the gas. Figure adapted with permission from \citet{Maloney1996}, \copyright AAS.}
    \label{fig:XDR_heating}
\end{figure}

\subsection{Gas Heating}
\label{sec:xdr_heating}
The XDR heating is produced by the degradation of the incident X-ray photon energy through several channels.
The first step is the production of primary photoelectrons and secondary electrons as follows:
The X-rays ionize heavy elements preferentially by removing a K-shell electron. The vacancy is then filled by a cascade of radiative (fluorescent) and non-radiative Auger transitions.
These primary photoelectrons are typically energetic enough ($E\sim$ keV) to induce secondary ionizations resulting in the ejection of (non-thermal, secondary) electrons ($E\sim$ 8 eV) which, in turn, play a key role in all the subsequent heating processes. Given that each primary photoelectron loses about 20-30 eV per H ionization  it produces  $\approx35$ secondary ionizations in addition to the initial ionization by the  X-ray photon. The next steps involve the
interactions of the primary and secondary electrons
with the gas
(see flowchart in {\bf Figure \ref{fig:XDR_heating}}).
\citet{Glassgold1973, Cravens1978, Dalgarno1999}, and \cite{Glassgold2012} carried out an extensive analysis of the energy loss of primary and secondary electrons in a H, He, $\rm H_2$ mixture. They derived the heating produced by elastic collisions with ambient thermal electrons (Coulomb heating), along with a detailed treatment of the dissociation, vibrational and rotational heating, and an in-depth analysis of the \emph{chemical heating} i.e., that resulting from exothermic dissociative recombination reactions between electrons and molecular ions. Each heating mechanism dominates in different conditions depending on the electron and \HH~fractions ($x_{e}$, and $x_{\rm H_2}$). The first branching is set by $x_{e}$ (see {\bf Figure \ref{fig:XDR_heating}}): if the electron fraction is relatively high ($x_{e}> 0.1$) nearly all the primary photoelectron energy goes into heating through Coulomb interactions between the secondary electrons and the ambient thermal electrons \citep[][]{Swartz1971, Dalgarno1999}.
By contrast,  in the low-ionization limit ($x_{e}\ll 1$) only $\approx10\%$ of the primary photoelectron energy goes into Coulomb heating while $\approx50\%$ is expended in excitation processes, and the remainder $40\%$ in ionization processes \citep{Glassgold2012}. 
To summarize, the XDR heating due to Coloumb interactions can be parameterized in terms of the energy deposition rate $H_{\rm X}$, as $n\Gamma_{\rm C} = \eta_{\rm c}\, n\, H_{\rm X}$, where $\eta_{\rm c} =\eta_{\rm c} (x_{e}, x_{\rm H_2})$ is the Coloumb heating efficiency \citep{Shull1985, Dalgarno1999, Meijerink2005, Glassgold2012}.
The excitation and/or ionization heating are instead mainly influenced by the \HH/\hi\ ratio  because $\rm H_2$, having a wide variety of energy levels, offers more channels for energy loss than those of pure atomic hydrogen. In atomic gas ($x_{\rm H2}\ll 1$), the secondary electrons collisionally excite Lyman-$\alpha$ \citep{Shull1985} and once their energy drops below $E_{Ly\alpha}=10.2\, \rm eV$, all the remaining electron energy goes into heating through elastic scattering with H and thermal electrons \citep{Dalgarno1999}. In this case the X-ray heating efficiency is only 12\%, whereas in molecular gas it can be up to 50\% at very high densities \citep{Glassgold2012}.
 If the \HH~fraction is high, $x_{\rm H2} \approx 0.5$, the initial energy is deposited as heat into the gas through several processes. Among the possible mechanisms there is the excitation of the rovibrational levels of the $\rm H_2$ molecules and the electronic excitation of \HH~followed by fluorescence to $v, J$ states. If the gas density exceeds the critical densities of such transitions ($n\gtrsim10^4 \rm \, cm^{-3}$), then the collisional de-excitation \citep{Tine1997} results in net heating \citep{Meijerink2005, Glassgold2012}. Additionally, heating is produced by the $\rm H_2$ excitation to dissociative states \citep{Dalgarno1999, Glassgold2012} which injects energetic H atoms that then thermalize. 
 A peculiarity of XDRs is the abundant presence of $\rm H_2^+$ produced in $\rm H_2$ secondary ionizations.
 These molecular ions 
  react with $\rm H_2$ to produce $\rm H_3^+$. The $\rm H_3^+$ may further undergo an exothermic dissociative  recombination with an electron or a proton-transfer reaction as e.g., in $\rm H_3^+ + O \rightarrow  OH^+ + H_2$.  The molecular ions can undergo
 an exothermic dissociative recombination thus adding to gas heating. 
 If the density is high enough for collisional de-excitation of 
 vibrationally excited \HH, then in the limit of  $x_{e}\approx 0$ and $x_{\rm H_2}\approx 1$, the total maximum heating is Q=18.7 eV per ion pair.

 \subsection{Gas Cooling}
\label{sec:xdr_cooling}
\begin{figure}[ht]
\begin{minipage}[c]{0.5\textwidth}
\includegraphics[width=\textwidth]{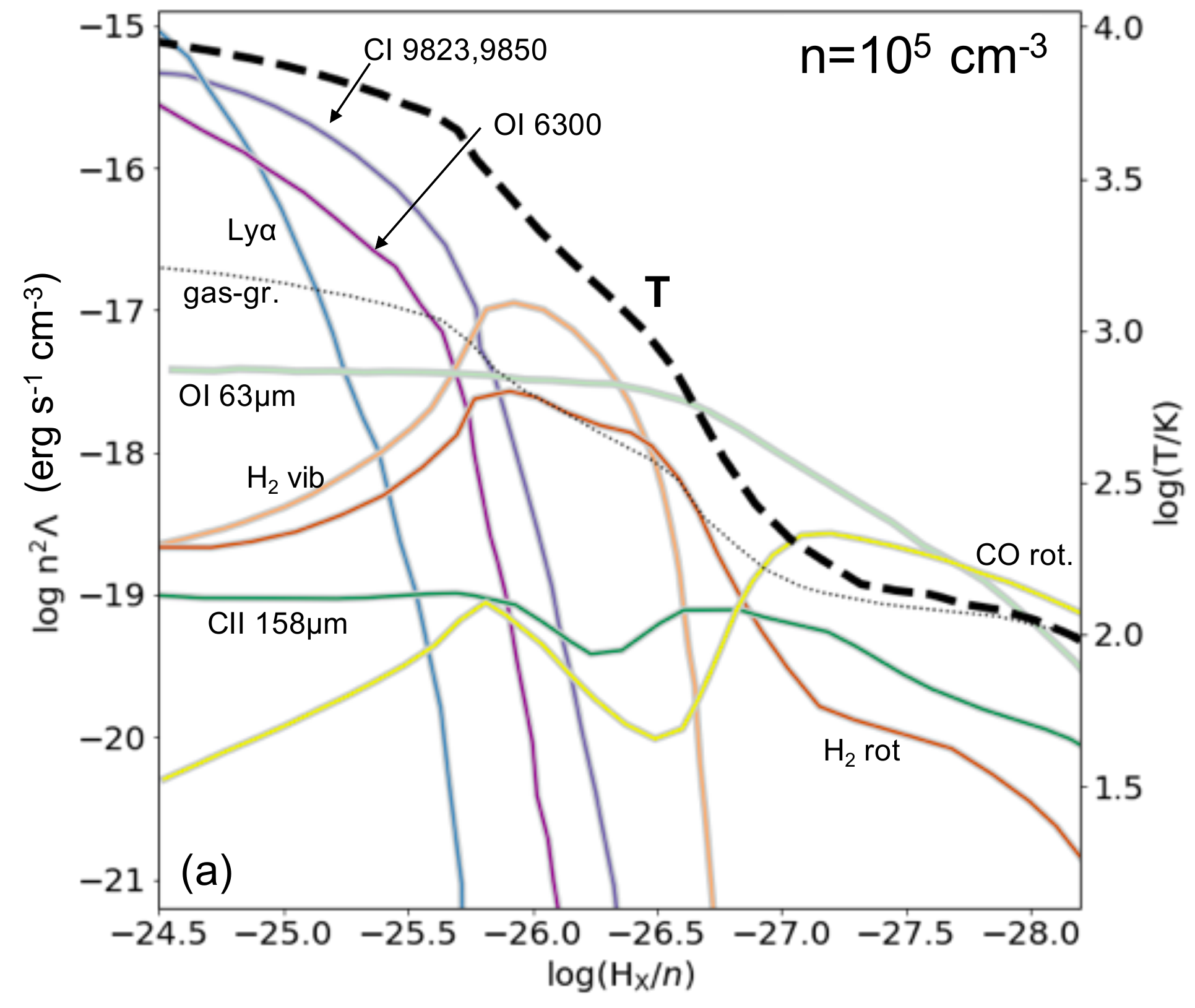} 
\end{minipage}\hfill
\begin{minipage}[c]{0.47\textwidth}
\includegraphics[width=\textwidth]{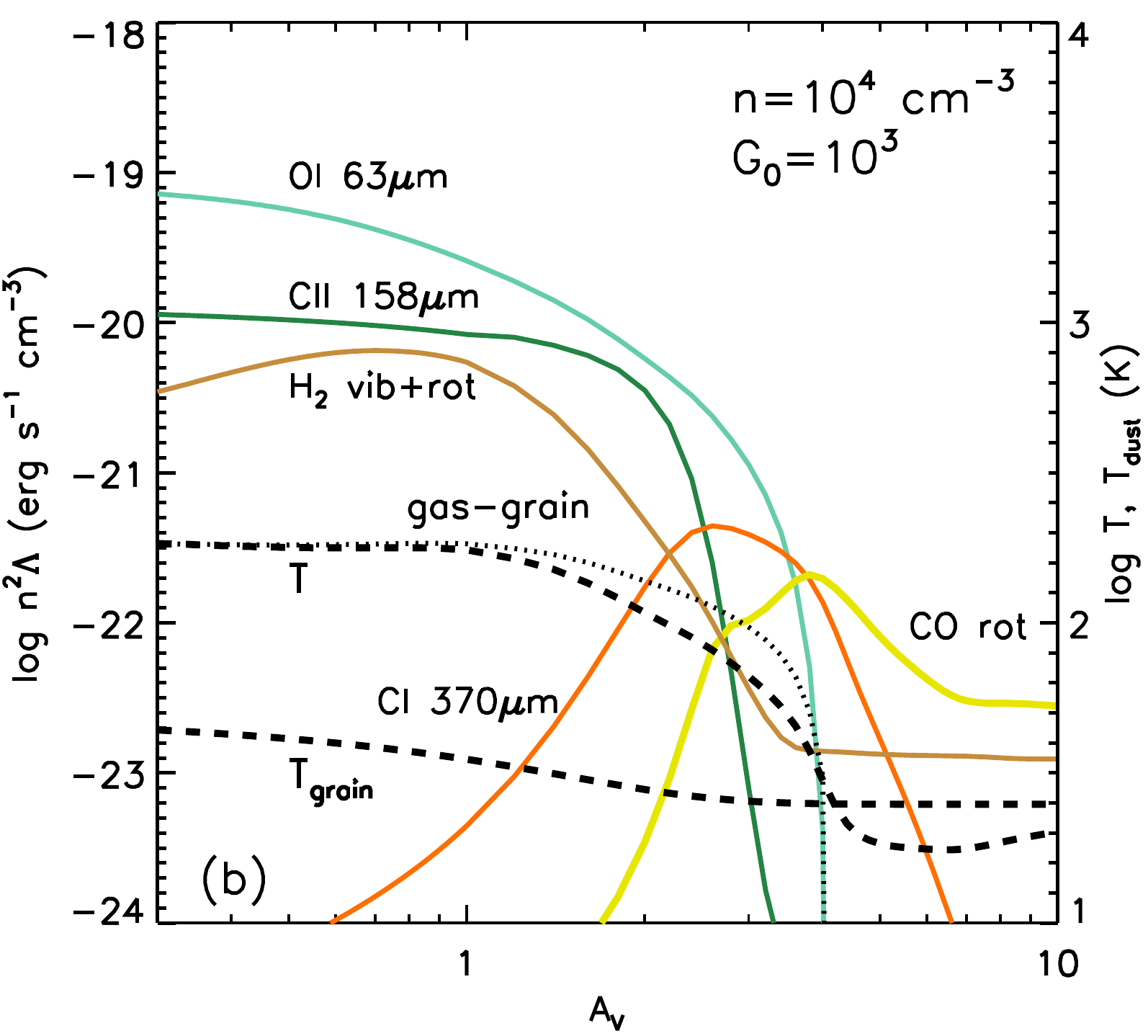}
\end{minipage}
\caption{($a$) Gas temperature profile (dashed line) of a typical XDR characterized by a gas density $\log(n/{\rm cm^{-3}})=5$ as a function of \Hxn. The major XDR gas coolants as a function of \Hxn~ are plotted with colored lines. Figure adapted with permission from \citet{Maloney1996}, \copyright AAS. ($b$) Gas and grain temperatures as  a function of $A_{\rm V}$ are
 shown as dashed lines for a PDR with $\log(n/{\rm cm^{-3}})=4$ and $G_0=10^3$. Major PDR gas coolants are indicated with colored lines. The gas temperatures are generally higher than the grain temperatures  and  grains
 act to cool the gas. At large depth  warm grains can heat the gas. PDR models from \url{https://dustem.astro.umd.edu} with freeze out
 turned off.
 }
 \label{fig:XDR_cooling}
 \vspace*{-1em}
\end{figure}

The deep penetration of X-ray photons heats the gas to high column densities. In thermal
equilibrium, the heating is balanced by cooling, and the
neutral and molecular gas stays warm to high column densities with the heating, cooling and temperature a function of \Hxn~(see {\bf Figure \ref{fig:XDR_cooling}a}. Also shown for comparison is a typical PDR in {\bf Figure \ref{fig:XDR_cooling}b}). 
One peculiarity of XDR cooling compared to PDRs is the overall higher ratio of the energy radiated as line emission over that emitted as infrared continuum \citep{Maloney1996}.
That is, gas cooling/grain cooling is higher than in PDRs.
While in PDRs almost all the FUV photon energy absorbed by dust is re-irradiated in the infrared, in XDRs about half of the energy is deposited into the gas and thus the cooling lines can carry a large fraction of the total deposited energy \citep{Maloney1996, Meijerink2007}.
For high \Hxn, the predominantly ionized gas ($x_e \gtrsim 0.1$)
is characterized by $T\approx 10^4\, \rm K$, as sketched in {\bf Figure \ref{fig:XDR_1D}} and shown by the temperature profile in {\bf Figure \ref{fig:XDR_cooling}a}. Thermal collisions can thus excite Ly$\alpha$ \citep[e.g.,][]{Sternberg1989}, and forbidden lines such as [\ci] $\lambda\lambda$9823,9850~\citep[e.g.,][]{Escalante1991}, and [\oi] $\lambda\lambda$6300,6363~ \citep[e.g.,][]{Stoerzer2000} that dominate the cooling.
The [\feii] 1.26 and 1.64\,\mic\ lines are also efficient coolants for high X-ray illumination, temperature, and density as the upper state of the 1.64\,\mic\ line lies about $10^4 \rm \, K$ above the ground state.
As \Hxn~decreases, the $\rm H_2$ abundance increases while the gas temperature remains warm ($T\approx 1000 \, \rm K$). These conditions favor the excitation of $\rm H_2$ rovibrational transitions that significantly contribute to the cooling \citep{Sternberg1989, Neufeld1993, LeBourlot1999, Spaans2008, Glover2008, Lique2015}.
The range of \Hxn~values for which vibrational lines dominate the cooling are rather narrow because  the first two vibrational levels ($v=1$ and $v=2$)  lie at $\approx 6000\rm \, K$ and $\approx 12000\rm \, K$ above the ground state, respectively \citep[see Fig.\ 3 in][]{Shaw2005}.
In particular, for $T>2000\,\rm  K$ the $\rm H_2$ 1-0 S(1) 2.122\,\mic\ transition significantly contributes to the total cooling, while once the temperature falls below $T\approx1000 \rm \, K$ the $\rm H_2$ cooling is dominated by rotational lines in $v=0$: 0-0 S(0) 28.22\,\mic, 0-0 S(1) 17.03\,\mic, 0-0 S(2) 12.28\,\mic, and 0-0 S(3) 9.66\,\mic.\footnote{
The \HH\ spectroscopic notation reads as follows: the first two numbers refer to the vibrational level transition, the letter indicates the branch (S corresponding to transitions between rotational states with $\Delta J=+2$), and the number in parenthesis is the rotational quantum number of the final state. For instance 1-0 S(1) stands for: ($v=1,J=3$) 
$\rightarrow$ ($v=0,J=1$).}
A recent analytic approximation of the $\rm H_2$ cooling function can be found in \citet{Moseley2021}.

Deeper into the XDR the temperature decreases ($T \lesssim 500\rm \, K$) and the [\oi] 63\,\mic, [\siii] 35\,\mic, and [\cii] 158\,\mic\ lines become the major coolants \citep{Maloney1996, Meijerink2007}. The [\oi] 63\,\mic\ cooling ($n_{\rm cr}\approx 5\times 10^5$
 cm$^{-3}$, $\Delta E/k=228\rm \, K$) is particularly efficient because warm temperatures ($T\gtrsim100$) remain at high column densities where the oxygen is neutral. XDRs are thus characterized by high [\oi]/[\cii] ratios ($\gtrsim 10$) as compared to values $\lesssim 10$ in PDRs \citep{Maloney1996, Hollenbach2009b}.  
XDRs also feature  high [\siii]/[\cii] ratios \citep{Meijerink2007} reflecting the high $\rm Si^{+}$ abundance that results both from X-ray induced secondary ionizations and from those produced by Lyman and Werner photons following  \HH~excitation. 
As \Hxn~declines further, the neutral carbon [\ci] 370\,\mic\ and 609\,\mic\ fine-structure transitions become important coolants \citep[][]{Bisbas2021}. Enhanced [\ci] emission as compared to PDRs is another peculiarity of XDRs as abundant C coexists with C$^+$ through a thick layer in XDRs (see Section~\ref{sec:xdr_structure}).
In the same range of \Hxn~significant cooling is produced by high-$J$ CO rotational transitions which are exceptionally bright in XDRs, boosted by the warm ($T\approx 200\rm \,K$) temperature at high column densities maintained by X-ray penetration \citep{Meijerink2007,Spaans2008}. 
By contrast, in typical PDRs most of the CO is present beyond the H/\HH~transition at much lower temperatures ($T\approx 20$ K), which do not allow for bright high-$J$ emission. Other coolants of the warm molecular gas in XDRs are $\rm H_2O$, $\rm HCN$, $\rm OH$ rotational lines.  
\citet{Goldsmith2017} found that electron excitation of HCN, HCO$^+$, CN, and CS might be important. This might be especially relevant when using HCN/HCO$^+$ ratios as a diagnostic for inferring the presence of XDRs in external galaxies (see Section~\ref{subsec:XDR_diagnostics}).
In the lowest \Hxn~regime the gas-grain collisions become an important cooling mechanism as the gas is warmer than the dust due to the effect of X-ray penetration and can therefore be cooled by collision with lower temperature dust grains \citep{Maloney1996}. 

\subsection{Chemistry}
X-ray dominated regions are characterized by a peculiar chemistry initiated by the secon\-dary ionizations resulting from the ejection of primary photoelectrons, and by the effect of internally generated Lyman and Werner photons arising from the decay of \HH~excited states. These processes are generally more important than the effect of direct X-ray ionizations and dissociations. Numerous studies have explored the effects of X-ray irradiation on the chemistry of the ISM \citep{Maloney1996, Meijerink2005, Stauber2005} assuming steady-state for the temperature and the abundances of the different species. \citet{Maloney1996} considered a chemical network focused primarily on carbon and oxygen,  whereas the \citet{Meijerink2005} network included all species with sizes up to 3 atoms and some of 4 atoms and adopted the UMIST database for the chemical reaction rates. The chemical signature of X-ray induced processes can be retrieved from the abundances of key species. We highlight below and in {\bf Table \ref{tab:XDR_chemistry}} the major formation/destruction pathways that differ from PDRs.

The Hydrogen and Helium chemistry can be summarized as follows: $\rm H_2^+$ (He$^{+}$), resulting from \HH~(He) collisional ionization produced by primary electrons trigger an X-ray characteristic chemistry which is similar to that induced by cosmic rays. These ions, whose abundance is particularly enhanced in XDRs, together with $\rm H_3^+$ resulting from reactions of \HH~with $\rm H_2^+$, efficiently exchange charge with neutral constituents (e.g., O, CO, OH, $\rm H_2O$) producing (molecular) ions.

The O-bearing molecular ions (OH$^+$, $\rm H_2O ^+$ and $\rm H_3O^+$) are overall enhanced in gas irradiated by X-rays \citep{Maloney1996, Meijerink2005}. Among them, $\rm OH^+$ is produced either through the reaction of atomic oxygen with $\rm H_3^+$, or between ionized oxygen and \HH. Note that the presence of ionized oxygen where the \HH~is abundant is due to charge transfer reactions between O and H$^+$ following X-ray induced ionizations of H. 
OH$^+$ can  also be formed in reactions between He$^+$ and $\rm H_2O$. The OH$^+$ forms $\rm H_2O^+$ and, subsequently, $\rm H_3O^+$  by reactions with \HH. For $\rm H_3O^+$, models predict an order of magnitude greater abundance in XDRs than in PDRs. 
The $\rm H_3O^+$/$\rm H_2O$ ratio is as large as $10^{-2}$ in XDR models, while the ratios in PDRs are generally $10^{-3}$ or less \citep{vanderTak2008}. This enhancement has been observed in relation to a strong X-ray illumination in the center of AGN host galaxy NGC\,1068 \citep[e.g.,][]{Aalto2011}.

 A  feature of carbon chemistry in  XDRs, which is also shared with environments characterized by high cosmic-ray fluxes \citep{Bisbas2021}, is the lack of the well defined C$^+$/C/CO stratification that characterizes PDRs \citep{Meijerink2005,Meijerink2007}. The C$^+$ and C coexist through the XDR layers because the internally produced FUV photons cause CO dissociation and C ionization much deeper than in PDRs. In XDRs there is another viable path for the production of C$^+$ through charge exchange reactions involving He$^+$ which, for high X-ray fluxes, dominates over the carbon photoionization induced by FUV photons. The abundance of CO$^+$ is enhanced in regions affected by strong UV and X-ray radiation fields \citep{Wolfire1995b}. At low $T$,  C is destroyed by reactions with abundant molecular ions such as $\rm HCO^+$ and  $\rm HOC^+$ which  are enhanced by the reactions of $\rm H_2^+$ and $\rm H_3^+$ with CO, and of C$^+$ with $\rm H_2O$  \citep{Lepp1996}. 
\label{sec:xdr_chemistry}
\begin{table*}[]
\centering
\caption{Major formation (left column) and destruction (right column) pathways for major chemical species in PDRs and XDRs.}
\label{tab:XDR_chemistry}
\resizebox{\textwidth}{!}{%
{\Huge
\begin{tabular}{c|l|l|}
\cline{2-3}
 &
  \multicolumn{1}{c|}{\textbf{\cellcolor{table-green}\color{table-text-green}MAJOR FORMATION PATHWAYS}} &
  \multicolumn{1}{c|}{\textbf{\cellcolor{table-orange}\color{table-text-red}MAJOR DESTRUCTION PATHWAYS}} \\ \cline{2-3} 
 &
  \multicolumn{2}{c|}{\textbf{\cellcolor{table-blue}Hydrogen and Helium chemistry}} \\ \hline
\multicolumn{1}{|c|}{H$_2$} &
  \begin{tabular}[c]{@{}l@{}}- Formation on dust grains\\ - Gas phase reactions (e.g. H$^-$ + H $\rightarrow$ \HH + $e^-$)\end{tabular} &
  \begin{tabular}[c]{@{}l@{}}- {\bf Secondary ionizations:} H$_2$+ $e^-\rightarrow$ H$_2^+$ + 2$e^-$\\ - Reaction with H$_2^+$ or O$^+$\\ - Reactions with O when T$>3000$ K\end{tabular} \\ \hline
\multicolumn{1}{|c|}{\cellcolor{table-yellow}H$_2^+$} &
  - {\bf Secondary ionizations:} H$_2$+ $e^-$ $\rightarrow$ H$_2^+$ + 2$e^-$ &
  - Reaction with \HH~to form H$_3^+$ \\ \hline
\multicolumn{1}{|c|}{\cellcolor{table-yellow}H$_3^+$} &
  - Reaction of H$_2^+$ with \HH &
  - H$_3^+$ + X $\rightarrow$ HX$^+$ +\HH~ where X is O, CO, OH, H$_2$O \\ \hline
\multicolumn{1}{|c|}{\cellcolor{table-yellow} He$^+$} &
  - {\bf Secondary ionizations:} He+ $e^-$ $\rightarrow$ He$^+$ + 2$e^-$ &
  - Charge exchange reactions: He$^+$ + XY $\rightarrow$ He + X + Y$^+$ \\ \hline
 &
  \multicolumn{2}{c|}{\textbf{\cellcolor{table-blue}Oxygen chemistry}} \\ \hline
\multicolumn{1}{|c|}{O$^+$} &
  - Charge exchange of O with H$^+$ &
  - Reaction with \HH: O$^+$ + \HH $\rightarrow$ OH$^+$ + H \\ \hline
\multicolumn{1}{|c|}{\cellcolor{table-yellow} OH$^+$} &
  \begin{tabular}[c]{@{}l@{}}- Ionized oxygen reaction with \HH: O$^+$ + \HH $\rightarrow$ OH$^+$ + H\\ - Neutral oxygen reaction with H$_3^+$: O + H$_3^+$ $\rightarrow$ OH$^+$ + \HH\\ - {\bf H$_2$O reaction with He$^+$}: He$^+$ + H$_2$O $\rightarrow$ He + OH$^+$ + H\end{tabular} &
  \begin{tabular}[c]{@{}l@{}}- Hydrogen abstraction: OH$^+$ + \HH $\rightarrow$ H$_2$O$^+$ + H \\ - Dissociative recombination: OH$+$ + $e^-$ $\rightarrow$ O + H\end{tabular} \\ \hline
\multicolumn{1}{|c|}{OH} &
  \begin{tabular}[c]{@{}l@{}}- H$_2$O + $e^-$ $\rightarrow$ OH + H\\ - Neutral-neutral reaction (high temperature): O + \HH $\rightarrow$ OH\\ - Photodissociation of H$_2$O {\bf by internally generated FUV photons}\end{tabular} &
  - At high temperature OH reacts with \HH~to form H$_2$O \\ \hline
\multicolumn{1}{|c|}{\cellcolor{table-yellow} H$_2$O$^+$, H$_3$O$^+$} &
  \begin{tabular}[c]{@{}l@{}}- Hydrogen abstraction\\ - Oxygen reaction with H$_3^+$\\ - H$_2$O ionization {\bf by internally generated FUV photons}\\ - HCO$^+$ + H$_2$O $\rightarrow$ H$_3$O$^+$\\ - H$_2$O$^+$ + \HH $\rightarrow$ H$_3$O$^+$\end{tabular} &
  - Dissociative recombination of H$_3$O$^+$ with $e^-$ \\ \hline
\multicolumn{1}{|c|}{H$_2$O} &
  \begin{tabular}[c]{@{}l@{}}- Recombination of H$_3$O$^+$ with $e^-$\\ - At high temperature OH reactions with \HH\end{tabular} &
  \begin{tabular}[c]{@{}l@{}}- Reaction with HCO$^+$ and H$_3^+$ for T$>$100 K\\ - At high temperature, reaction with H atoms\\ - Dissociation by FUV photons\end{tabular} \\ \hline
\multicolumn{1}{l|}{} &
  \multicolumn{2}{c|}{\textbf{\cellcolor{table-blue}Carbon chemistry}} \\ \hline
\multicolumn{1}{|c|}{\cellcolor{table-yellow}C} &
  - Photodissociation of CO by {\bf internally generated FUV photons} &
  \begin{tabular}[c]{@{}l@{}}- Photoionization by {\bf internally generated FUV photons}\\ - Reaction with O$_2$ and HCO$^+$, at T$<100$ K\end{tabular} \\ \hline
\multicolumn{1}{|c|}{C$^+$} &
  \begin{tabular}[c]{@{}l@{}}- C ionization by internally generated FUV photons\\ - {\bf CO reactions with He$^+$:} He$^+$ + CO $\rightarrow$ He + C$^+$ + O\end{tabular} &
  \begin{tabular}[c]{@{}l@{}}- C$^+$ + $e^-$ $\rightarrow$ C\\ - C$^+$ + H$_2$O\end{tabular} \\ \hline
\multicolumn{1}{|c|}{\cellcolor{table-yellow}CO$^+$} & \begin{tabular}[c]{@{}l@{}}
  -  Electron impact ionization of CO\\ - {\bf CO$_2$ reaction
with He$^+$} \\ - Reactions of C$^+$ with OH and O$_2$,
respectively \end{tabular} &
 - Reactions of CO$^+$ with \HH \\ \hline
\multicolumn{1}{|c|}{\cellcolor{table-yellow}HCO$^+$} &
  \begin{tabular}[c]{@{}l@{}}- Reaction of CO with H$_3^+$\\ - Reaction of C$^+$ with H$_2$O\\ - Reaction of CO$^+$ with \HH\end{tabular} &
  \begin{tabular}[c]{@{}l@{}}- Reaction with H$_2$O\\ - Electron recombination to form CO + H and OH + C\end{tabular} \\ \hline
\multicolumn{1}{l|}{} &
  \multicolumn{2}{c|}{\textbf{\cellcolor{table-blue}Sulfur chemistry}} \\ \hline
\multicolumn{1}{|c|}{S$^+$} &
  - Photoionization of S due to the {\bf internally generated FUV photons } &
  - Reaction  with \HH, OH, O$_2$ \\ \hline
\multicolumn{1}{|c|}{\cellcolor{table-yellow}S$^{2+}$} &
  - X-ray photoionization &
  - Reaction with \HH \\ \hline
\multicolumn{1}{|c|}{\cellcolor{table-yellow}SO$^{+}$ SH$^{+}$} &
  \begin{tabular}[c]{@{}l@{}}- Reaction of S$^+$ with \HH\\ - Reaction of S with H$_3^+$ and HCO$^+$\\ - Reaction of S$^{2+}$ with \HH\\ - Reaction of S$^{+}$ with OH and O$_2$\end{tabular} &
  - Electron recombination reactions \\ \hline
\multicolumn{1}{l|}{} &
  \multicolumn{2}{c|}{\textbf{\cellcolor{table-blue}Nitrogen chemistry}} \\ \hline
\multicolumn{1}{|c|}{N$^+$} &
  \begin{tabular}[c]{@{}l@{}}- {\bf X-ray ionization and secondary ionizations} of atomic N\\ - {\bf N$_2$ reaction with He$^+$}: He$^+$ + N$_2$ $\rightarrow$ He + N$^+$ + H\end{tabular} &
  \begin{tabular}[c]{@{}l@{}}- N$^+$ reactions with \HH~to produce NH$^+$\end{tabular} \\ \hline
\multicolumn{1}{|c|}{\begin{tabular}[c]{@{}c@{}}HCN\\ HNC\end{tabular}} &
  \begin{tabular}[c]{@{}l@{}}- Dissociative recombination of HCNH$^+$ \\ (with approximately equal branching between HCN and HNC)\\ - Reaction of CN with \HH\end{tabular} &
  \begin{tabular}[c]{@{}l@{}}- HNC + H $\rightarrow$ HCN + H (when T$>200$ K)\\ - Reactions with H$_3^+$, H$_3$O$^+$, HCO$^+$\end{tabular} \\ \hline
\end{tabular}%
}}
\begin{tabnote}
Processes which are peculiar to XDRs are highlighted in boldface, species that are enhanced in XDRs are highlighted in yellow. References: \cite{Maloney1996}; \cite{Yan1997}; \cite{Stauber2005}; \cite{Meijerink2005} \cite{Abel2008} \cite{Notsu2021}.
\end{tabnote}
\end{table*}

The sulfur chemistry is also affected by X-rays because the internally produced FUV field dissociates SH, followed by direct X-ray photoionization of sulfur, which enhance S$^{2+}$ column densities with respect to PDRs. 
 Reactions of S$^+$ with OH, and $\rm H_3^+$ or HCO$^+$ with OH produce SO$^+$ and SH$^+$ respectively, which are also particularly abundant in XDRs. As outlined by \citet{Abel2008} and \citet{Godard2012}, SH$^+$ can be formed also via another channel involving S$^{2+}$ and \HH. If the branching ratio of S$^{2+}$ + \HH $\rightarrow$ SH$^{+}$ exceeds 1\%, compared to other reaction products, then the double ionized chemistry will be the dominant pathway to SH$^{+}$.
 
 X-ray induced FUV photon production and the presence of He$^+$ in XDRs influence the nitrogen chemistry by  $\rm N_2$ dissociation.
 The atomic nitrogen initiates a series of reactions which ultimately enhance HCN abundance. In particular N reacts with OH to form NO, which subsequently reacts with C to form CN. HCN and HNC are formed in almost equal amounts through the dissociative recombination of HCNH$^+$, and through reactions of CN with \HH. The abundance difference between HCN and HNC are thus largely determined by the selective destruction pathways of HNC which have relatively high activation barrier. For this reason low HNC-to-HCN line ratios generally indicate warm temperatures \citep[e.g.,][]{Hacar2020}.
 In XDRs, for high \Hxn, HNC/HCN is lower than in PDRs. At high column densities, where \Hxn~is low (and so is the temperature), HNC/HCN ratios are equal or somewhat higher than those of PDRs \citep{Meijerink2005}.
   
  The most recent and comprehensive discussion concerning the physics and chemistry of water in star forming regions \citep{vanDishoeck2021} is also relevant for those interested in XDR chemistry because  the impact of X-ray irradiation on the H$_2$O abundance is also addressed. 
  In astrophysical environments three pathways lead to $\rm H_2O$  \citep[see Fig.\ 4 in][and references therein]{vanDishoeck2021}. The first route is through neutral-neutral reactions $\rm O+H_2 \rightarrow OH + H$, $\rm OH + H_2 \rightarrow H_2O +H$ that contribute to the water formation only at high temperatures  \citep[$T\gtrsim250$\,K;][]{vanDishoeck2013}, the second route involves ion-molecule reactions \citep[e.g.,][]{Hollenbach2009, Stauber2005} that are effective in diffuse and moderately ionized gas, and the third route is by grain surface reactions \citep{Notsu2021} because at low $T\approx 10-30$ K water is efficiently formed by hydrogenation of oxygen atoms sticking onto dust grains \citep{Cuppen2010}. Thus, accounting for thermal and non-thermal gas-grain interactions and for grain-surface reactions is fundamental \citep{Meijerink2012, Notsu2021}. 
  
  Gas-phase destruction of molecules by X-ray chemistry and X-ray-induced photodesorption are important processes as well \citep[e.g.,][]{Dupuy2018, Notsu2021}. Gas-phase water is mainly destroyed by ion-molecule reactions (with X-ray boosted HCO$^+$, H$^+$, $\rm H_3^+$, He$^+$) and by X-ray induced photodissociation \citep[to H+OH;][]{Meijerink2012}. This might explain the low gas-phase abundance of warm water in the inner regions of protostellar envelopes \citep{Notsu2021}. \citet{Meijerink2012} instead, studied the possibility of enhanced gas phase $\rm H_2O$ abundance in X-ray exposed environments using (bare) carbonaceous dust grains as a catalyst. Outside the
snowline, the gas-phase abundance of water is likely increased
by X-ray photodesorption from icy grains, although
 results are very sensitive to the photodesorption rates assumed \citep[see][which acknowledge that they might be overestimated]{Notsu2021}.

 Thermal equilibrium and chemical balance are often assumed when comparing PDR vs XDR che\-mi\-cal composition but time-dependence and non-equilibrium conditions are often relevant in many astrophysical environments. For instance, short-term X-ray flaring is common in young solar mass stars, and AGN activity varies due to strong fluctuations in the super massive black hole accretion rate. \citet{Meijerink2013} included time dependence in their XDR code and found that tracers such as the HCO$^+$/HCN ratio (see Section~\ref{subsec:XDR_diagnostics}) are strongly time-dependent \citep[see also][]{Harada2013}. Strong evolutionary trends, occurring over time scales $0.01 -100$ Myr are also found for $\rm H_3O^+$, CO, and $\rm H_2O$. These species reflect time dependent effects in the ionization balance, the transient nature of the production of molecular gas, and the freeze-out/sublimation of water (key to much of the grain surface chemistry), respectively. \cite{Viti2017} also addressed time-dependence in XDR-like environments with UCLCHEM, focusing on the four most observed species (CO, HCN, HCO$^+$, and CS). Among them, HCN is the most affected by time-dependence followed by CS. 
 Recently, \citet{Waggoner2019} addressed the effect of time-dependence on water chemistry reporting a significant but short-lived (days) boost in gas-phase $\rm H_2O$ abundance. \citet{Mackey2019} presented a non-equilibrium XDR chemistry calculation using a simplified chemical network of 17 species to study the time-dependent response of a molecular cloud to X-ray flares, emphasizing the faster destruction of CO by an internally generated FUV field as compared to \HH. 
 Using KROME \citep{Grassi2014}, \citet{Liu2020} presented a time-dependent study of the molecular chemistry of the Galaxy as resulting from the putative past AGN activity of Sgr$^*$. In particular, H$_2$O, CH$_3$OH, and H$_2$CO abundances are enhanced with respect to the baseline model without X-ray irradiation both in the gas phase and on the dust grain surface up to 10 Myrs after the turn off of the X-ray source.

\section{Using Models to Analyze Observations}
\subsection{Overview}
\label{sec:Overview}

Since the mid 70's \cite[e.g.,][]{Glassgold1975, Black1977}  several groups have developed PDR codes that are either directly available for download, or make their output available on-line. These include the model derived from the original TH85 code \citep{Kaufman1999, Kaufman2006, Neufeld2016}, the UCL-PDR code \citep{Papadopoulos2002, Bell2006, Priestley2017}, CLOUDY \citep{Abel2005,Ferland2013, Ferland2017}, the Meudon PDR code \citep{LePetit2006, LeBourlot2012, Bron2014} and \textsc{KOSMA-$\tau$} \citep{Rollig2006,Cubick2008,Rollig2013}. \cite{Meijerink2005} emphasized   modelling XDRs, but we note that now also CLOUDY \citep{Ferland2017} explicitly handles XDR calculations. 
The majority of these models are 1-dimensional and steady-state. 
We discuss multi-dimensional and time-dependent models in Section \ref{sec:multiD_PDR}. Specific PDR and XDR codes have also been developed for distinct applications, especially for the study of protoplanetary disks. We will not review the disk  models here \citep[see][]{Bergin2007,Oberg2021}.

The considerable heterogeneity among  models (including their geometry, physical and chemical structure, choice of parameters) makes the comparison between them challenging. Nevertheless, a number of benchmark models have been created to understand where different results originate and, as much as possible, converge on a common solution for a common input. \cite{Rollig2007} give a detailed report of the 2004 Lorentz Center workshop comparing 10 different PDR codes. An additional Lorentz Center workshop on the CO ladder from both PDR and XDR models was held in 2012 with results that can be found on-line at
\url{https://markusroellig.github.io/research/CO-workshop/}.
Even using consistent inputs and similar microphysics there is considerable range in the outputs in the gas temperatures, abundances,
and line intensities. From the first workshop, for typical PDR conditions and
neglecting the obvious outliers, the [\cii] line
is most consistent between models and varies within a factor of ${\sim}2-3$ depending on $G_0/n$. The next most consistent are the [\oi] lines (varying
by a factor ${\sim}2-5)$ and the least consistent are
the [\ci] lines. From the second workshop, the CO ladder line intensities can vary by 10-100 between models for similar inputs. 
The largest  differences are likely due to differences in gas temperature where the lines are produced. The gas temperatures typically
vary between models by factors of 2-3 at $A_{\rm V} \approx 1$, but  are seen to vary by as much as a factor  4-10 at
$A_{\rm V}<1$, but also at $A_{\rm V}\approx 1$ between models with unconstrained microphysics. The most extreme variations occur
for the test run $\chi=10^5$, $n=10^{5.5}$ ${\rm cm^{-3}}$
 \citep[see also][]{Bruderer2012}. 
The temperature differences could be the result 
of a thermal instability that rapidly drives gas
temperatures from $T{\sim}2000$\,K to ${\sim}8000$ K  with only minor differences in the heating/cooling rates. In addition, the details
of the ${\rm H_2}$ formation rates,  vibrational heating rates,
and photoelectric heating rates can cause large differences in gas
temperature and are especially important for
the prediction of ${\rm H_2}$ and high-J CO line intensities.
The conclusion from \cite{Rollig2007} is that the model outputs for
specific densities and radiation fields should not be
considered exact results but should instead be used
as guides to the physical conditions. 
The choice of a specific code to compare to observations should be motivated by the physics and chemistry included in the code but also the characteristics (in terms of geometry, density profile, etc.) most adapted to the emission
source. See \cite{Rollig2007} for a guide to PDR model characteristics while noting that many
codes have been in continuous 
update since the workshops and another 
workshop
would be well worth revisiting. 

\subsection{Main input parameters}
\label{subsec:PDRXDR_inputparams}
An important parameter of the models is the incident radiation field. The
field can be adjusted in several ways including  setting the shape, strength, and geometry (isotropic, uni-directional, external vs internal source for spherical clouds, or
one-sided vs two-sided for slabs). 
Several scalings of the FUV field are in use based on estimates of the local Galactic interstellar radiation field (see {\bf Table \ref{tab.FUVnotation}}). 
For comparison, it is convenient to refer to the integrated energy density $u(\lambda) = 4\pi J(\lambda)/c$, where $J(\lambda) = 1/(4\pi)\int I(\lambda) d\Omega$ is the mean intensity and $I(\lambda)$ the specific intensity. The value of $J$ equals $I$ for an isotropic field only. 
When integrating over the surface of a semi-infinite cloud illuminated over $2 \pi$ and taking into account backscattering from grains, $J \sim 0.54 I$ \citep{LePetit2006, Rollig2013}.

The Habing field \citep{Habing1968}, noted as $G_0$, when integrated over a range of energies 6 eV $< h\nu < 13.6$ eV, has an energy density $u=5.33\times 10^{-14}$ erg ${\rm cm^{-3}}$ corresponding to an isotropic intensity of $I = J =  1.3 \times 10^{-4}$ erg ${\rm cm^{-2}}$ ${\rm s^{-1}}$ ${\rm sr^{-1}}$. The 1-D flux  is the  unidirectional flux equivalent to the total flux
incident on a sphere in the isotropic
radiation field $= 4\pi J = cu =
1.6 \times 10^{-3}$ erg ${\rm cm^{-2}}$ ${\rm s^{-1}}$. Other fields that are in use are the Mathis field ($U$; \citealt{Mathis1983, Weingartner2001}) 
and the Draine field ($\chi$; \citealt{Draine1978}). 
In terms of the Habing field, the median local Galactic interstellar radiation field is estimated to be $G_0{\sim}1.6$ \citep{Parravano2003}, comparable to the Draine field, while $G_0{\sim} 10^5$ for the PDR behind the Trapezium cluster. 

For XDR codes the incident radiation flux between 1-100 keV, $F_X$ (in $\rm erg\,s^{-1}\,cm^{-2}$) is instead the required input parameter. For historical reasons \citep{Maloney1996}, the impinging radiation is generally assumed to follow a powerlaw distribution, $F_{\rm X}(E)=F(0)(E/100 {\rm \,keV})^{-\alpha}$, with a low energy cut-off at 1 keV. In \citet{Meijerink2007}, $\alpha=0.9$ is chosen, while CLOUDY has a default $\alpha=0.7$.  This is a good approximation of the X-ray regime of the spectral energy distribution (SED) of AGN.
Note that the 1-100 keV range is only a fraction of the total AGN SED, which is also bright in the mid-infrared, optical and ultraviolet. XDR models including the total AGN SED instead of the standard power-law for the incident radiation, produce different kinetic temperatures, mid-infrared line intensities, and low level populations of $\rm H_2$ despite being normalized to have the same $F_{\rm X}$ between 1-100 keV \citep{Ferland2013}. 
XDR models are also used to derive the gas conditions in 
the vicinity of young stellar objects. The incident photon flux is then expressed in terms of the spectrum of a thermal plasma $F_{\rm X}(E,r)=F_0(r) {\rm exp}(-E/kT_X)$ in units of photons s$^{-1}$ cm$^{-2}$ eV$^{-1}$, where $r$ is the radius in the envelope from the central protostar, and $T_X$ is the temperature of the X-ray
emitting plasma. In this case, the spectrum goes as an exponential and \Hxn~falls off more steeply at high column than a power-law.

\begin{table}[h]
\tabcolsep7.5pt
\caption{Conversion between FUV fields$^{\rm a}$}
\label{tab.FUVnotation}
\begin{center}
\begin{tabular}{@{}l|c|c|c|l|l@{}}
\hline
& Energy Density & Isotropic Intensity & 1-D Flux &  & Examples \\
Notation & (erg ${\rm cm^{-3}}$) &
(erg ${\rm cm^{-2}}$ ${\rm s^{-1}}$ ${\rm sr^{-1}}$) &
(erg ${\rm cm^{-2}}$ ${\rm s^{-1}}$) & Ref. & of codes\\
\hline
$G_0$  & $5.3\times 10^{-14}$ & $1.3 \times 10^{-4}$ & $1.6 \times 10^{-3}$ & [1] & UCL-PDR\\
$\chi$& $8.9 \times 10^{-14}$ &$2.1\times 10^{-4}$ &$2.7\times 10^{-3}$ &[2] & \textsc{KOSMA-$\tau$}\\
$U$ & $6.1\times10^{-14}$ &$1.4\times 10^{-4}$ &$1.8\times 10^{-3}$ &[3] & Meudon$^{\rm b}$\\

\hline
\end{tabular}
\end{center}
\begin{tabnote}
$^{\rm a}$Energy density, Isotropic Intensity, and 1-D flux for unit values of radiation fields $G_0$, $\chi$, and $U$. Quantities are integrated over 6\,eV to 13.6\,eV unless otherwise noted. The Meudon code assumes limits of 912 \AA\ to 2400 \AA\ or
13.6\,eV to 5.166\,eV, resulting in energy densities of $5.6\times 10^{-14}$ (erg ${\rm cm^{-3}}$), $1.05\times 10^{-13}$ (erg ${\rm cm^{-3}}$), and $6.8\times 10^{-14}$ (erg ${\rm cm^{-3}}$) for 
$G_0$, $\chi$, and $U$ respectively.; $^{\rm b}$Can also use the
Draine field. 
References: [1] \cite{Habing1968}; [2] \cite{Draine1978}; [3] \cite{Mathis1983}.
\end{tabnote}
\end{table} 

In addition to the radiation field, the cosmic-ray ionisation rate is also a necessary input. This is usually given
as the total (including secondaries) ionization rate per
\HH, $\zeta_{\rm H_2}$, but may also be given as the primary ionization rate per H,
$\zeta_{\rm p}$,
where the rate per \HH\ is higher by a factor ${\sim} 2.3$. 

The abundances of gas phase metals, PAHs, and large
grains affect the cooling rates
heating rates, and conversion between $A_{\rm V}$
and column density, $N$. These initially vary linearly with
metallicity, \Zsun, where \Zsun\ is the metallicity with respect
to solar, but are
observed to have different dependencies
for \Zsun$\lesssim 0.2$ \citep{Gordon2003,Sandstrom2010,RemyRuyer2014}
and the model results will depend on the relative variations (see e.g., \citealt{Rollig2006}, and \citealt{Jameson2018}).
In light of other uncertainties a linear dependence with 
\Zsun\ is often assumed. 

The choice of geometry is also important to consider. Many models adopt a plane-parallel slab of fixed
width in terms of \Av\ (or with a stopping condition provided as input), while others have spherical clouds with central cavities, 
spherical clumps, or distributions of clumps.
Several density structures can be adopted, such as constant density, constant pressure, or a user specified density law.
The adopted model for the photoelectric heating is also important and can result in
significant temperature variations.
Finally, different models include different chemical networks and elemental abundances. The  species, reactions, and rates that are used have an impact on the resulting predictions.

\subsection{Diagnostic plots}

Combinations of line intensities and dust continuum can be used to constrain the physical conditions of the gas in PDRs and XDRs, as well as to distinguish between excitation sources (PDR, XDR, shocks). We describe here how some of the commonly observed line intensities and line ratios lead to predictions of the incident radiation field strength and gas density in the PDRs and XDRs, and help to distinguish between different heating processes.

\subsubsection{Diagnostic to determine the gas density $n$, and radiation field strength, $G_0$}
\label{subsec:PDR_diagnostics}

\begin{figure}
\begin{minipage}[c]{0.5\textwidth}
\includegraphics[width=\textwidth]{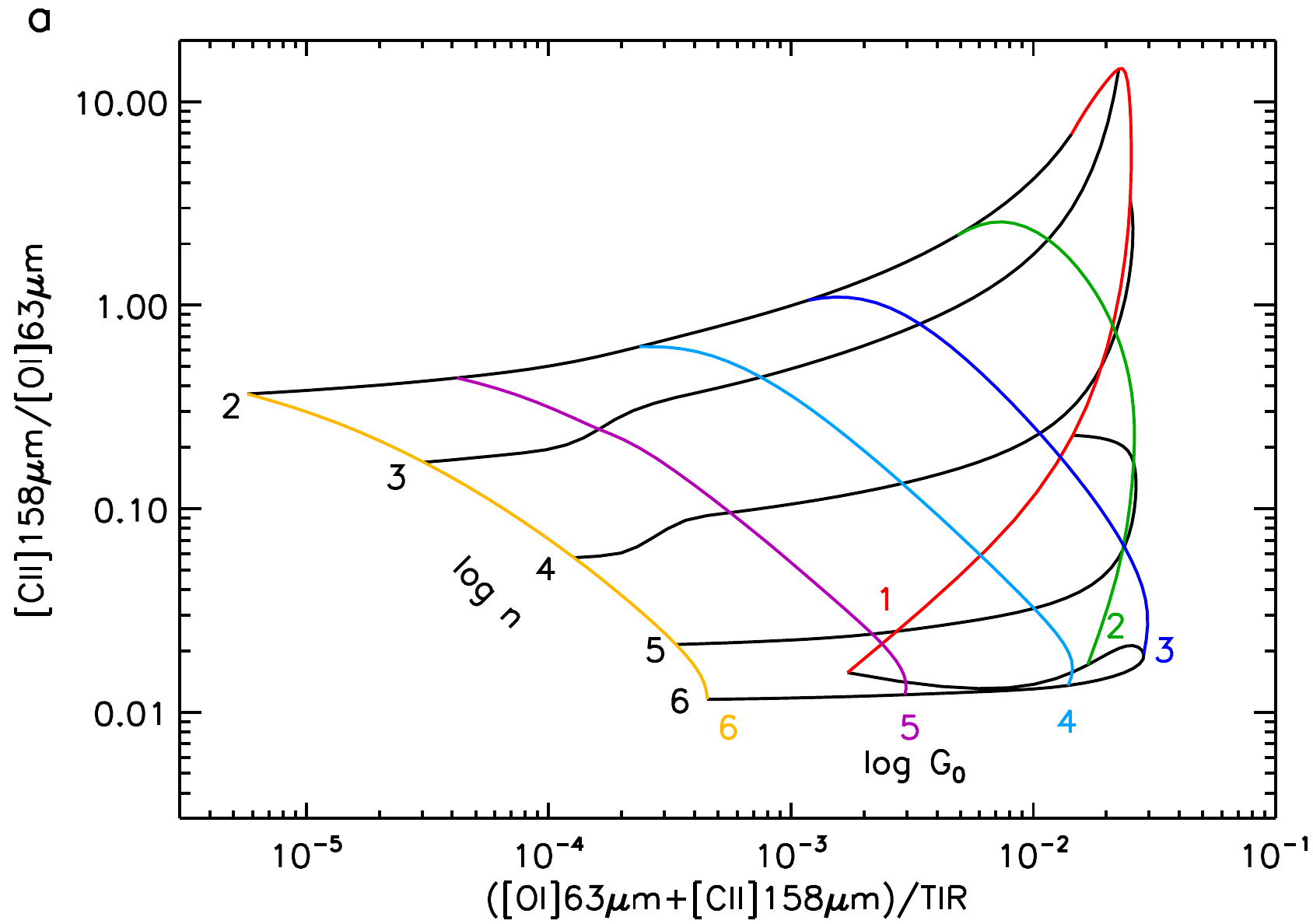} 
\end{minipage}\hfill
\begin{minipage}[c]{0.47\textwidth}
\includegraphics[width=\textwidth]{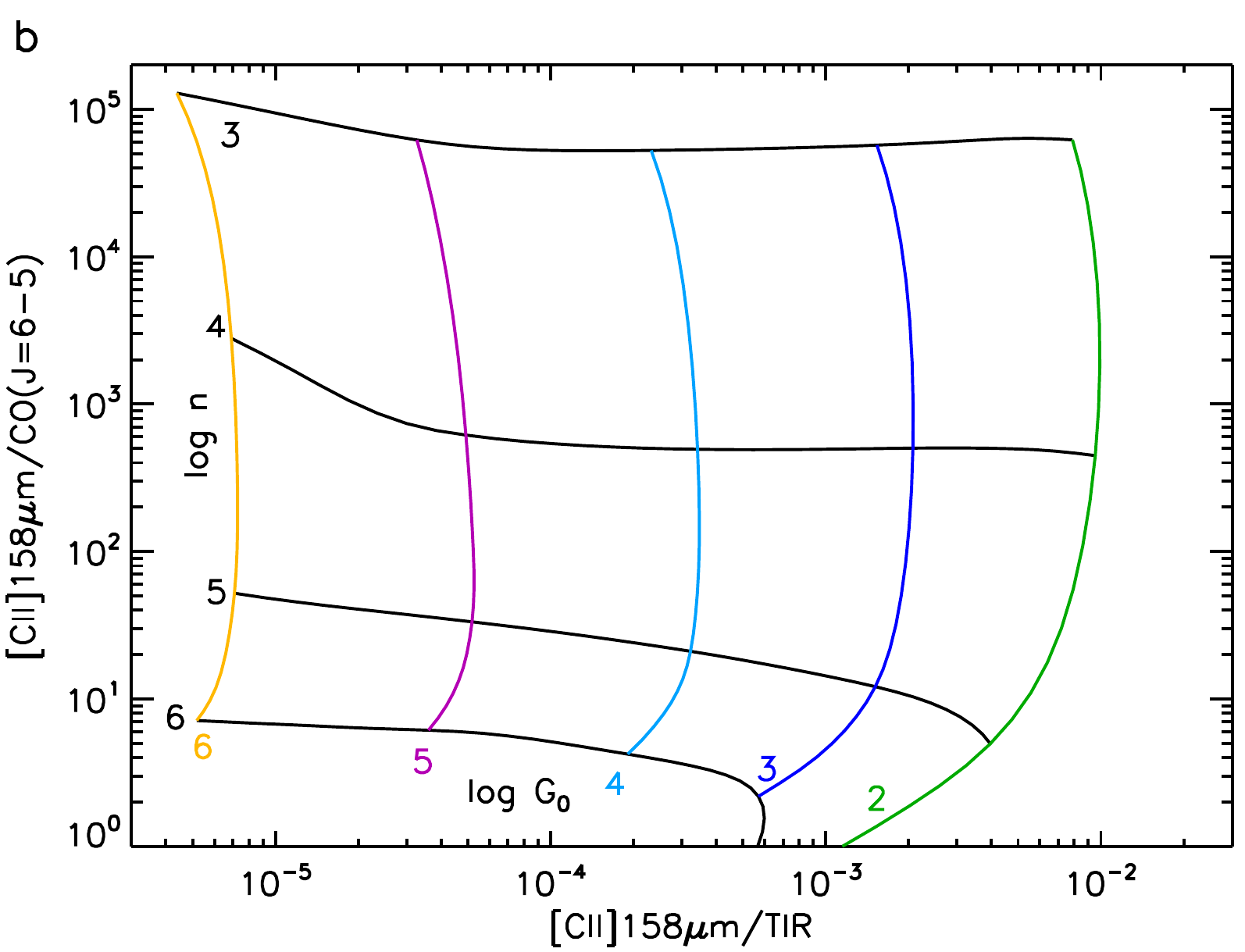}
\end{minipage}
\caption{Diagnostic contour plots of model integrated line intensities and
continuum ratios. The intensity units are erg ${\rm cm^{-2}}$ ${\rm s^{-1}}$ ${\rm sr^{-1}}$ (rather than K  km ${\rm s^{-1}}$).  ($a$) [\cii]/[\oi] ratio versus the ([\oi]+[\cii])/TIR ratio where [\oi] refers
to the 63\,\mic\ transition.
The [\cii]/[\oi] ratio is mainly sensitive to density
while the ([\oi]+[\cii])/TIR ratio is sensitive to
$G_0/n$. ($b$) [\cii]/CO(6-5) versus [\cii]/TIR. The [\cii]/CO(6-5)
ratio is a strong density indicator while [\cii]/TIR is
mainly sensitive to $G_0$. PDR models from \url{https://dustem.astro.umd.edu}. 
\label{fig:PDRDiagnostics} 
\vspace*{-1em}
}
\end{figure}
The PDR model used here is based on that
described in \cite{Wolfire2010}, \cite{Hollenbach2012}, and
\cite{Neufeld2016}. We use a maximum $A_{\rm V}=7$,
freeze out turned off, and a power-law dependence of the cosmic-ray ionization
rate $\zeta_{\rm p}=2\times 10^{-16}(1+ N/2\times 10^{21}\, {\rm cm}^{-2})^{-1}$. The [\cii]/[\oi] 63\,\mic\ ratio is
a good diagnostic for density since for $n\gtrsim 
3000$ ${\rm cm^{-3}}$, the critical density for [\cii],
this ratio will be a  strong function
of density, and for $G_0\gtrsim 100$ [\oi] cooling is important.
The ratio 
([\cii] + [\oi])/TIR is a measure of the heating
efficiency (Section \ref{sec:PDR_heating}) where TIR is
the total (3\,\mic-$1.1\,{\rm mm}$) infrared integrated dust
continuum. 
The combination
of these two ratios  ({\bf Figure \ref{fig:PDRDiagnostics}a}), are
useful to obtain physical conditions (gas density and radiation field strength) directly from plotted observations. {\bf Figure \ref{fig:PDRDiagnostics}b}
shows [\cii]/CO ($J=6-5$) versus [\cii]/TIR and
is an example that is especially useful for 
high density regions where CO ($J=6-5$) emission is produced.
The CO ($J=6-5$) line is often near the peak of the CO ladder ({\bf Figure \ref{fig:COLadder_extragalactic_PDR_XDR}}) in 
extragalactic observations and is often the brightest
(or even the only) line observed. For the model plots, we assume the TIR intensity is $2\times G_0$ where the factor of 
2 accounts for dust heating by stellar optical radiation and
by EUV radiation emitted by the star that
is converted by line emission to optical radiation. 
The observed continuum intensity to plot on the model grids  should ideally be the TIR.
However, in practice, 
the integrated intensity depends on the 
wavelength range of the available observations. For $\lambda \gtrsim 40$\,\mic, generally referred to as the FIR continuum, ${\rm TIR/FIR}{\sim}2$ depending on the grain temperature. See e.g., 
\cite{Dale2002} to convert between TIR and FIR.
Additional on-line plots and model results can be found at
\url{https://dustem.astro.umd.edu} 
or \url{https://ism.obspm.fr} for the models
included here and for the Meudon code respectively.
Both also include the \textsc{KOSMA-$\tau$} models. 
The on-line tools can analyze both pointed observations and maps. 

Diagnostic plots typically use ratios of intensities  rather than absolute values. This is justified because the model outputs are
for unit beam filling factor, $f_b=1$. For unresolved sources,
$f_b<1$ and using line ratios
has the advantage of cancelling the beam
filling factor.  However, if the emitting regions
for different lines have substantially different
filling factors, then the intensities used in the ratio should
account for these differences as best as possible
(see e.g., \citealt {Wolfire1990, Kaufman1999}, for a
typical procedure).
Once the physical conditions are estimated
from diagnostic plots, the beam filling factor can
be derived by comparing the model with 
the observed intensity, $f_b = I^{\rm obs}/I^{\rm model}$.
Note that the filling factor is less than 1 for unresolved sources but can be greater than 1 for several PDRs along the line of sight. The covering factor, discussed by 
\cite{Cormier2019}, is the fraction of the \hii\ region 
surrounded by neutral PDR gas. This can be calculated using
models that include both the ionized and neutral gas (e.g., CLOUDY). In a sample of low-metallicity galaxies, \cite{Cormier2019} find PDR covering factors $f_b = 0.2 - 1$, with a median value of 0.4, and show that the PDR covering factor decreases with metallicity (see Section~\ref{subsec:LowMetallicityEnv}).

Simple, model independent estimates can be made for $G_0$, $n$, and $T$ \cite[e.g.,][]{Pabst2017}. From the distribution  of OB stars on the sky, the maximum $G_0$ is ${\sim}0.5L/(4\pi d^21.6\times 10^{-3})$ with $d$ the projected distance in
cm between the source and PDR, and $L$ the stellar luminosity in erg ${\rm s^{-1}}$. This maximum $G_0$ and the $G_0$ estimated from a PDR model
are compared 
to determine the true (rather than projected) distance between an FUV source and the illuminated PDR and thus reveal the 3-dimensional geometry of a region. This method has been used for example to establish that the PDRs in 30 Doradus that seem close to the central cluster in projection are actually located at more than 40\,pc away from it \citep{Pellegrini2011, Chevance2016}, and to measure the  deprojected distance of the Trapezium stars to the Orion bar
\cite[0.33-0.45 pc,][]{Salgado2016}.
From the physical distance, $d$, between [\cii] and CO
emission peaks  observed in a resolved edge-on PDR, and assuming
an $A_{\rm V}{\sim} 2$ between the ${\rm C^+/C}$ and CO
transitions, the density can be estimated from $n=1.9\times 10^{21}A_{\rm V}/d = 3.8\times 10^{21}/d$ ${\rm cm^{-3}}$ where $d$ is measured in cm. 
 \cite{Pabst2017} use this method to find $n{\sim}3\times 10^3-4\times 10^4$ ${\rm cm^{-3}}$ in the Horsehead PDR and surrounding region.  Finally, the gas temperature can be inferred from the
peak [\cii] line brightness and estimate of gas density, and from
\HH\ pure rotational lines. Typical gas temperatures found are between 100 and 500\,K \citep[e.g.][]{YoungOwl2002, Pabst2021a}.
 
\subsubsection{Diagnostics to distinguish X-rays from other heating mechanisms}
\label{subsec:XDR_diagnostics}

Differences in the temperature and chemical abundance structure between PDRs and XDRs (Sections \ref{sec:pdr} and \ref{sec:xdr}) can be leveraged to infer the presence of an XDR, despite the challenge -- as we will discuss later in this Section -- of distinguishing X-ray from cosmic-ray or shock heating.
\begin{figure}
    \centering
    \includegraphics[scale=0.4]{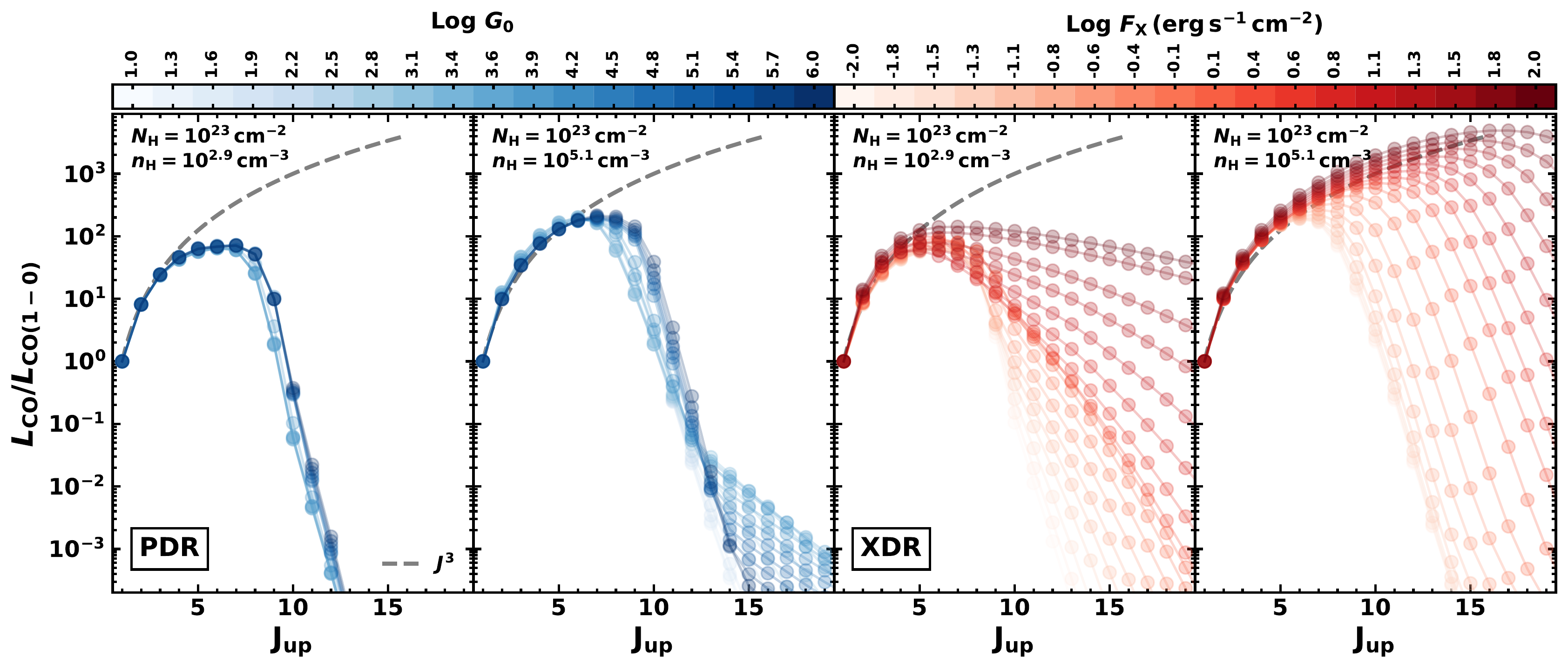}
    \caption{Normalized CO(1-0) ladder models from CLOUDY in the PDR (left panels) and XDR regime (right panels) for $n=\rm 10^{2.9}$ and $10^{5.1}\,\rm cm^{-3}$. The variation in $G_0$ and $F_{\rm X}$, respectively, are indicated in the colorbar. The dashed line indicates a thermalized CO ladder. Figure adapted with permission from \citet{Pensabene2021}, \copyright ESO.}
    \label{fig:COLadder_extragalactic_PDR_XDR}
\end{figure}

The [\siii] 35\,\mic/[\cii] 158\,\mic, [\oi] 63\,\mic/[\cii] 158\,\mic, and [\feii] 26\,\mic/[\cii] 158\,\mic, ratios
are all larger in XDRs than in PDRs \citep{Meijerink2007}. The same holds true -- albeit with caveats on similar results from an enhanced cosmic-ray rate \citep[][]{Bisbas2017, Bisbas2021} -- for [\ci]/CO and [\ci]/[\cii]. These ratios are often used as diagnostics in the circumnuclear disk of AGN \citep[e.g][]{Izumi2020} and in high-$z$ quasar host galaxies \citep[e.g.][see Section~\ref{sec:high-z}]{Venemans2017}. 

The whole CO ladder of AGN host sources can be another valuable tool to infer the presence of an XDR component contributing to the gas heating \citep[][]{Bradford2009, vanderWerf2010, Haley-Dusheath2012, Pozzi2017, Mingozzi2018, Valentino2021} as XDRs are characterized by large column densities of warm molecular gas, where high-$J$ CO lines ($J\gtrsim8$) can be efficiently excited. For this reason the CO Spectral Line Energy Distribution (SLED) resulting from XDR models peaks at increasingly higher $J$ for increasing X-ray flux \citep[][and see \textbf{Figure \ref{fig:COLadder_extragalactic_PDR_XDR}}]{Vallini2019}, even though CO emission can be suppressed by X-ray induced CO dissociation (Section~\ref{sec:xdr_chemistry}) at intermediate column densities \citep{Kawamuro2020}.
As a caveat, note that mechanical heating from shocks can boost high-$J$ CO lines either on galactic scales \citep[e.g.][as resulting from outflows and/or merger activity]{Meijerink2013, Bellocchi2020} or in resolved molecular complexes \citep{Lee2016}. \citet{Kazandjian2015} show that the temperature of the molecular gas can be significantly increased by a small amount of mechanical heating from stellar feedback in the form of stellar winds and supernovae,
although we note that the mechanical energy input may be fairly localized in the star-forming regions and may be dissipated in neutral atomic or ionized gas \citep{Lancaster2021,Lancaster2021b}, thus having much less an impact on the bulk of the molecular gas. 
While the emission from low-$J$ CO lines only moderately increases with mechanical heating, high-$J$ CO line emission can increase by several orders of magnitude in clouds with $n {\sim} 10^5$\,\cm\ and a galactic star formation rate of 1\,\Msun\,yr$^{-1}$. 
For spatially resolved observations towards dense ($n \gtrsim 10^5$\,\cm) clumps in molecular clouds, the differentiation between PDRs and XDRs based on the peak of the CO ladder breaks down as high-density PDRs produce bright high-$J$ CO lines \citep[][]{Burton1990, Joblin2018,Wu2018}.
Note however that on galactic scales, high density PDR contributions to high-$J$ CO emission are generally diluted by low filling factors \citep{Indriolo2017}.

\begin{figure}
    \centering
     \includegraphics[scale=0.45]{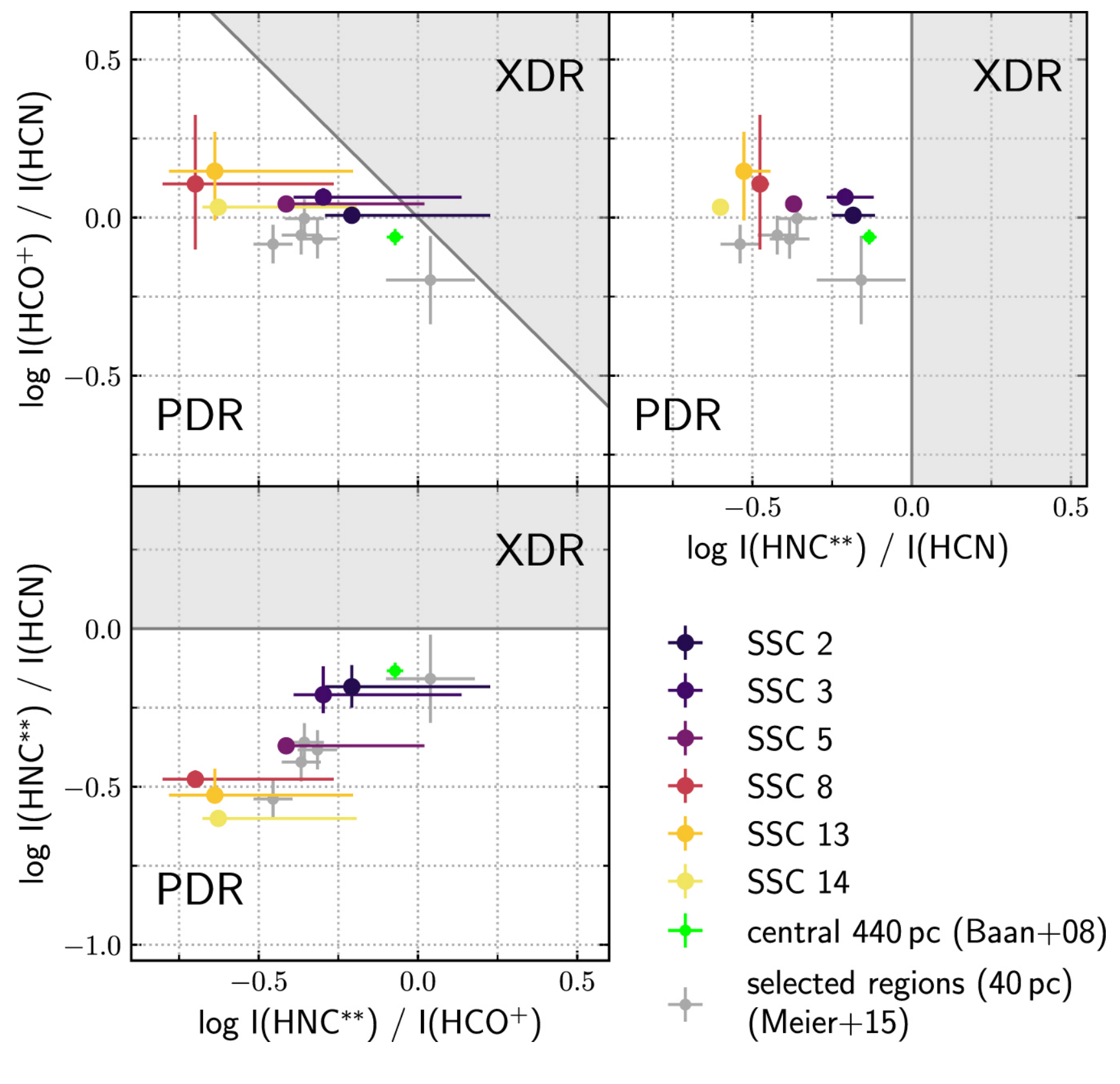}
    \caption{HCN, HNC, HCO$^+$ intensity ratios used to infer
    excitation mechanisms. Points with error bars are super star
    clusters in NGC\,253 and indicate that PDR chemistry is favored
    over XDR chemistry. Figure adapted with permission from \cite{Krieger2020}, \copyright AAS.}
    \label{fig:HCNdiagnostic}
\end{figure}

The use of line ratios involving high density tracers (such as HCN and HCO$^+$ lines) as XDR diagnostics can be valuable but it is still debated given the non-trivial effects of gas density, temperature, opacity, and time-dependence \citep[see  Fig.1 in][]{Viti2017}.
XDR models \citep{Meijerink2007} indicate that HCN/HCO$^+$ abundance ratios exceed that of PDRs only when $F_X \gtrsim100 \rm \,erg\, s^{-1}\, cm^{-2}$, and $n\approx10^4\,$\cm. Below these limits, the enhanced abundance of HCO$^+$ in XDRs (see Section~\ref{sec:xdr_chemistry}) drives the ratio below that of dense PDRs.
Note that HCN emission can be boosted by IR pumping \citep{Costagliola2011, Martin2015, Vollmer2017, Harada2018} and electron impact excitation in XDRs \citep{Goldsmith2017}. A combination of HCN/HCO$^+$ ratios with HNC/HCN represents another set of valuable diagnostics, as outlined in \citet{Baan2008,Loenen2008} and recently in \citet{Krieger2020} where such diagrams have been exploited to study the heating mechanisms in star forming regions within NGC\,253 ({\bf Figure \ref{fig:HCNdiagnostic}}). In particular the higher HNC/HCN abundance ratios in XDRs as compared to PDRs at large column densities \citep[][and discussion in Section~\ref{sec:xdr_chemistry}]{Baan2008,Loenen2008} make HCN/HNC intensity ratios in XDRs somewhat higher than those in PDRs. A caveat is that, as for HCN/HCO$^+$, IR-pumping \citep{Aalto2007} and shock heating could produce similar effects \citep{Canameras2021}. \citet{Meijerink2007} and \citet{Harada2013} also suggest CN/HCN as a potential XDR-vs-PDR diagnostic because the column density ratios for PDRs and XDRs are very different, ranging from 40-1000 in the XDRs, to 0.5-2 in PDRs. Nevertheless, only relatively modest CN enhancements have been reported in circumnuclear disks of three AGN \citep{Ledger2021}.

Observations of CH/CO ratios in four nearby AGN host galaxies \citep{Rangwala2014} reveal values $\approx 10$ times higher than expected in normal galaxies, thus suggesting that CH/CO could have some potential as an XDR diagnostic. \citet{Rangwala2014} supported this conclusion by noting that \citet{Meijerink2007} models return CH abundances that are significantly enhanced relative to other molecules in XDRs.
A discussion concerning the PDR vs XDR impact on the abundances of hydrides (CH$^+$, OH$^+$, H$_2$O$^+$, CH) in the context of young stellar objects is presented in \citet{Benz2016}. By studying CH$^{+}$/OH$^{+}$, OH$^{+}$/H$_2$O$^{+}$, and C$^+$/CH$^+$ \citet{Benz2016} find no chemical evidence for X-ray induced chemistry and put tight constrains on the (negligible) contribution of an XDR component to the observed emission.

Distinguishing an XDR from an environment characterized by high cosmic-ray rates is instead rather difficult, both in external galaxies where observations encompass very extended gas patches \citep{Meijerink2006}, and in the envelopes of young stellar objects \citep{Bruderer2009c}. 
The non-thermal electrons resulting from cosmic-ray
and X-ray ionizations interact in a similar way with the atomic or molecular gas \citep{Glassgold2012}, thus producing comparable heating and chemistry (e.g., high [\ci]/CO and [\ci]/[\cii] ratios; \citealt{Bisbas2017, Bisbas2021}).
For this reason, enhanced cosmic-ray flux is sometimes used to ``mimic" XDR-like conditions \citep[][]{Bayet2011, Viti2017}. The X-ray chemistry can be reproduced by means of an enhanced cosmic-ray ionization rate
with deviations by less than 25\% \citep{Bruderer2009c}. \citet{Meijerink2011} suggested that OH$^+$/OH and $\rm H_2O^+$/H$_2$O ratios might help in distinguishing very high cosmic-ray rates from the typical XDR conditions at $\approx 150$ pc from an AGN of $L=10^{44}\, \rm erg\, s^{-1}$. A cloud of density $n=10^{5.5}\rm \, cm^{-3}$ at such distance from the galaxy center experiences an impinging $F_X\approx 29 \rm \, erg\,s^{-1}\, cm^{-2}$, and is expected to have OH$^+$/OH and $\rm H_2O^+$/H$_2$O column density ratios as high as $(250-7.5)\times10^{-4}$ and $(64-3)\times10^{-3}$, compared to $(3-5)\times10^{-4}$ and $(2-14)\times10^{-4}$ for high CR rates $\zeta_{\rm H_2} \approx (10^{-14}-10^{-13})\,\rm s^{-1}$ (for $N=3\times 10^{22}-10^{24}\, \rm cm^{-2}$).
However no attempts were made to model the resulting line emission. \citet{Vallini2019} found that cosmic rays can boost the mid/high-$J$ end of the CO ladder but not at the level observed in XDRs \citep[see also][for Arp 220]{Rangwala2011}.

The effect of mechanical heating needs to be considered before determining physical parameters of the gas even though estimating the contribution of stellar winds/supernovae to the mechanical heating in external galaxies is challenging. \citet{Meijerink2013} suggest that shock dominated ISM yields a much larger (up to $\approx 10$ times) CO-to-continuum ratio than XDRs because shocks heat the gas directly, with little  heating of dust, and hence do not contribute appreciably to the IR luminosity \citep[][]{Pellegrini2013}. Note that shock velocities $v_{\rm sh}<30$ km/s are required to preserve the CO from dissociation  \citep{Hollenbach1989}. Fast ($v_{\rm sh}>50$ km/s) shocks destroy the CO (although it can reform in post shock gas to some extent before getting too cold) and also produce optical and FUV photons which heat the grains, providing IR continuum.
\subsection{Origin of [\cii]}
\label{subsec:OriginofCII}

The ionisation potential of neutral carbon C (11.3\,eV) is lower than that of neutral hydrogen (13.6\,eV). As a result, the [\cii] line can in principle originate both from the ionized and the neutral gas. 
In the Milky Way and in nearby galaxies, most studies have concluded that the vast majority of the [\cii] originates from the neutral regions \citep[e.g.,][]{Oberst2011, Pineda2014, Goicoechea2015,  Croxall2017, Pabst2017, Pabst2021a}. 
A low contribution from the ionized gas to the [\cii] emission is also found at high-redshift, both observationally \citep[e.g.,][]{Stacey2010, Gullberg2015} and in simulations \citep[e.g.,][]{Olsen2015, Katz2017, Pallottini2017, Lagache2018}.
Typically, observations find the ionized gas contributes 
${\sim} 5-30$\% to the [\cii] emission. 
    
The component of the  neutral  phase of the ISM that
dominates  the [\cii]  emission is still a matter of debate. Early suggestions were that [\cii] is associated with molecular clouds \citep{Stacey1985,Shibai1991} or the diffuse ISM \citep{Bennett1994}. 
The dominant component must depend on the observing
scale  and emission source \citep{Mookerjea2016}.
Observations concentrating on Galactic star-forming
regions generally find the
dense, intensely  illuminated gas, dominating
the emission.  Analysis of the line emission from unresolved  normal galaxies finds that [\cii] arises from PDRs of  moderately high FUV fields $G_0=10^2-10^{4.5}$ and densities  $n=10^2-10^{4.5}$ ${\rm cm^{-3}}$ \citep{Malhotra2001} with similar range
found in resolved ($0.2-2.1$ kpc) star forming regions in local galaxies \citep{Sutter2021} and in low-metallicity dwarf galaxies  \citep{Cormier2019}. These are usually interpreted as the bright classic PDRs. Extended
mapping of [\cii] in Orion \citep{Goicoechea2015,Pabst2017, Pabst2021a}, and a detailed PDR analysis of extragalactic \hii\ regions 
\citep{Abdullah2017} finds that
the extended,
moderately illuminated, molecular cloud surfaces 
can contribute significantly or dominate the emission.
An analysis of the COBE [\cii] emission 
in ${\sim} 400$ pc beams from the Orion-Eridanus superbubble \citep{Abdullah2020} finds that molecular cloud surfaces
up to ${\sim} 10$ pc from the stellar clusters
dominate the emission with a moderate field
strength of $G_0 {\sim} 100$ and density $n{\sim}10^3$ ${\rm cm^{-3}}$. Because these 
projected beam
sizes are comparable to those of the \herschel\ Space Observatory observations in nearby galaxies, \cite{Abdullah2020} suggest that 
the [\cii] emission in these observations may be similarly
dominated by moderately illuminated extended cloud surfaces.
Such moderate field strengths are also found
for the Galactic [\cii] 
emission using the
all-sky COBE [\cii] survey \citep{Cubick2008},
the pointed GOT C+ survey \citep{Pineda2013},
and for dust SEDs in nearby galaxies in which only  ${\sim} 10-15$\% of the infrared luminosity arises from dust that is illuminated by strong,
$G_0 \gtrsim 400$, radiation fields \citep{Aniano2020}. The result that Galactic and
extragalactic observations point to different 
sources of [\cii] emission still needs to be reconciled, and is perhaps the result of mixing various gas components within the beam (e.g., surfaces of molecular clouds, diffuse atomic gas, moderately strong PDRs, dense PDRs, ionized gas).

One way of separating the different components of Galactic [\cii] emission is to use velocity resolved observations, where the \hi, CO, and
[\cii] can be separated along a line of sight due to Galactic rotation. Relying on such observations, \cite{Pineda2014} find that 
30\% of Galactic [\cii] luminosity comes from 
dense PDRs, 25\% from CO-dark gas, 25\% from cold \hi, and
20\% from ionized gas.
For extragalactic observations the  multiphase ISM mixed in the beam can be partially separated by comparing velocity resolved
\hi, CO, and fine-structure line  profiles.
\cite{Tarantino2021} find about equal [\cii] contributions from atomic and molecular gas in ${\sim}500$\,pc
beams, 
while separate  CO and CO-dark gas contributions  could not be easily separated
at these spatial resolutions. \cite{Lebouteiller2019} and \cite{Okada2019}
observing the LMC find most of the [\cii] arises in
CO or CO-dark gas. 
The fraction of  [\cii] arising in the CO-dark gas is generally seen to increase
with lower metallicity (see Section \ref{subsec:LowMetallicityEnv}).

\subsection{Caveats in using models to analyze observations}

As with any comparison between observations and models, the interpretation must be done carefully. Several potential difficulties can alter the accuracy or validity of the conclusions. Below is a non-exhaustive list of common points requiring careful consideration.

\begin{itemize}
    \item{Differences in model gas temperature.} PDR model
    workshops have shown variations in gas temperatures by factors of 2-3 at $A_{\rm V}\approx 1$  
    with the same input parameters and microphysics but can vary by 4-10 when the microphysics are unconstrained (see Section \ref{sec:Overview}, \citealt{Rollig2007}, and CO ladder workshop
    \url{https://markusroellig.github.io/research/CO-workshop/}).
    These differences can strongly affect the predicted line intensities that are sensitive to temperature such as the mid-infrared ${\rm H_2}$ lines and the CO ladder. The exact density or radiation field strength predicted by the models will vary but they are still good guides for
    the general physical conditions and dominant
    physical processes. 
    
    \item {Contribution from ionized gas.} 
     A correction must be applied to subtract the fraction of the [\cii] emission arising in the ionized gas
     \cite[e.g.,][]{Rubin1985, Abel2005, Kaufman2006}.
     This can be  done using the theoretical [\cii]/[\nii]~205\,\mic\ ratio that
     is nearly constant (${\sim} 5$)  with $n_{e}$
     \citep[e.g.,][]{Oberst2011, Langer2016}.
     If the [\nii]~205\,\mic\ line is unobserved, 
     the [\cii]/[\nii]~122\,\mic\ ratio may be used with an estimate
     of $n_{e}$ from the 
      [\suiii]~18\,\mic/[\suiii]~33\,\mic\ ratio 
      (sensitive to 
    $10^2 \lesssim n_{e} \lesssim 10^5$ ${\rm cm^{-3}}$)
      or from the
     thermal pressure in the PDR/\hii\ region \citep[e.g.,][]{Seo2019}.  
     Alternatively, the fraction of [\cii] can be determined by comparing the resolved velocity distributions of tracers of the ionized, neutral and molecular gas \citep[e.g.,][]{Pineda2013, Anderson2019, Lebouteiller2019, Seo2019}.
     The fraction of [\cii] originating from the ionized gas typically amounts to ${\sim}$5-30\% and should be
  subtracted before analysis.
   
    \item {Mixing of components in the beam.} The origin and spatial distribution of [\cii] and [\oi] (or any other diagnostic line) may not be similar. Especially in the case of extragalactic observations, multiple gas components with a variety of physical properties are mixed in a single, large beam. The [\cii] emission tends to be widespread while [\oi] tends to trace more compact, dense, and warm regions \citep[e.g.,][]{Lebouteiller2019, Okada2019}. 
   
    \item {Origin of the FIR emission.}  Velocity-resolved observations may show multiple emitting components and thus a fraction of the FIR continuum needs to  be 
    assigned to each \citep{Schneider2018}. The fraction of
    [\oi] (or [\cii]) emission may be used as a rough guide.
    
    \item Absorption of [\oi] {and} [\cii]. The PDR line emission might be absorbed along the line-of-sight \citep[e.g.,][]{Abel2007, Guevara2020}
    leading to incorrect line ratios. Typically [\oi] 63\,\mic\ is most affected and some caution (or correction) should be used (see Section \ref{subsec:PDRsIntermediateFUV}).

 \item {Edge-on effects.} A PDR viewed edge-on can have different
emerging line intensities than one viewed face-on and the intensities may vary across the source as deeper layers are observed (e.g., see \citealt{Hogerheijde1995}
for the edge-on Orion Bar PDR).
The intensities depend on the column densities along the line-of-sight. For face-on PDRs,
[\oi] 63\,\mic\ and [\cii] are optically thick or marginally thick but other lines (e.g., [\oi] 145\,\mic, \HH\ rovibrational) are optically thin, so these  
increase in intensity with increasing column density compared to [\oi] 63\,\mic\ and [\cii].
The FIR continuum is 
directly proportional to the line of sight column density. 
Several models
(e.g., Meudon code) provide the intensities viewed at several angles. 
Calculations of edge-on models are described and shown in \cite{Pabst2017}.

    \item {Radiation pressure on grains.} Solutions with
    $G_0/n \gtrsim 5$ might be excluded since radiation pressure, 
    photoelectric emission, and photodesorption forces
    would drive grains through the gas and are therefore not consistent
    with a steady-state model solution \citep{Weingartner2001b, Hollenbach2012}. However, if the magnetic field is perpendicular to the radiation, and charged grains are tied to
    the field, then this may not be an issue.
    
    \item Degeneracy of diagnostics. Model fits using a combination of [\oi], [\cii] and FIR often present a degeneracy between a low $n$-high $G_0$ and high $n$-low $G_0$ solution, while model fits using CO and [\ci] are degenerate between 
    low $n$-low $G_0$ and high $n$-high $G_0$ solutions, highlighting the need for additional, independent, observational constraints \citep[e.g.,][]{Okada2019}.
\end{itemize}

\section{Time Dependent and Multi-Dimensional Models}
\label{sec:multiD_PDR}

The 1-D picture of PDRs  described in section
\ref{sec:PDR_structure} needs further context since for O and early B stars, both  \hii\ region
evolution 
and molecular cloud photodissociation and photoevaporation modify the PDR boundary conditions and internal structure. The \hii\ region is initially embedded within the molecular cloud and expands rapidly into
the surrounding gas but shortly slows down and proceeds as
a ``D-type" ionization front (IF) 
\citep[e.g.,][]{Spitzer1978}. 
The high thermal pressure in the ionized
gas compared to the molecular cloud produces a shock wave that sweeps up and compresses the ambient gas so that
there is pressure equilibrium 
between the ionized and neutral gas. In addition, the FUV radiation
dissociates the \HH\ in the compressed layer in a dissociation front (DF) \citep[e.g.,][]{Hill1978}.
It is this layer and the ambient cloud beyond that are the PDR.
 When the \hii\ region breaks out of the cloud, the hot gas
escapes into the ISM in a ``champagne flow'' also
called a blister \hii\ region \citep[e.g.,][]{Tenorio-Tagle1979}.  Several forces may be acting
to expand the \hii\ region and PDR and to disrupt the cloud
that are collectively known as stellar feedback 
\citep[e.g.,][]{Lopez2014, Krumholz2019}.  These
include thermal pressure from the ionized gas, stellar radiation
pressure, photoevaporation, and supernovae shock waves. Strong stellar winds during the embedded phase produce  a bubble of hot $T{\sim} 10^6$ K shocked gas which further compresses the \hii\ region  \citep[e.g.,][]{Castor1975, Weaver1977}, 
although e.g., \cite{Rosen2014}  and \cite{Lancaster2021,Lancaster2021b}  suggest that, due to
turbulent mixing, the effects of stellar winds are weaker than predicted by Castor and Weaver.
After breakout, the hot gas escapes and the wind shocks at the \hii\ region and propagates inward towards the star (in a reverse shock).
The post-shock hot gas 
expands and imparts some momentum to the \hii\ region and PDR. 
The EUV photon heating of the ionized gas and the FUV heating of the neutral gas (if it becomes sufficiently warm) can evaporate the cloud resulting in both cloud dispersal and an
additional pressure on the neutral layer.  We assess the stellar feedback terms including
PDR observations in Section \ref{subsec:PDRsHighFUV}.
\begin{marginnote}[]
\entry{IF}{Ionization front where
EUV photons ionize hydrogen at the
interface between the \hii\ region and the PDR. }
\end{marginnote}
\begin{marginnote}[]
\entry{DF}{Dissociation front where \HH\ is
photodissociated by the external FUV radiation that
penetrates into the PDR.}
\end{marginnote}
\begin{marginnote}[]
\entry{Advection}{Transport of material, (in this case \HH )  by
the mean fluid flow.}
\end{marginnote}

 Time-dependent PDR models with non steady-state chemistry are
 warranted if  chemical time scales exceed the
dynamical time scales.  The time for \HH\ to
come to chemical balance, $t_{\rm H_2, chem}$,  
is  often the slowest and therefore the most important to consider, while dynamical time scales depend on the physical process. 
Time dependence could be important in several cases:\,(1)there is rapid 
transport to a different radiation field or density compared to
$t_{\rm H_2 , chem}$, such as
resulting from advection towards the IF or DF, or advection between thermal phases in a turbulent medium,
(2)the radiation field or the density change faster than 
$t_{\rm H_2,chem}$ such as in expanding 
shells in planetary nebulae, 
compression of an atomic cloud, or due to turbulent compression
and rarefaction, and (3)chemical
time scales are long compared to cloud lifetimes (${\sim}10-30$\, Myr, \citealt{Chevance2020a}) such as can
occur
for grain-surface chemistry.

The time scale for \HH\ 
to achieve chemical balance  is $t_{\rm H_2,form} = 1/(2 nk_{\rm H_2} + D)$ \cite[e.g.,][]{Bialy2017}, where $k_{\rm H_2}$ is the formation rate coefficient
and $D$ is the local
photodissociation rate including \HH\ self-shielding. 
In mainly atomic regions, where there is relatively little 
H$_2$ shielding and $D$ is large 
compared to $2nk_{\rm H_2}$, then $t_{\rm H_2, chem}$ can be quite short 
($t_{\rm H_2, chem} {\sim} 1/D {\sim} 10^{3}$ yr) and \HH\ rapidly approaches the steady-state abundance. 
However, 
in atomic regions that are suddenly shielded so
that $2nk_{\rm H_2}$ is much larger than $D$, $t_{\rm H_2, chem}$ can be quite long ($t_{\rm H_2, chem}=1/(2nk_{\rm H_2}) {\sim} 10^9/(2n)$ yr to reach predominantly molecular gas).
Molecular regions may also be suddenly
illuminated by intense FUV radiation
as in the case of massive star formation. Although $D$ is large at the
surface, the interior is still shielded
so that $t_{\rm H_2, chem}\sim 10^9/(2n)$ and is long  to
reach steady state. 
We note that $k_{\rm H_2}$ can be slower at lower 
metallicity due to a lower dust abundance thereby increasing $t_{\rm H_2, chem}$ \citep[e.g.,][]{Hu2021}.  
In contrast, 
$k_{\rm H_2}$  may be a few
times faster in warm PDR surfaces compared to diffuse
gas, perhaps due  to the ER mechanism and \HH\ formation on PAHs,  
and  $R_{\rm H_2}=nn_{\rm H}k_{\rm H_2}$ can effectively increase in turbulent gas
due to positive density fluctuations  
\citep{Glover2007}. \cite{Sternberg2021} considered
time scales in a dust free environment where the formation of \HH\ is dominated by  gas-phase processes. Similar to the dusty case,
for $f_{\rm H_2, shield}{\sim}1$, 
$t_{\rm H_2, chem} {\sim} 10^{3}$ yr, but
for dense $n{\sim} 10^6$ cm$^{-3}$, well shielded regions where cosmic-ray chemistry
dominates $t_{\rm H_2, chem} {\sim} 10^{7}$ yr.

\subsection{One-dimensional time dependent models}
\label{subsec:1Dtimedependent}

The propagation of the IF and DF into the PDR and the  time-dependent effects on the PDR structure and emission are considered by e.g., \citet{Hollenbach1995,Bertoldi1996, StoerzerHollenbach1998}, and \citet{Natta1998}. In the frame of the IF, the \HH\ is advected towards
the edge of the PDR with a maximum speed of 
$v_{\rm adv}=c^{2}_{\rm PDR}/(2c_{\rm II})~\sim 0.5-1$ km ${\rm s^{-1}}$ where $c_{\rm PDR}$ and $c_{\rm II}$ are the isothermal sound
speeds in the PDR and \hii\ region respectively. 
The maximum speed occurs during the blister stage where a photoevaporated flow drives gas off the PDR and into the \hii\ region and is at a minimum for an embedded
\hii\ region within a cloud where the evaporated flow speed is much smaller. 

As the advected \HH\ nears the FUV illuminated surface, 
the \HH\ 
abundance can be enhanced over steady state  leading to enhanced FUV pumping and enhanced vibrational populations that
affect both heating and chemistry. Advection becomes increasingly
important as the travel time decreases, and the IF front and DF
will merge for a travel time less than the dissociation time.
The travel time is 
$t_{\rm H_2,travel} {\sim} N / (n\, v_{\rm adv})$,
where $N$ is the column density for an optical
depth of one in the FUV ($N{\sim} 10^{21}$ ${\rm cm^{-2}}$), 
leading to the constraint for a merged IF and DF,  $G_0/n < 0.1 v_{\rm adv}$ (with  $v_{\rm adv}$ in ${\rm km\, s^{-1}}$ and $n$ in ${\rm cm^{-3}})$. A smaller value of $G_0/n$ or larger $v_{\rm adv}$ is required if chemical reactions destroy \HH , such as reactions with
${\rm H_2^*}$. 
If advection is important then the PDR
 is known as a non-stationary 
PDR, although a steady-state structure is established in the frame of the IF. 
The CO has a much shorter photodissociation timescale than \HH,
so that the ${\rm C^+/C/CO}$ transition and the [\cii] and [\oi] emission are less affected relative to a stationary case. Since typically $v_{\rm IF}{\sim} 1$ km 
${\rm s^{-1}}$ and $G_0/n {\sim} 0.1-1$ (see {\bf Figure \ref{fig:PthGUV}} with 
$T\sim 300$ K) then advection is expected to  be of marginal importance, however, for a rapidly
expanding \hii\ region and 
for FUV illuminated clumps where the density is higher, advection may become 
significant (see Section \ref{sec:PDRsGalacticObs}). In expanding shells surrounding planetary nebulae, 
the advection of \HH\ is rapid, and the FUV and EUV radiation, as well as the shell density,  change
on sufficiently short time scales so that time dependent
\HH\ chemistry is required \citep{Goldshmidt1995, Natta1998}.

 \cite{Bron2018} followed the evolution of the IF and DF including photoevaporation in a plane parallel code for a range of incident radiation fields, densities, and stellar spectra. 
 Photoevaporation, either by EUV or FUV photons can increase the thermal pressure at the cloud (or clump) edge by a factor of 2
\cite[e.g.,][]{Gorti2002, Tielens2005}.
\cite{Bron2018} find that the PDR remains nearly isobaric and thus, in the 
molecular layer, efficient cooling
and decreased heating due to dust opacity results in a 
temperature drop and gradual compression by factors
of $10-100$. The compression will be less if turbulent or magnetic pressures contribute \citep[e.g.,][]{PerezBeaupuits2015} and individual clumps of high
thermal pressure would not arise out of a surrounding  medium of lower thermal pressure.
Instead of a gradual density increase, individual high
density clumps can arise from FUV 
heating and compression of existing over-dense structures  \citep{Gorti2002,Decataldo2019}.

1-D codes in spherical geometry account for the
divergence of the radiation field for central sources,
or for surface illumination of spherical clouds, clumps, or
disks.
Dynamical \hii\ region models in spherical geometry have been combined with PDR models 
to predict the time dependent IF and DF, the line cooling, and 
line emission across the \hii\ region and PDR. Examples include \cite{Hosokawa2006}, the WARPFIELD-EMP code \citep{Pellegrini2020} and the
MARION code \citep{Kirsanova2020}. The WARPFIELD-EMP code
uses CLOUDY to calculate the \hii/PDR properties while the MARION code uses
an updated network from \cite{Rollig2007} for the PDR. In general, 
models that follow the \hii\ region dynamics
do not include a detailed calculation of the PDR
chemistry and thermal balance, however,
the combined \hii/PDR codes  offer 
a coherent treatment of both the dynamics and physical
conditions giving rise to both the \hii\
region and PDR emission. 
Spherical PDR codes have also been used
to model irradiated disks such as the proplyds seen
in Orion \citep{Johnstone1998, Storzer1999}.

Depending on $n$, $A_{\rm V}$, and $G_0$, the chemical time
scales for grain surface reactions 
can become comparable to cloud life times for gas at moderately high 
$A_{\rm V}\gtrsim 5$,
and time
dependence becomes important \citep{Bergin2000,Hollenbach2009}. 
Estimates of time scales for various processes are given in \cite{Hollenbach2009}. For example, the adsorption time for
species $i$ is $t_{\rm ad}\approx 8\times 10^4 (m_{\rm O}/m_i)^{1/2}
(10^4\,{\rm cm^{-3}}/n)(30 {\rm K}/T)^{1/2}$ yrs where $m_{\rm O}$ is 
the mass of oxygen. For CO at cloud densities $n\lesssim 10^3$
${\rm cm^{-3}}$, and $T=10$ K the adsorption time is $t_{\rm ad} \gtrsim 2\times 10^{6}$
yrs. At long time scales $\gtrsim 10^{7}$ yrs, and large depths $A_{\rm V}\gtrsim 7-8$ the gas-phase C/O ratio
can become greater than 1 \citep{Hollenbach2009}. This is because of the slow dissociation of
CO by ${\rm He^+}$ produced by cosmic-ray ionizations. The oxygen produced will
then freeze out in ${\rm H_2O}$ and ${\rm CO_2}$ ice leaving a high abundance of gas-phase
carbon. This can occur for grain temperatures $\gtrsim 20$ K and 
$\lesssim 100$ K where CO ice is thermally desorbed but ${\rm H_2O}$ ice can still form.
A similar process occurs for colder grain temperatures where CO ice is desorbed 
by cosmic rays. Steady-state models will predict a large gas phase C abundance
(and strong [\ci] line intensities) if there are few chemical paths to reduce
the gas-phase C abundance, for example, in the form of ${\rm CO_2}$ ice
\cite[e.g.,][and M.\ Kaufman, private communication]{Esplugues2019}. In general, time dependent surface chemistry is
most important at large $A_{\rm V}$ and low $G_0$. 

Additional time-dependent processes have been
modelled in 1-D including fractionation of C, D, and N, species \citep{Roueff2015}, cosmic-ray production
of H in cloud interiors (\citealt{Goldsmith2005}, see also \citealt{Padovani2018} for steady-state),
and the time dependent 
photoevaporation of an externally illuminated GMC containing a distribution of clump and interclump
gas \citep{Vallini2017}.

\subsection{Multidimensional codes with steady state chemistry}

In some cases, a multidimensional geometry is more appropriate, rather than the plane-parallel or 1-D spherical geometry described above. For example, in 1-D geometry, the evaporated gas remains along a line between the star and cloud.
This is especially problematic in the \hii\ blister phase where the 1-D radial symmetry is broken and the
evaporated gas streams into the ISM. 
Variations in the density structure from clumps or turbulence can lead to 
multiple pathways for the FUV field to enter the cloud, creating
a range of
physical conditions in the same telescope beam  \cite[e.g.,][]{Nagy2017}. 
There may also be multiple internal sources of FUV radiation.
To address these problems, PDR models for multidimensional clouds 
have been developed (e.g.,\ \textsc{KOSMA}-$\tau$, \citealt{Rollig2013}; 3D-PDR, \citealt{Bisbas2012}). 
The \textsc{KOSMA}-$\tau$  models consist of a single or distribution of 1-D spherical clumps. 
This was  carried further by \citet{AndreeLabsch2017} who constructed a 3-D ensemble 
of pixels, each containing a distribution of clumps
to simulate the structure of the Orion Bar. 
Multidimensional PDR codes have also been developed to
model protostellar envelopes and outflows \citep{Visser2012,Bruderer2009b,Lee2014}.
For example, \cite{Visser2012} found that the UV illumination of
outflow walls can dominate the observed mid-$J$ CO line emission. 
\cite{Spaans1994} constructed a 2-D PDR model with arbitrary geometry and density that uses a Monte Carlo approach for the radiation transfer.
A common technique is to use a time-dependent hydrodynamic simulation to obtain the 
density and velocity fields and then to ``post-process" it with a PDR code
to obtain the steady-state chemical abundances and 
thermal equilibrium gas temperature 
\citep[e.g.,][]{Levrier2012}.  
The 3D-PDR code is highly flexible  and well suited
to post-process the complex geometries and density
distributions obtained with time-dependent magnetohydrodynamics (MHD) codes. 
  The 3D-PDR code has been used, for example, to obtain the emission diagnostics 
from MHD simulations of a molecular cloud for
a range of cosmic-ray ionization
rates, UV fields and densities \citep{Bisbas2021}.

The H/\HH\ transition, and abundance of molecular ions, have been examined in 
simulations of FUV illuminated, turbulent, diffuse clouds by \cite{Bialy2017,Bialy2019b}. 
The simulations were post-processed
to find the steady-state abundances of \HH\ and from a steady-state PDR model, the molecular ion abundances. The density fluctuations broaden the probability distribution functions
of column densities although the mean $N_{\rm H}$ is well fit by the analytic
solution of \cite{Sternberg2014} for uniform density. Comparing the model distribution of column densities
to observations
constrains the characteristic driving scale and Mach number of the turbulence. An approach by
\cite{Bisbas2019} uses  lognormal  distributions in  $A_{\rm V}$ coupled with an $A_{\rm V}-n$ 
relation from simulations as inputs to a 1-D PDR code to 
 simulate the H/\HH\ and ${\rm C^+}$/C/CO fractions in the ISM.

\begin{marginnote}
\entry{Mach number}{Usually noted
as ${\cal M}=\sigma_{\rm turb}/c_{\rm PDR}$, is the ratio of the turbulent velocity dispersion
over the sound speed in the gas.}
\end{marginnote}

\subsection{3-D Magnetohydrodynamic codes with time dependent chemistry}

The codes discussed in the previous section post-process the density and velocity fields with steady-state PDR models. Here we consider  
MHD codes with incident radiation that calculate  simultaneously the time-dependent chemistry, and the density and velocity
fields produced from turbulence. Due to large computational times, the
chemistry is generally not as detailed as in a 1-D PDR code 
nor is the resolution as fine, but the density and velocity 
are physically motivated and lead to processes that can not
be accounted for in a steady-state code.
One technique is to follow only 
the H/\HH\ chemistry and assume the remaining chemistry is in balance with the \HH. Another is
to follow a more complete chemistry but with a number of pseudo reactions to limit and close the network \citep[e.g.,][]{Glover2010, Gong2017,Hu2021}.  The KROME package 
can be used to integrate chemical networks with simulations \citep{Grassi2014}.
Time dependent chemistry is important if 
 $t_{\rm H_2,chem}$  exceeds 
the turbulent crossing time $t_{\rm turb} = L/\sigma_{\rm turb}(L)$ where 
$L$ is a characteristic length scale and
$\sigma_{\rm turb}(L)$ is the (1D) velocity dispersion over that length. 
Observations show that for Galactic molecular clouds 
$\sigma_{\rm turb}(L) \approx 1\, {\rm km\, s^{-1}}(L/{\rm pc})^{1/2}$ and $t_{\rm turb}{\sim} 1$ Myr $(L/{\rm pc})^{1/2}$.
Using the volume averaged density of a molecular cloud ($n{\sim} 10^2$ ${\rm cm^{-3}}$) would yield $t_{\rm H_2, chem}$ long
compared to $t_{\rm turb}$ and would indicate 
that a fully molecular gas is difficult to establish. However, 
turbulent compression leads to higher densities than the volume average and more rapid \HH\ formation.
In a shocked, compressed, PDR layer,
the turbulence may well be 
more related to the turbulence in the adjacent \hii\ region and local
feedback processes than
to the larger cloud \citep[e.g.,][]{Lancaster2021}.  Observations of spectrally resolved  [\cii] lines 
show a range of Mach numbers from weakly supersonic \citep{Goicoechea2015} to 
$\sigma_{\rm turb}/c_{\rm PDR} \sim 5$ \citep{PerezBeaupuits2015}.  Using 
a typical turbulent dispersion velocity of [\cii] found in Orion PDRs, $\sigma_{\rm turb} = 1.7$ km ${\rm s^{-1}}$ \citep{Pabst2020},
a density of $n=10^3$ ${\rm cm^{-3}}$, and a column density for the ${\rm C^+}$ layer
of $N\sim 2\times 10^{21}$ ${\rm cm^{-2}}$, then $t_{\rm H2, chem}/t_{\rm turb}\approx 1$
and turbulence may become important for the chemistry.
Similarly, for typical diffuse clouds, these two time scales can be comparable for moderate Mach numbers
\citep[e.g.,][]{Bialy2017}. As the metallicity decreases, $t_{\rm H2, chem}$ increases, and time dependent \HH\ chemistry becomes increasingly important
\citep{Hu2021}. 

An important aspect for PDR chemistry 
is the 3-D penetration of the external FUV field through the 
turbulent gas, as well as the line transfer for the gas cooling and line emission. Typically, the column density to a point is estimated by averaging over several directions and the local FUV field is found by the normally incident attenuation by that column (equation \ref{eq:photorate}). Similarly the CO and \HH\ self-shielding
are found using the estimated column density. The averaging is done
over a fixed number of angles or using a sophisticated algorithm 
such as TreeCol \citep{Clark2012}
to reduce the number of lines-of-sight. The excitation
of atoms and molecules and the subsequent line-transfer is often handled in post-processing 
with a 3-D radiation transfer code such as RADMC-3D\footnote{https://github.com/dullemond/radmc3d-2.0}. 

Simulations for the evolution of single or colliding 
clouds have been carried out by several groups
\citep[e.g.,][]{Clark2019,Seifried2017}.
As a result of turbulent
compression, the molecular formation time is enhanced with
${\sim} 50$\% conversion to molecular gas within a few Myr, a factor
of 10 faster than with no internal dynamics \citep{Goldsmith2007}.
At early times, when the \HH\ fraction is small, the [\cii] emission traces
mainly \hi\ and is a poor tracer of CO-dark gas, while [\ci] and
CO mainly trace \HH. At later times, the fraction of [\cii] emission tracing the \HH\ likely increases \citep{Franeck2018,Clark2019}. The density fluctuations 
can produce both
CO-bright and CO-dark gas at the same depth, and \cite{Seifried2020}
fit a relation for the \HH\ mass 
based on the CO line intensity and ${A_{\rm V}}$. The  distribution of [\ci] emission has long been a matter of debate.
In a layered PDR the [\ci] should arise in a thin region on the cloud surface
tracing the CO, yet on molecular cloud scales, the [\ci] is widespread
and correlates better with $^{13}{\rm CO}$ in the interior 
than $^{12}{\rm CO}$ \citep[e.g.,][]{Keene1985,Plume2000,Burton2015}. The observed distribution
might be the result of a complex turbulent geometry that produces many internal surfaces and allows for
greater FUV penetration (\citealt{Spaans1994, Glover2015}, see also \cite{Szucs2014} for ${\rm ^{12}CO/^{13}CO}$ in a turbulent cloud).

3-D MHD simulations describe a dynamical process for the
formation of the thermal phases of the diffuse ISM: the hot ionized medium (HIM, $T{\sim} 10^6$ K), the warm neutral medium (WNM, $T{\sim} 8000$ K), and cold neutral medium (CNM, $T{\sim} 100$ K), as well as the gas flows between them.  These phases coexist within a range of thermal pressures between $P_{\rm min}/k {\sim} 10^3$ K ${\rm cm^{-3}}$ and $P_{\rm max}/k\sim 10^4$ K ${\rm cm^{-3}}$.
Steady-state calculations have found that PDRs play an important
role in producing and maintaining the WNM and CNM  multiphase 
medium (\citealt{Wolfire1995,Wolfire2003}, see also \citealt{Wolfire1995b}, \citealt{Bialy2019}, and \citealt{Hu2021} for low metallicity).
Both phases are heated by the FUV radiation from 
the interstellar radiation field via grain photoelectric heating, while the CNM is cooled mainly by [\cii], and the WNM is
cooled by [\cii], [\oi], and H Ly$\alpha$ radiation. Thermal instability 
 caused by efficient fine-structure line cooling drives the separation
between phases. 
The observed mass fractions in the local Galaxy are  
approximately 30\% CNM, 50\% WNM, and 20\% in the thermally unstable
regime \citep[e.g.,][]{Murray2018, Heiles2003}. \cite{Kalberla2018} find a larger unstable fraction (${\sim}40$\%) but with large systematic uncertainties.
The 3-D MHD simulations of 2-phase (WNM+CNM) and 3-phase ISM, find that thermal instability and turbulence act together
 to continuously drive gas between phases \citep{Audit2010,Seifried2011,Walch2015,Kim2017,Hill2018,Bellomi2020}. A 
 multiphase ISM is produced with phases at similar thermal pressures, with a fraction of thermally unstable gas that is passing through phases. 
 In global models of
 the ISM, \citep[]{Ostriker2010} a feedback loop between the FUV radiation generated by star formation, the turbulence injected by SN, 
 and the pressure of the gas in the Galactic midplane, maintains the thermal pressure in the range for a multiphase medium. 
 
3-D MHD simulations \citep[e.g.,][]{Bellomi2020} have also 
reproduced the distribution of 
\HH\ versus $N$
as observed in the diffuse gas \citep{Shull2021}.
The simulations result in a distribution of gas
densities where low densities are correlated with
lower $N$. Similarly,
steady-state models with high $G_0/n$ values 
lead to low \HH\ column densities
 \citep{Wolfire2008}.
In diffuse or low density gas within molecular clouds, \HH\ can have a significant abundance 
due to turbulent mixing or
mass transfer between phases
  \citep[e.g.,][]{Glover2007,Valdivia2016, Seifried2017,Bellomi2020}.
This is an abundance that is high compared to the steady-state value and is partially shielded from the dissociating
FUV by intervening columns in the complex geometry.
A small fraction of the diffuse \HH\, can be FUV pumped 
or collisionally excited in warm gas phases
and drive endothermic reactions such as those required
to produce ${\rm CH^+}$ (e.g., \citealt{Lesaffre2007, Valdivia2016}, Godard et al.\ 2022, in preparation).
See also \cite{Gerin2016} for 
additional observations and models of the diffuse gas.

\section{Galactic Observations}
\label{sec:PDRsGalacticObs}

Since the previous reviews, Galactic PDRs have been
observed with space based (\spitzer,
\herschel), suborbital (Stratospheric Observatory for Infrared Astronomy; SOFIA, Stratospheric Terahertz Observatory 2; STO2), and
ground based (many including Atacama Large Millimeter/submillimeter Array; ALMA) observatories, with increasing spatial and velocity resolution in pointed 
and mapping modes. Here we discuss 
results driven mainly by new observations.

\subsection{High FUV field PDRs}
\label{subsec:PDRsHighFUV}
\begin{figure}[ht]
\includegraphics[width=5in]{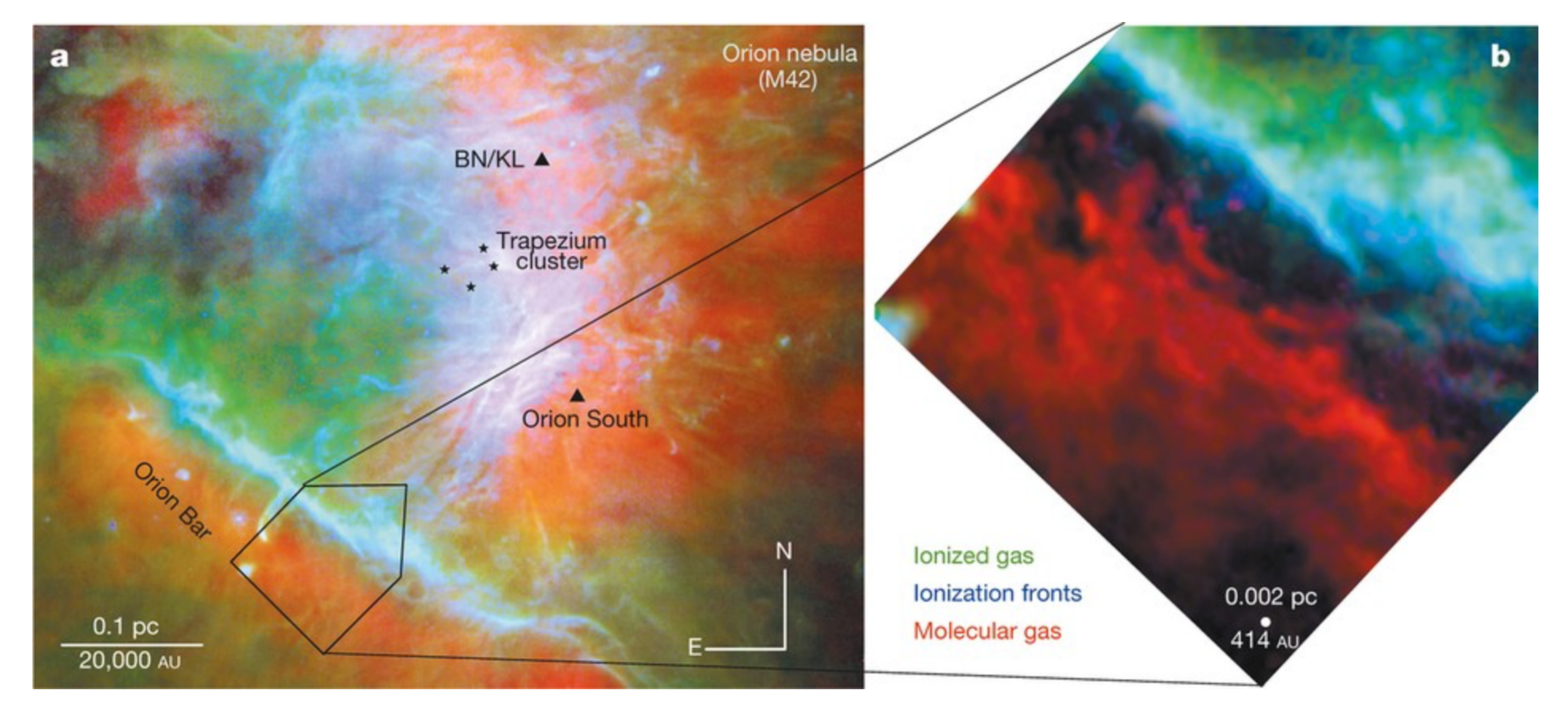}
\caption{($a$) Overview of the Orion nebula with 
the edge-on Orion Bar PDR to lower left. 
The hot ionized gas produced by the Trapezium stars is seen in [\sii] 6731 \AA\ (green),  
the ionization front  is seen in [\oi] 6300 \AA\ (blue), and the molecular PDR in ${\rm HCO^+}$ ($J=3-2$)
(red). Not shown is the atomic
[\cii] and [\oi] 63\,\mic\ emission, and PAH emission,
that lies between the 
ionized and molecular gas. ($b$) Close up of the
ALMA ${\rm HCO^+}$ ($J=4-3$) emission (red) showing
the ionization front  and molecular dissociation front. Figure published with permission from 
\cite{Goicoechea2016}, \copyright Nature.}
\label{fig:GoicoecheaOrionBar}
\end{figure}

{\bf Figure \ref{fig:GoicoecheaOrionBar}a} shows an
overview of the Orion nebula. The PDR behind
the Trapezium cluster is the one modeled by 
TH85. The Orion Bar is the prototypical edge-on PDR in which the PDR layers are spread across the sky \citep[e.g.,][]{Tielens1993, Tauber1994,Hogerheijde1995}.
The stellar winds
and \hii\ region  powered by the stars in the Trapezium cluster have created  a bowl in the molecular cloud
and the Orion Bar is the FUV illuminated edge of the bowl.
The IF traced by [\oi] 6300 \AA\ emission is seen 
in blue and the DF of the molecular PDR traced 
by ${\rm HCO^+}$ in red. Although not shown in this figure, the 
[\cii],  [\oi] 63\,\mic, and PAH emission is bright between
the ionized and molecular gas. {\bf Figure \ref{fig:GoicoecheaOrionBar}b} shows
a close-up view of the bar in ${\rm HCO^+}$ ($J=4-3$)
emission in red  taken
with ALMA, and clearly showing the separation between ionization
and dissociation fronts filled by warm, neutral, atomic gas.
Emission from FUV pumped ${\rm H_2^*}(v=1-0)$ is also observed 
at the 
edge of the molecular gas (\citealt{Walmsley2000}, Le Gal et al.\ in preparation). 

 High density ($n{\sim} 10^5-10^7$ ${\rm cm^{-3}}$) and high pressure ($P_{\rm th}{\sim} 10^8$ K ${\rm cm^{-3}}$) clumps are a common feature  in high FUV field PDRs,
as demonstrated by a number of observations
including interferometry \citep[e.g.,][]{YoungOwl2002,Lis2003},
\HH\ line emission \citep{Sheffer2011}, and [\oi] and high-$J$ CO molecular line emission (\citealt{Ossenkopf2010,Wu2018,Joblin2018}, see also \citealt{Visser2012} for discussion of the CO ladder in the context of protostellar envelopes). The clumps
are embedded in a lower density medium mainly responsible for the  [\cii] and a portion of the [\oi] line emission and the low- to mid-$J$ CO  
line emission. 
For the Orion Bar, an incident radiation field of $G_0 = 3\times 10^4$ and interclump density of $n=5\times 10^4$ ${\rm cm^{-3}}$  
is consistent with the [\oi] and [\cii] line emission as well
as the separation between the IF and DF \citep{Tielens1993,Hogerheijde1995,Marconi1998}.
The ALMA
${\rm HCO^+} (J= 4-3)$ observations close to the DF indicate clump densities of $n=10^6$ and sizes of $0.004$ pc. The high-$J$ 
CO ladder also indicates high density clumps \citep{Joblin2018}.  
Deeper into the bar, 
larger ($0.01-0.02$ pc) clumps are observed.
In general, thermal pressures in clumps are higher than in the 
interclump medium and could be self-gravitating, or transient, turbulently compressed
features or compressed by FUV photoevaporation \citep{Gorti2002,Lis2003}. 
In the Orion Bar, thermal pressures
are $P_{\rm th}{\sim} 3\times 10^8$ K ${\rm cm^{-3}}$ for the clumps, and  $P_{\rm th}{\sim} 1.5\times 10^7$ K ${\rm cm^{-3}}$ for the interclump gas which is comparable to the magnetic pressure $P_{\rm B}{\sim} 3\times 10^7$ K ${\rm cm^{-3}}$ \citep{Goicoechea2016, Pabst2020}.

The layered structure in the Orion Bar is apparent in
many atomic and molecular tracers \citep[e.g.,][]{Tielens1993, Walmsley2000, Lis2003,vanderWiel2009,BernardSalas2012,Joblin2018,Parikka2018} demonstrating an edge-on geometry. Nevertheless,
on small scales a more complex structure is seen at the DF.   
Molecular emission in the form of globules
or plumes extends into the atomic gas, indicating advection of the molecular gas through the DF \citep{Goicoechea2016}. In addition, the
${\rm HCO^+} (J= 4-3)$, high-$J$ CO \citep{Parikka2018}, and ${\rm H_2^*}(v=1-0)$ are nearly coincident,
suggesting a merging of the \HH\ and CO dissociation layers
  \citep{Goicoechea2016,Kirsanova2019}. However, we note that this may also be due to high densities and resulting small 
scale sizes, with vibrationally excited \HH\  driving the carbon chemistry to produce both CO and ${\rm HCO^+}$.  We also note that
the FUV pumping rate producing ${\rm H_2^*}$ is proportional to the
destruction rate which, in steady state,  is equal to the formation rate ($R_{\rm H_2}\propto n_{\rm H}n$). Thus, both \HH\ formation and FUV pumping are proportional 
to $n_{\rm H}$ and the ${\rm H_2^*}$ peaks where the gas is atomic rather than molecular. 
In the (isobaric) [\oi]
and [\cii] emitting regions, the temperature varies by a factor
of a few and the gas density is relatively constant.
In the Orion bar and NGC\,7023, \cite{Joblin2018} find
that the  rate of \HH\ formation is enhanced by a factor of 3-4 over that in diffuse gas, thereby drawing the \HH\ closer to the surface 
and leading to warmer gas by collisional de-excitation
of exited \HH\ (see {Section \ref{sec:PDR_chemistry}}). 
Endothermic carbon chemistry (${\rm H_2^*} + {\rm C^+} \rightarrow {\rm CH^+} + {\rm H}$) driven by FUV pumped 
and warm \HH\ also draws ${\rm CH^+}$ and  CO into warm gas regions \citep{Goicoechea2019}. However, \citep{Goicoechea2021} find that endothermic
reactions with ${\rm H_2^*}$ are not sufficient to explain the observed
column densities of sulfur species in the Orion Bar. 
Although sulfur chemistry is still poorly understood, they
suggest that surface chemistry can make up the difference.

Velocity resolved line profiles of [\cii] and [\oi] 
used in combination with other tracers such as [\ci] and CO 
reveal outflows \citep[e.g.,][]{Schneider2018},
thermal and
non-thermal pressures \citep[e.g.,][]{Goicoechea2015, PerezBeaupuits2015},
multiple PDR components 
\citep[e.g.,][]{Dedes2010,Seo2019}, 
and can separate the neutral and ionized gas. For example, line widths  suggest 
an equipartition of thermal, turbulent, and magnetic pressures in the PDRs associated with the Orion region and RCW\,49 \citep[][]{Tiwari2021,Pabst2021a}. \cite{Anderson2019} find that the (velocity resolved) [\nii] 205\,\mic\ emission in S235 is clearly velocity shifted from the [\cii] emission and only 
${\sim}10$\% of the [\cii] comes from the neutral gas\footnote{Many of the estimates of the ionized gas contribution to [\cii] come from spectrally unresolved [\nii] observations (Section \ref{subsec:OriginofCII}). A limited number of resolved [\nii] 205\,\mic\ observations have been carried by
\herschel\ HIFI, and by
SOFIA GREAT. Large scale mapping of the spectrally resolved [\nii] 205\,\mic\ line will be carried by GUSTO.}. In contrast,  \cite{Seo2019} find in the Trumpler 14/Carina region that
most of the [\cii] comes from the ionized gas along lines of
sight that pass through the blister \hii\ region with
embedded, 
high density, neutral cores also contributing.

\begin{figure}[ht]
\includegraphics[width=\textwidth]{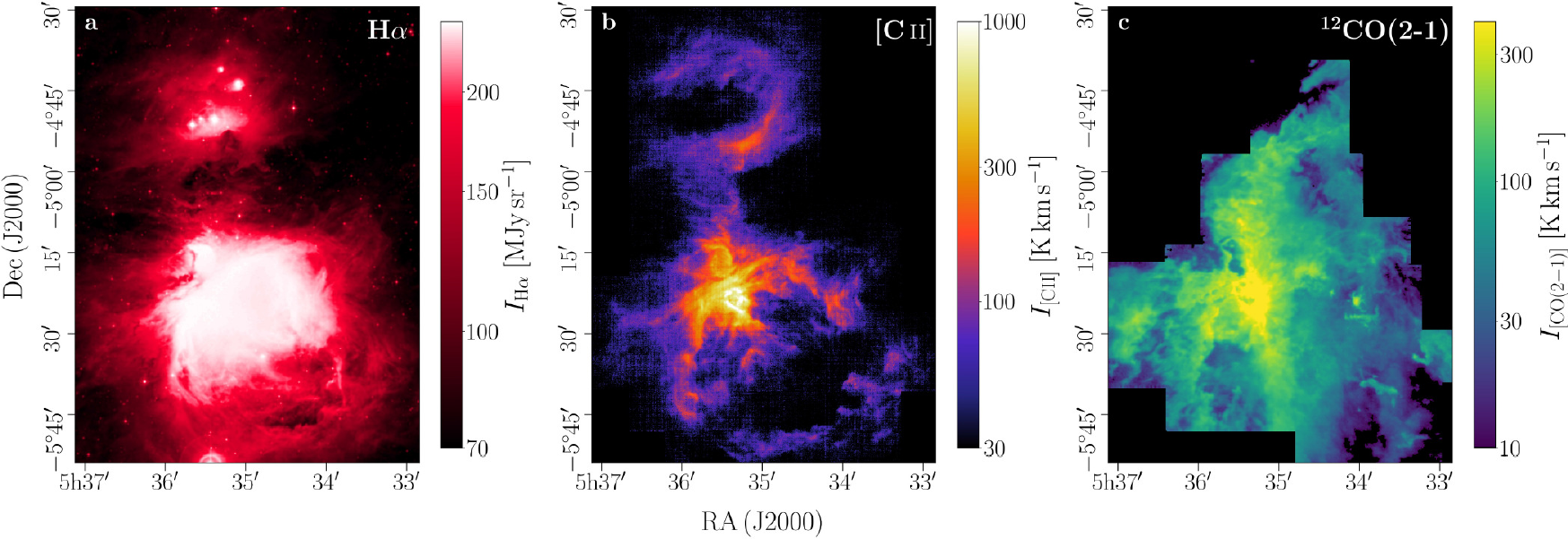}
\caption{Square-degree-sized maps of the Orion region. ($a$) H$\alpha$ emission showing the distribution of ionized gas. ($b$) Integrated [\cii] showing the distribution of neutral atomic and molecular CO-dark gas 
in the PDR. The data are taken by SOFIA/upGREAT at $\Delta v =0.2$ km ${\rm s^{-1}}$ and $16''$ spatial resolution.  ($c$) CO(2-1) emission showing the distribution of molecular gas; regions in black have not been observed. 
Bubbles are clearly seen in [\cii] with ionized gas interiors.
CO shows the molecular ridge with on-going star formation.
Figure adapted with permission from \cite{Pabst2020}, \copyright ESO. The
\href{https://www.astro.umd.edu/~mwolfire/images/Orion.mp4}{\bf Supplemental Video}
steps through the [\cii] velocity channel maps showing the dynamical structure of the entire region.  Video from  Universit\"at zu K\"oln/NASA/SOFIA,
see also \cite{Higgins2021}.
\label{fig:PabstCIIOrion}}
\end{figure}

\begin{figure}[ht]
\includegraphics[width=4.0in]{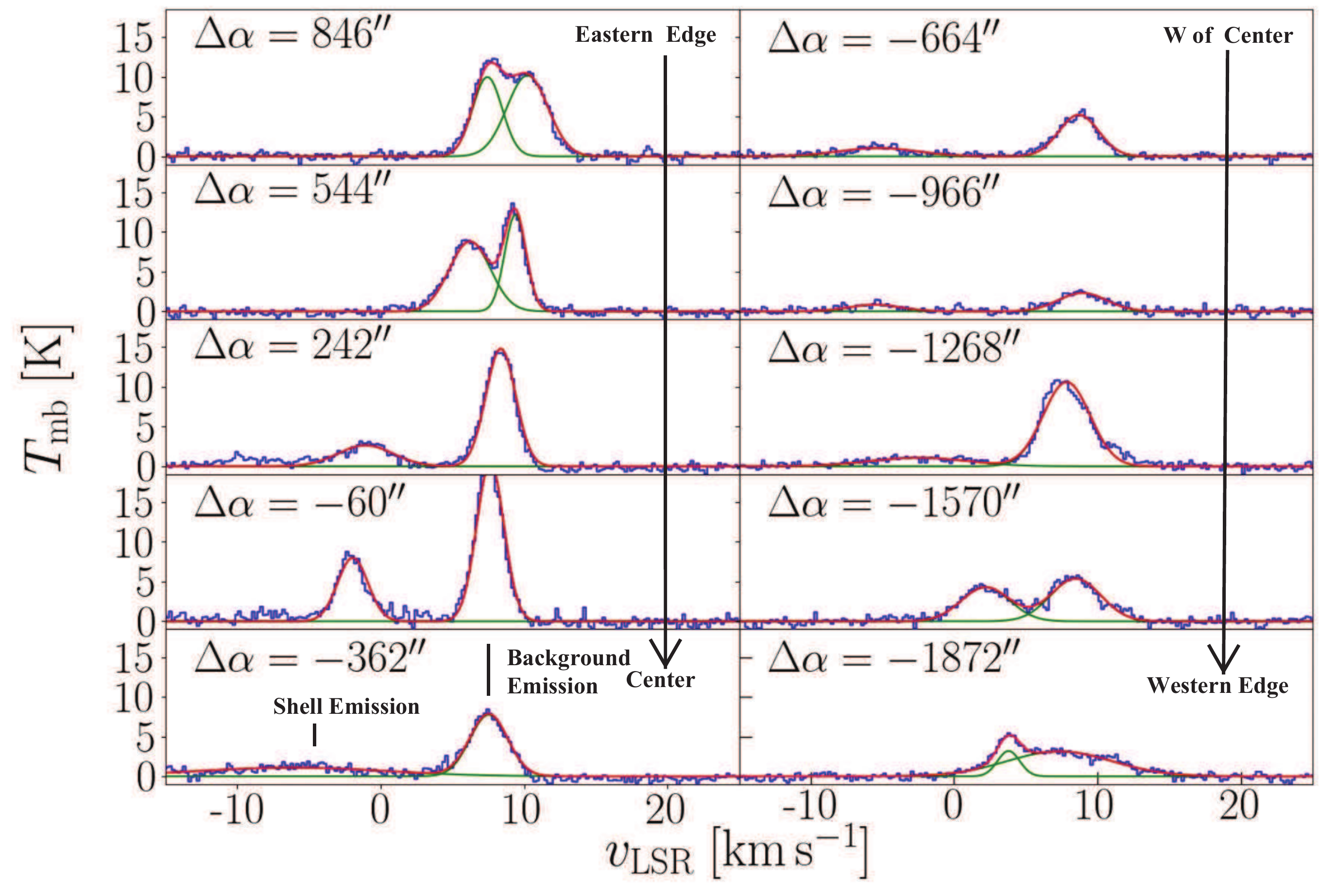} 
\caption{Velocity resolved profiles of [\cii], 
 along an East-West  cut across the Orion Veil shown in Figure \ref{fig:PabstCIIOrion}. 
 The profiles show
that the [\cii] emission comes from an expanding shell of gas. In  going from east to west ($\Delta \alpha=846''$ to $\Delta \alpha=-1872''$), there is a sequential shift in the peak of the blue shifted shell component
relative to the background at ${\sim}8$\,km\,${\rm s^{-1}}$. At the shell edge, the line merges with the background while at the shell center the blue shifted velocity is at a maximum ($\Delta v{\sim}13$\,km\,${\rm s^{-1}}$)
relative to the background.
Figure adapted with permission from \cite{Pabst2020}, \copyright ESO.
\label{fig:PDRLineProfiles} 
\vspace*{-1em}
}
\end{figure}

The fine-structure line profiles, along with high spatial resolution large-scale mapping ({\bf Figure  \ref{fig:PabstCIIOrion}}) have been used to examine
the kinematics of the gas and stellar feedback processes \citep[see][for an overview of the SOFIA FEEDBACK project]{Schneider2020}. 
Analyzing the [\cii] line spectra along cuts across the source ({\bf Figure \ref{fig:PDRLineProfiles}}) 
and position-velocity diagrams, delineates expanding shells
of neutral gas and gives the radius, velocity, and mass of these expanding 
shells, which provides the energetics required to drive the expansion. An analysis of the Orion Veil [\cii],  associated with the  M42  \hii\ region \citep{Pabst2019, Pabst2020} and seen 
in the Orion map ({\bf Figure \ref{fig:PabstCIIOrion}} to the south west)
find the neutral shell is
expanding at $13\, {\rm km\, s^{-1}}$, and, along with estimates of 
the various feedback pressures using observations of H$\alpha$, CO, and
X-rays, is interpreted as a consequence of a wind-blown bubble as discussed in Section \ref{sec:multiD_PDR}.
A similar result was found analyzing the [\cii]
emission in RCW\,120 \citep{Luisi2021}. The bubbles in the M43 and NGC\,1977 regions to the 
north of M42 are powered by early B stars, with weak winds, and are
found instead to be dominated by the thermal expansion of the \hii\ region \citep{Pabst2020}. The [\cii] emission in RCW\,49 \citep{Tiwari2021}
suggests the bubble was powered initially by the winds 
from the Wd2 cluster, but currently, in late stage
evolution, is likely driven by a Wolf-Rayet star wind.

Analytic models \citep[e.g.,][]{Krumholz2009}, simulations 
\cite[e.g.,][]{Walch2012,Kim2018}
and observations \citep[e.g.,][]{Lopez2014,Barnes2020, Chevance2022} have been used to estimate the dominant feedback
processes.
It is clear that
pre-SN feedback mechanisms dominate in destroying molecular clouds but there is disagreement between mainly observational determinations 
and hydrodynamical simulations as to the dominant 
process. Infrared radiation pressure is unlikely to be 
important since each thermal re-emission from dust is shifted to
increasingly longer wavelengths where 
clouds are optically thin 
\citep{Wolfire1986,Reissl2018}. 
The [\cii] and pressure results from Orion and RCW\,49 suggest that 
winds dominate, especially in the early phases but can
also dominate in the later phases. 
Photoionization can also be important, especially for early B stars which lack strong stellar winds. 
The analysis of feedback timescales in nearby galaxies \citep{Kruijssen2019,Chevance2022}  suggests indeed that stellar winds and photoionization are the two dominant  feedback mechanisms responsible for dispersing molecular clouds. In contrast, hydrodynamic simulations
suggest that rapid admixing of cold shell material into the hot
gas in a turbulent mixing layer leads to rapid radiative cooling  and  thereby diminishes
the effects of winds \citep{Lancaster2021,Lancaster2021b}. Additional simulations and observations are called for in  a
range of evolutionary stages and environments. 

Bubbles are seen
throughout the Galactic plane in images from \spitzer\ at 8\,\mic\ \citep{Churchwell2006} mainly arising from PAH emission in PDRs. However, most have yet to be analyzed in velocity resolved PDR lines. 
The Galactic/Extragalactic ULDB Spectroscopic Terahertz Observatory; GUSTO, balloon project \citep{Goldsmith2022}, will map large areas of the inner Galactic plane  (and LMC) at angular resolution
${\sim}0.6'-0.9'$, and spectral resolution greater than 1 ${\rm km}$ ${\rm s^{-1}}$
in [\cii],  [\oi] 63\,\mic\ and 
[\nii] 205\,\mic. GUSTO will map 
the kinematics of large scale [\cii] structures driven by feedback processes, the [\cii] association with ionized gas and dense PDRs, and 
the CO-dark gas fraction.

Small hydrocarbons and fullerenes observed
near the surfaces of PDRs illuminated by high
radiation fields could be the result of PAH
fragmentation \citep{Pety2005,Berne2012,Guzman2015}. In addition, 
the observation of COMs near the surface, could be the result
of the better mobility of molecules in warmer ice mantles
that produce some COMs that are then photodesorbed \citep{Guzman2014}.
\cite{LeGal2019} suggest, instead,  that high C/O ratios (for example as a result
of ${\rm H_2O}$ freeze-out) result in hydrocarbons, e.g., ${\rm HC_3N}$  produced by
gas-phase chemistry alone. 

A greater understanding of PDR 
structure, chemistry, and thermal balance will come from 
{\it James Webb Space Telescope} (JWST) observations. With
a resolution of ${\sim} 100\,{\rm  AU}$ in nearby Galactic PDRs, observations of
the \HH\ line emission, for example, greatly constrain the
 PDR temperature profile, ${\rm H_2}$ formation mechanism, and the ${\rm H_2^*}$ chemistry
(see e.g., the PDRs4ALL program targeting the Orion Bar; 
\citealt{Berne2022}).

\subsection{Intermediate to Low FUV fields}
\label{subsec:PDRsIntermediateFUV}

A large number of PDRs have been observed in both
pointed observations and mapping modes giving the
integrated line strengths of the
dominant cooling lines. These include e.g., observations
from \spitzer, \herschel, and SOFIA and are too numerous
to discuss separately and only a few have  been mentioned throughout this review. They provide the inputs needed
to analyze the physical conditions in many sources covering a wide range in densities, radiation fields, and temperatures, and lead to mapping 
the physical 
conditions over the source. The observations allow for an assessment of the dominant 
energy input and its magnitude, and to better
understand the chemical and thermal processes in the ISM.

It has long been suspected that [\oi] 63\,\mic\ absorption along the
line-of-sight
 can diminish the integrated
[\oi] 63\,\mic\ emission and affect the interpretation of line ratios \citep[e.g.,][]{Kraemer1998,Liseau2006}.
\begin{marginnote}
\entry{Self-absorption}{A line profile 
is self-absorbed if there is dip at line
center, generally caused by foreground absorption or
a temperature gradient.}
\end{marginnote}Sources with high 
[\oi] 145 $\mu$m/[\oi] 63 $\mu$m ratios $\gtrsim 0.1$ cannot be
explained by an externally illuminated PDR model with face-on geometry, but 
presumably results from foreground absorption of the [\oi] 63\,\mic\ line.
Note that PDR models do account for optical 
depth effects in the line within the PDR, with [\oi] 63\,\mic\  typically being optically thick, [\oi] 145\,\mic\  optically thin, and [\cii] marginally thick. \cite{Guevara2020}  used observations of the [$^{13}$\cii] hyperfine transitions  to determine [$^{12}$\cii] optical depths of 1-2 in several PDRs.  \cite{Goldsmith2019} considering [\oi], and earlier \cite{Wolfire1993}
for CO, note that subthermal excitation at the edge of a cloud 
or decreasing gas temperatures can
result in self-absorbed line profiles,  but since models account for sub-thermal
excitation and a variable temperature distribution, the emergent integrated intensity calculated by the models is unaffected
by self-absorption within the PDR.
The models, however, do not include foreground absorption due to cold or low density gas along the
line-of-sight which would absorb  the observed line intensity emitted 
at the surface.  Generally, a self-absorbed line refers to the observed shape
of the profile and not where it is produced.
With velocity resolved line observations 
it has become clear
that [\oi] 63 $\mu$m towards PDRs can be self-absorbed  \cite[e.g.,][]{Ossenkopf2015, Leurini2015,Schneider2018,Mookerjea2021}. The [\cii] line has also been observed to be self-absorbed towards a few PDRs \citep[e.g.,][]{Graf2012, Guevara2020,Mookerjea2021} as well as both
[\cii] and [\oi] seen in diffuse gas \citep{Gerin2015, Wiesemeyer2016}. Although it takes only a small
column of cold oxygen to fit observations with $\tau_0{\sim} 1-2$, $N({\rm O}){\sim} 2-4\times 10^{17}(\Delta v_{\rm FWHM}/{\rm km\, s^{-1}})$ ${\rm cm^{-2}}$, the required column of cold ${\rm C^+}$
can be quite large \citep{Graf2012,Guevara2020}.
For example, the average foreground column towards M17SW is
$N({\rm C^+}){\sim} 2\times 10^{18}$ ${\rm cm^{-2}}$,  requiring an
$A_{\rm V}{\sim} 6$ in the ${\rm C^+}$ layer, and is larger
than expected from a single foreground PDR. 
The physical location of the O and ${\rm C^+}$ absorbing layers are not well determined but must be closely associated with the PDR due to velocity coherence of the absorption and
emission components. Absorption might result from a line-of-sight that
passes through  a cloud to a PDR on the far side \citep{Kraemer1998,Goldsmith2021}, or to a PDR on the inner
edge of a neutral shell surrounding an \hii\ region \cite[][]{Kirsanova2020}. 
The correction to the [\oi] 63\,\mic\ line intensity due to self-absorption is typically estimated to be a factor
${\sim}2-4$ \citep[e.g.,][]{Schneider2018,Goldsmith2021}. The [\oi] 145\,\mic\ line can be used
as a diagnostic if the 63\,\mic\ line is self-absorbed, since it
is not seen in absorption due to the high
 $\Delta E/k = 228$ K height above ground of the 
 lower level. 

Using gamma-ray observations, \cite{Grenier2005} found gas
that was not seen in either \hi\ nor CO emission.
It is now considered to be mainly molecular \citep{Murray2018b}
and is known as the 
CO-poor or CO-dark molecular gas that was observed 
earlier in translucent clouds and explained by theoretical
models \citep{Lada1988, vanDishoeck1987, vanDishoeck1990}. 
The CO-dark layer is located at an
$A_{\rm V}{\sim} 1$ in PDRs, where \HH\ 
self-shielding creates a layer of \HH, 
while CO forms deeper into the cloud
\citep{vanDishoeck1988}. 
 \cite{Wolfire2010} constructed theoretical 
models of the surfaces of illuminated molecular clouds and found, for local Galactic conditions, a CO-dark mass
fraction of ${\sim} 30$\%. 
Values of ${\sim} 30-50$\% are roughly consistent with Galactic observations from
gamma-ray observations \citep{Grenier2005}, dust emission from 
{\it Planck} \citep{PlanckDarkGas2011}, 
extinction from 2MASS \citep{Paradis2012},
 [\cii] line emission from \herschel\ \citep{Pineda2013}, and from numerical
 simulations of galactic disks where the CO-dark gas is found in interarm filaments (e.g., \citealt{Smith2014}, see
 also \citealt{Kalberla2020}). 
 There is also
a significant CO-dark fraction in the diffuse gas discussed in Section \ref{subsec:PDRsDiffuseGas}.
The [\cii] and [\ci] emission can in principle trace the CO-dark gas but must
be calibrated since the [\cii] also arises in \hi\ gas, and [\ci]
misses the ${\rm C^+/H_2}$ regions.
Models \citep[e.g.,][]{Wolfire2010, NordonSternberg2016, Madden2020} predict the dark gas fraction should increase 
at lower metallicity due to reduced shielding of the CO, although \cite{Hu2021} suggests the
effect of metallicity is not as large as in previous models. 
The trend is confirmed by observations in
the Galactic disk  where the metallicity decreases by a factor ${\sim}2.6$ between 4 kpc and 10 kpc while the CO-dark gas fraction increases by a factor of ${\sim}4$ \citep{Pineda2013, Langer2014} and also by observations in low metallicity galaxies (Section \ref{subsec:LowMetallicityEnv}). 

Observations of several PDRs illuminated by low FUV fields
suggest 
that grain photoelectric
heating may not be sufficient to explain the
observed line emission \citep{Goldsmith2010, Pon2016}. The \HH\ S(3)/S(1) 
line ratio observed in Taurus can not be achieved by
FUV heating alone, and the mid-$J$ CO lines observed in several infrared dark clouds
are stronger than can be produced by PDRs. \cite{Pon2016}  suggest that weak shocks ($v_{\rm sh}{\sim} 3$ km
${\rm s^{-1}}$) plus PDRs can excite both the low and mid-$J$ CO lines, although uncertainties in 
the heating rate and temperature distribution may allow for sufficient CO emission without  the addition of shock heating.

\subsection{Diffuse Gas}
\label{subsec:PDRsDiffuseGas}
The \herschel\ HIFI GOT C+ program carried out a pointed [\cii] line survey in the Galactic plane\footnote{The GOT C+ program consists of 500 lines-of-sight around the Galactic plane in a volume weighted sparse survey. Although dense molecular gas was detected, most of the volume is filled with
diffuse gas.}. The $\Delta v{\sim} 0.8$ km s$^{-1}$ velocity resolution allowed for the separation of
[\cii] clouds along the line of sight due to Galactic rotation
and for the  determination of  kinematic distances \citep{Pineda2013}. An estimate of the \HH\ fraction was made by comparing
the observed [\cii] line emission with that expected from the  observed \hi\ column density and assuming any additional [\cii]  emission is the result of collisions
with \HH. 
\cite{Langer2014} find a significant mass fraction of CO-dark \HH\ gas that varies considerably over cloud type, from ${\sim}75$\% for diffuse molecular clouds to ${\sim} 20$\% for dense molecular clouds 
and with an average mass fraction of ${\sim} 44\pm 28$\%. \cite{Kalberla2020} find a mass fraction of ${\sim}46$\% in diffuse molecular clouds by comparing observed \hi\ column densities with extinction.

In addition to the inferred \HH\ fractions,   observations have directly detected molecules in
diffuse molecular clouds 
in absorption e.g., HCO$^+$, HCN \citep{Hogerheijde1995b,Lucas1996}, HF, ${\rm H_2O}$  \citep{Flagey2013,Sonnentrucker2015},  
${\rm OH^+}$, ${\rm H_2O^+}$  \citep[e.g.,][]{Wyrowski2010,Gerin2010}, ${\rm ArH^+}$
\citep{Schilke2014,Jacob2020}, CH \cite{Sheffer2008}, SH \cite{Neufeld2015},
and in both absorption and emission e.g., OH \citep{Liszt1996,Busch2019}, with CH, HF and ${\rm H_2O}$ being
particularly good (linear) tracers of the \HH\ fraction. 
Although not strictly CO-dark in the original sense
since the bulk of the gas may be detectable in
\hi , nevertheless these observations detect a portion of molecular gas
that is not seen in CO emission. 
The abundances of several carbon
 (e.g., ${\rm CH^+}$, ${\rm HCO^+}$, and CO)  
 and sulfur (e.g., SH, ${\rm SH^+}$) species
 are under produced in PDR models and additional processes are required to drive the initial endothermic reactions.
These could be turbulent dissipation regions \citep{Godard2014,Myers2015,Moseley2021},
ion-neutral drift \citep{desForets1986, Visser2009}, or warm \HH\ as a result of turbulent mixing \cite[e.g.,][]{Valdivia2016}.
Observations of molecular ions
have been used to constrain the cosmic-ray
ionization rate in diffuse gas (see Section~\ref{sec:PDR_heating}).
See also the HyGAL survey using SOFIA/GREAT \citep{Jacob2022}, and reviews by \cite{Snow2006} and \cite{Gerin2016} for additional observations and models.

\section{Extragalactic Observations}
Observations in the Milky Way have the advantage of a high spatial resolution. However, the range of environmental conditions that can be probed is limited and the confusion on the line of sight might be important when looking at regions within the disk. Extragalactic observations are therefore a crucial step towards understanding the physical processes at play in the ISM and how these vary with environmental properties such as gas pressure, temperature, and metallicity. The mixing of different regions within a beam of finite resolution of the observation remains a limitation, but recent and future telescopes are shedding a new light on extragalactic PDRs and XDRs. We present new findings from observations of the nearby Universe in this Section. The particular case of the high-redshift galaxies is discussed in Section~\ref{sec:high-z}.

\subsection{Observations on galaxy scales}
\label{sec:obsgalaxy_scale}
With recent infrared observatories such as  \herschel\ and SOFIA, it is possible to resolve the ISM down to a few pc in the most nearby galaxies. However, most extragalactic observations remain unresolved, hindering the detailed studies possible in the Milky Way (see Section~\ref{sec:PDRsGalacticObs}). 
The JWST will allow high-resolution observations of nearby galaxies in the
near infrared 
(${\sim}0.25''$ at 8\,\mic\, or 3.6 pc at 3 Mpc distance), complementing the sub-mm view of ALMA,
now routinely mapping the molecular gas disks of nearby galaxies \citep[e.g.,][]{Leroy2021}, and bridging high-resolution Milky-Way observations and the large range of environments covered by extragalactic observations.
We describe below commonly observed properties of extragalactic PDRs and challenges related to the interpretation of these unresolved observations.

While the [\cii ] intensity has been thought to correlate well with the FIR intensity, already with the Infrared Space Observatory  observations a deficit in [\cii ] relative to FIR was observed at high FIR \citep{Malhotra1997, Luhman1998, Malhotra2001}.  
More recently, using the PACS instrument on board \herschel, the SHINING survey \citep[][]{Herrera-Camus2018a}
has obtained observations of the six main FIR atomic and ionized gas lines in the range $\sim55-200$\,\mic\ for 52 galaxies. This sample includes star-forming galaxies, AGN dominated systems, as well as luminous and ultra-luminous infrared galaxies (ULIRGS). This line deficit relative to FIR seems to affect all observed fine-structure PDR and ionized gas lines \citep[][see also \citealt{GraciaCarpio2011}]{Herrera-Camus2018a}. 
In environments with a FIR luminosity $\gtrsim 10^{12}$\Lsun , line intensities relative to FIR can be more than an order of magnitude fainter than in lower FIR environments. 
\cite{Smith2017},
using the KINGFISH survey of galaxies \citep{Kennicutt2011} with an
average spatial resolution of ${\sim}500$ pc, 
found a
decreasing [\cii]/FIR ratio with increasing star formation surface density.
It is important to note that this line `deficit' is defined empirically by comparison with the [\cii ]/FIR ratio observed in normal galaxies. PDR models indeed predict that the correlation between [\cii ] and FIR weakens towards high radiation fields and high densities.
There exist several possible explanations for this phenomenon \citep[e.g.,][and references therein]{Herrera-Camus2018b}. PDR models show that an increase in the grain charge  parameter with increasing FIR results in the charging of small grains and PAHs, and therefore a decrease in the photoelectric efficiency, plausibly causing the [\cii] line (as well as other fine-structure lines from the PDRs) to cease tracing the FUV radiation field \citep{Malhotra2001, Croxall2012}. 
Alternatively, the fraction of EUV and FUV photons absorbed by dust increases with the ionization parameter in dusty star forming regions, increasing FIR, but leaving a comparatively smaller fraction of both EUV and FUV photons to ionize the \hii\ region and heat the neutral gas \citep[e.g.][]{GraciaCarpio2011}. We note however,
that \cite{Draine2011} suggests that radiation pressure will push grains to the outer edge of an
\hii\ region and the dust absorption within the ionized
gas will not be large.
Line deficits (also relative to the SFR) have also been reported in high-redshift starburst galaxies \citep[e.g.,][]{Maiolino2009, Stacey2010, Brisbin2015} and is discussed in more detail in Section~\ref{sec:high-z}.

Another commonly observed phenomenon is the deficit of [\cii] emission towards galaxy centers \citep[e.g.,][]{Parkin2013, Herrera-Camus2015, Smith2017, Pineda2018}. This is not always linked to the presence or the influence of an AGN, but likely to the different physical conditions in the nucleus compared to the disk (warmer temperatures, higher densities, [\oi] rather than [\cii] becomes the main coolant of the gas). In the specific case of an AGN, the increased hardness of the radiation field changes the C$^{++}$/C$^{+}$ ratio (in the ionized gas), contributing to the [\cii] deficit \citep{Langer2015}.

The [\cii ] emission is often used as a star formation rate indicator in external galaxies.
The low ionisation potential of C (11.3\,eV) means that [\cii] emission can arise from both the ionized and the neutral gas  (Section~\ref{subsec:OriginofCII}). 
The comparison of [\cii] with the ionized gas lines [\nii] 122\,\mic, [\nii] 205\,\mic, and [\oiii] 88\,\mic\ shows that the majority of the [\cii] emission originates from the neutral gas, and that this proportion increases in the most active star-forming regions (from 60\% to 90\% in the SHINING sample; \citealt{Herrera-Camus2018a}). Moreover, the fraction of the [\cii ] emission originating from the ionized gas likely arises from low-ionisation, diffuse gas, or from the outer parts of
\hii\ regions, where in both cases the ionization is driven by stellar EUV photons.

In the neutral gas, and at moderate densities, [\cii] is the main coolant and, assuming thermal equilibrium, the [\cii] emission is therefore a measure of the heating rate. For normal galaxies, the heating is dominated by the photoelectric heating effect, which results from the interaction of the FUV radiation from (young, high-mass) stars with small dust grains and PAHs \citep[e.g.,][and Section~\ref{sec:PDR_heating}]{Hollenbach1999RvMP}. 
 There is therefore a direct, expected link between [\cii] emission and star formation activity, although this does not imply a linear relation, as evidenced above.

However, despite the encouraging correlations observed between FIR lines and the star formation rate (SFR) on galactic scales \citep[e.g.,][]{Stacey1991, Boselli2002, DeLooze2011, DeLooze2014, Herrera-Camus2015} an accurate calibration for their use as star formation tracers is still lacking. [\cii], [\oi] 63\,\mic\ and [\oiii ] 88\,\mic , commonly observed in the nearby Universe with \herschel, are considered to trace relatively well the SFR (determined via H$\alpha$, FUV, 24\,\mic, TIR or a combination of there; see e.g. \citealt{Herrera-Camus2015}) with uncertainties of about a factor of 2 in normal, starburst and AGN galaxies. There is an offset of this relation for ULIRGS due to the line deficit mentioned above \citep[e.g.,][]{DeLooze2014}. 
The scatter of the SFR--[\cii] relation strongly increases for metal-poor dwarf galaxies and in the high-$z$ Universe \citep[with a dispersion of $\sim 0.5$\,dex, approximately two times larger than observed for normal galaxies in the local universe,][see also discussion in Section \ref{sec:high-z}]{Carniani2018}. This suggests that in such environments, with lower metal abundance, warmer dust temperature, and higher ionisation state of the gas, other lines might dominate the cooling (see Section~\ref{subsec:LowMetallicityEnv}).
This scatter can be reduced by combining the emission from multiple (ideally all) lines contributing to gas cooling.

\begin{marginnote}
\entry{SFR}{Star formation rate, often derived from observed
intensities (or surface brightness) in units of
${\rm M}_\odot\,{\rm yr^{-1}}\,{\rm kpc^{-2}}$ or from observed luminosities
in ${\rm M}_{\odot}\,{\rm yr^{-1}}$.}
\end{marginnote}

Another challenge regarding the use of PDR models to interpret extragalactic observations, is that even in the most nearby galaxies, the finite resolution mixes several environments in one beam. The various sources of [\cii ]  emission can have multiple origins that might not be co-spatial and can originate from different physical regions with distinct physical properties. 
While PDR models have been originally developed to explain the emission from single, nearby, Galactic regions, such as the Orion nebula, they have also been successfully applied to larger scales and even full galaxies \citep[e.g.,][]{Wolfire1990, Malhotra2001}. Nonetheless, it is important to keep in mind that even PDR models considering an ensemble of clouds represent very idealized cases, especially when applied to extragalactic conditions, where observations result from a mix between (diffuse) ionized and neutral gas, and filamentary, highly structured molecular clouds. With the limited amount of data available, it is often only possible to constrain an average PDR model. Only high-sensitivity, high (spatial and velocity)-resolution, multi-wavelength observations enable disentangling the different components and help distinguishing between an extended component or a collection of dense clouds \citep[e.g.,][]{Kramer2005}. Even when restricting the analysis to the molecular gas alone, there is evidence for a diversity of environments from which emission lines arise. One example is the ratio $^{12}$CO/$^{13}$CO, which is observed to be high in nearby galaxies \citep[${\sim} 10-20$; ][]{Schulz2007, Gallagher2018}, in contrast with high-resolution Galactic studies where it is typically ${\sim}7$ \citep{Burton2013}. Although both lines are emitted from the same regions, these observations suggest the presence of a diffuse molecular gas component, where $^{12}$CO is marginally
thick and radiative trapping drives levels towards thermal excitation, while
$^{13}$CO is thin and remains subthermally excited.

Finally, by contrast with early PDR models suggesting that [\ci] is emitted from a thin layer around molecular clouds, both observational studies (\citealt{Plume1994, Ikeda1999} and more recently \citealt{Popping2017, Nesvadba2019, Valentino2020}) as well as theory and simulations \citep{Papadopoulos2004, Offner2014, Glover2015, Gaches2019, HeintzWatson2020} have shown a strong correlation between [\ci] emission and that from CO isotopologues, likely resulting from turbulence within clouds. The turbulence produces a complex geometry with many internal
surfaces and FUV pathways to dissociate the CO and excite [\ci].

Despite being less luminous than [\cii], both [\ci] lines at 370 and 609\,\mic\ therefore show a strong potential as tracers of the total mass of molecular gas, especially at low metallicity or at high redshift, where the fraction of CO-dark molecular gas is expected to increase \citep[e.g.,][]{Glover2016, Madden2020,Hu2021}. With \herschel\ SPIRE and now ALMA, [\ci] observations of nearby to high-redshift galaxies are becoming more common, but further investigation will be required to better understand the conditions under which [\ci] lines are emitted (also see Section \ref{subsec:XDR_diagnostics}) and establish an accurate calibration of the \ci-to-H$_2$ conversion factor.

\begin{figure}[h]
\includegraphics[width=4in]{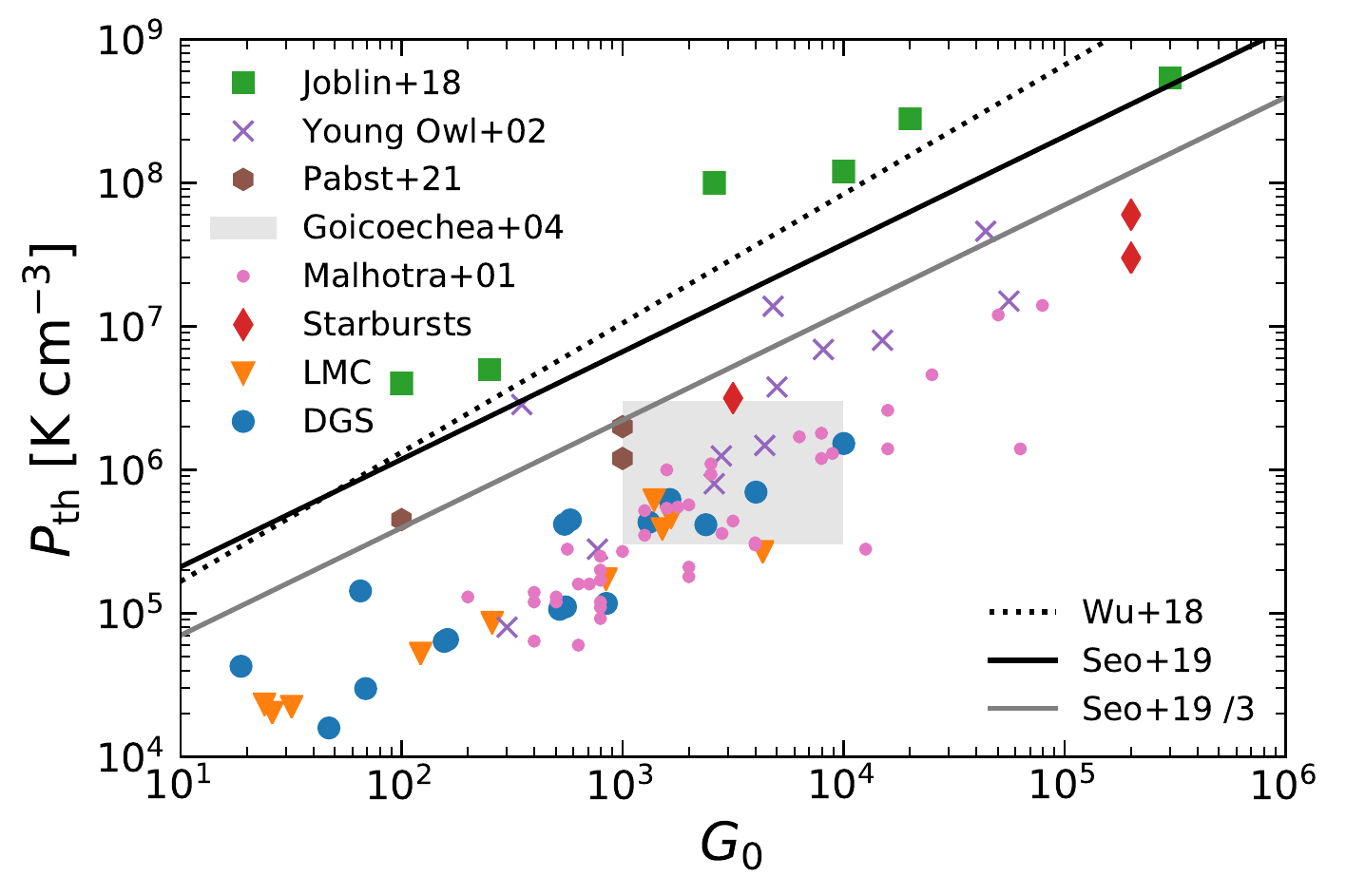}
\caption{Thermal pressure as a function of the incident radiation field in different environments. Both quantities are constrained from PDR models based on the observed emission of fine-structure lines or CO lines. When necessary, the thermal pressure was estimated as the product of the gas density and the gas temperature (constrained from the observations or assumed as noted below). The data points represent the following observations: the Milky Way \citep[][and references therein]{Joblin2018, YoungOwl2002, Pabst2022}, Sagittarius B2 \citep[assuming a temperature of 300K;][]{Goicoechea2004}, normal and starburst galaxies \citep[][and references therein]{Malhotra2001, Nagy2012} and low metallicity galaxies (LMC, \citealt{Chevance2016PhDT}; Dwarf Galaxy Survey, \citealt{Cormier2019}). For the Dwarf Galaxy Survey (DGS), we assume a temperature of 150\,K. The dashed line is the fit from \cite{Wu2018}, the black solid line is the model from \cite{Seo2019}, using $f=2.18$ and $\Phi_{\rm EUV} = 2.2 \times 10^{50}$\,s$^{-1}$ (see text), and the grey line is for the same model divided by 3, accounting for pressure equipartition. Figure
adapted with permission from \cite{Pabst2022}, \copyright ESO.}

\label{fig:PthGUV}
\end{figure}

\subsection{Mapping and velocity resolved observations applied to galaxies}

A resolution of ${\sim} 50-100$\,pc is typically required to separate the emission from individual giant molecular clouds. At the same time, covering areas of several kpc will reveal the influence of galactic structure (rotation curve, spiral arms, radial dependencies). In order to achieve both the required large scale coverage and the high spatial resolution within a feasible amount of time, galaxy surveys have typically been restricted to the most nearby ones, and to the brightest emission lines ([\cii], [\ci], CO, but HCN and other dense gas tracers are now becoming possible to map at high resolution thanks to ALMA), in combination with broadband infrared emission from dust (see e.g., \citealt{Schruba2017} for observations of NGC\,6288 at 2 pc resolution). Nonetheless, even in the most nearby galaxies, the sub-pc resolution attained by the Milky Way observations is unachievable.  High-velocity resolution observations can be an asset to try and disentangle the different components mixed into the beam, and start bridging Galactic and extragalactic observations. We explore below some new results lead by recent observations at high (spatial and/or spectral) resolution.

The Antennae galaxies have been fully mapped in three CO transitions
which were used in a PDR model to 
estimate the total molecular gas mass of the galaxies \citep{Schulz2007}. While the large size of the beam ($\gtrsim 1$\,kpc) makes it impossible to disentangle the different components of the ISM, \cite{Schulz2007} argue that this spatial averaging has little influence on the derived total hydrogen column density, as well as on the temperature and density structure of the clouds (which determine the relative CO line intensities).

At higher spatial resolution, the HERM33ES survey \citep{Kramer2010} obtained \herschel/PACS observations of M33, with a projected beam of ${\sim} 50$\,pc. 
This is sufficient to associate most of the [\cii ] emission with a PDR, start probing the physical conditions around a single \hii\ region, and suggest the presence of CO-dark molecular gas \citep{Braine2012}.
Based on the comparison of [\cii], [\oi] and TIR emission with PDR models, \citet{Kramer2020} infer a relatively homogeneous medium on these scales in M33, with the large-scale average gas properties being similar to the average of the gas properties measured in four individual regions; namely: a moderate density ($n {\sim} 2\times 10^2$\,cm$^{-3}$) and radiation field ($G_0 {\sim} 60$), with a relatively low beam filling factor of the PDRs, defined as the ratio between the radiation field constrained from the modeling and the radiation field inferred from the observed TIR emission ($f_b {\sim} 1$ by contrast to a larger value enabled by the presence of several PDRs on the line of sight). No difference was detected between the inner and outer parts of the galaxy. 

In moderately-resolved galaxies, multi-wavelengths observations can help identify different components within the gas. \cite{Schirm2017} has mapped M51 at a spatial resolution of 40\,pc with the SPIRE FTS instrument on \herschel\ and ground based observatories. They detect in total seven CO transitions and both [\ci] transitions and   identify two gas components: a cold molecular gas component residing in PDRs and a warm component requiring additional source of heating, potentially from supernovae, stellar winds, or shocks.
For that same galaxy, \citet{Pineda2020} has made use of the high velocity resolution provided by the GREAT instrument on SOFIA to identify large-scale velocity gradients in \hi , CO and [\cii], at 700\,pc spatial resolution. These observations reveal a clear offset in the position-velocity space between these lines at the location of the spiral arms, with CO tracing the upstream molecular gas and [\cii] tracing the downstream star-forming regions. They also confirm the presence of shocked gas in the spiral arms of M51 and in the arm connecting the galaxy to its companion M51b. We note however that the interpretation might be complicated by the presence of strong tidal interactions, and it would be very valuable to perform a similar study in an isolated galaxy.

With these types of observations spanning a large range of environments, a correlation between the thermal pressure and the strength of the incident radiation field can be identified over several orders of magnitude. This has been observed for Galactic regions \citep[][]{YoungOwl2002, Joblin2018, Wu2018, Pabst2022} and LMC and SMC star forming regions \citep{Chevance2016PhDT}. \textbf{Figure~\ref{fig:PthGUV}} shows that this relation also holds for entire galaxies, from low metallicity dwarfs to starburst galaxies, although more observations are needed to confirm this trend. The lack of homogeneity between the various data sets makes it indeed difficult to quantitatively characterize this correlation. Different methods have been used to constrain the PDR models and the gas properties, such as the analysis of fine-structure line emissions or CO ladders, possibly creating a displacement in the $P_{\rm th}$ -- $G_0$ plane. In particular, the results from \citet{Joblin2018} and \citet{Wu2018}, based on high-$J$ CO lines (with high critical density) are offset towards higher pressures in \textbf{Figure~\ref{fig:PthGUV}}, compared to studies based on fine-structure lines. Nonetheless, it is a puzzling result that, despite the differences in methodology and the wide range of sources, some that are described by an \hii\ region/PDR interface and some that are not (reflection nebula, embedded clumps, large beams encompassing diffuse gas and star forming regions, unresolved galaxies), all these observations fall on a similar trend.

One possible explanation of the $P_{\rm th}-G_0$ relation comes from balancing \hii\ region and PDR thermal pressures at the St\"omgren radius
\citep{YoungOwl2002, Seo2019}:
\begin{equation}
P_{\rm PDR} = 4.6 \times 10^4 f^{-3/4} \left(\frac{\Phi_{\rm EUV}}{10^{51} \rm s^{-1}} \right)^{-1/4} G_0^{3/4} {\rm ~K~cm}^{-3}\, ,
\end{equation} 
 where $f$ is the ratio of the FUV photon luminosity  to the hydrogen ionizing photon luminosity, $\Phi_{\rm EUV}$. We show this relation as a black solid line in \textbf{Figure~\ref{fig:PthGUV}} using $f=2.18$ and $\Phi_{\rm EUV} = 2.2 \times 10^{50}$\,s$^{-1}$, following \citet{Seo2019}. 
 However, if there are substantial contributions by magnetic and turbulent pressures in the PDR, then the thermal pressure required to balance the \hii\ region is lower.
 We therefore also show in \textbf{Figure~\ref{fig:PthGUV}} the above relation divided by a factor of three, assuming equipartition between thermal, turbulent, and magnetic pressure (as measured in Orion by \citealt{Pabst2020}). There is a qualitative agreement between the observations and the relatively simple theoretical model of pressure equilibrium between PDR and the ionized gas, assuming equipartition between thermal, turbulent, and magnetic pressure. We note nonetheless that entire galaxies have lower pressures for a given incident field, which could be at least partially explained by averaging over various environments (and in particular by significant diffuse gas reservoirs).
 We also note that the high pressure results derived from 
 high-$J$ CO observations  \citep{Joblin2018} lead to a slightly steeper dependence. 
 
In the future, bridging Milky-Way and extragalactic observations will be key to interpret the mix of environments present in one beam of unresolved observations, and understand the nature of the correlation described above. A promising project is the LMC+ survey (PI: S. Madden), which will cover a wide field-of-view ($1.3^{\circ} \times 0.5^{\circ}$) in the LMC, at high spatial resolution (${\sim} 2.5$\,pc) with SOFIA/FIFI-LS. 
In the coming years, JWST observations will also contribute to this effort, by increasing the achievable spatial resolution for many near-IR PDR emission lines. 

\subsection{Galactic centers}
Cold gas in the center of galaxies fuels both star formation and -- in active galaxies -- the accretion onto the central supermassive black hole. For this reason both PDRs and XDRs can exist in the center of galaxies. 
The high gas density and pressure, the presence of strong magnetic fields, highly turbulent gas, and strong radiation make it difficult to disentangle the different heating sources. The multiple diagnostics proposed for identifying the dominant heating mechanism(s) in extragalactic observations are not always decisive (see Section \ref{subsec:XDR_diagnostics}). Detailed, high resolution observations of a close-by galactic center, the one of our own Milky Way, can bring a better understanding of these extreme environments.

Several mechanisms are potentially responsible for the heating in the central $\sim 500$\,pc (the Central Molecular Zone, CMZ) of the Milky Way \citep[see e.g.,][for a recent review]{Mills2017}. We provide here a brief description of these different sources. The center of the Milky Way does not exhibit (at the moment) strong mid-infrared ionized lines which are typically associated with the presence of an AGN or XDR. Contrary to some other galaxy centers, X-rays are not thought to be the current dominant source of heating in the CMZ. However, past events could have triggered intense X-ray radiation, orders of magnitude higher than present values \citep[][]{Baganoff2001, Inui2009, Ponti2010}. X-ray irradiation can enhance (by $\approx 2$ orders of magnitude) the abundances of molecular species such as ${\rm H_2O}$, ${\rm CH_3OH}$, and ${\rm H_2CO}$ \citep[e.g.][]{Liu2020}, and may remain visible several million years after the X-ray sources turns off, in particular in high-density molecular clouds in the vicinity of the galactic center.

FUV radiation participates in the heating of the gas in the CMZ, but most likely not as the only source \citep{Rodriguez-Fernandez2004}. The [\cii] emission
can be accounted for by dense PDRs and ionized gas \citep{Langer2017, Harris2021}, but FUV photons do not penetrate deeply into the dense molecular gas,
and cannot be responsible for the high temperatures (up to $\gtrsim100$\,K) found there \citep[e.g., as traced by H$_{2}$CO;][]{Ao2013}. 
By contrast, cosmic rays penetrate deep into the clouds, and can lead to the difference in temperature observed between gas and dust \citep{Ao2013, Clark2013, Krieger2017}. The column density of H$^+_3$ (as well as that of OH$^+$, H$_2$O and H$_3$O$^+$) is well reproduced with a cosmic-ray ionisation rate of $\zeta_{\rm H_2}{\sim}10^{-15}-10^{-13}$\,s$^{-1}$ \citep[e.g.,][]{Oka2005, LePetit2016, Oka2019}, tracing a warm and diffuse gas component, in which the formation rate of \HH\ is likely enhanced, compared to local diffuse clouds. However, some low temperatures found in CMZ clouds ($< 50$\,K; e.g., \citealt{Nagai2007, Krieger2017}) indicate that a high cosmic-ray ionisation rate cannot be the globally dominant source of heating in the CMZ \citep{Ginsburg2016}.
Turbulence and shocks are another likely heating source \citep[e.g.,][]{RequenaTorres2012}. Dissipation of supersonic turbulence can also explain the observed difference in temperature between dust and gas. Based on the emission of fine-structure lines of neutral species ([\si], [\oi]) and of \HH\ rotation lines, \cite{Rodriguez-Fernandez2004} conclude that low velocity C-shocks due to turbulence are good heating source candidates. These shocks are likely driven by the large-scale dynamics of the CMZ (e.g., tidal forces, shear, gas inflow along the bar; \citealt{Kruijssen2014, Kruijssen2019, Krumholz2015, Tress2020}) rather than local processes (e.g., feedback ejecta). Detailed modelling of the molecular lines have led to the conclusion that turbulence is an important heating source in other galactic centers \citep[e.g.,][]{Rosenberg2014a, Rosenberg2014b}.

Around SgrA*, PDR models do not provide a good fit to the high-$J$ CO emission lines, suggesting the presence of another heating mechanism for the hot molecular gas in the inner central parsec of the Galaxy \citep{Goicoechea2013}. Low-density shocks (accompanied by supersonic turbulent dissipation and magnetic viscous heating) likely contribute to the heating of the gas. Interestingly, \citet{Goicoechea2013} point out similarities between the FIR spectrum of the hot gas around SgrA* and that of the starburst galaxy M82 \citep{Kamenetzky2012}, where shocks and turbulent heating are also found to be necessary to reproduce the high-$J$ CO emission. 

Turning now to nearby galaxies, the presence of several ISM components is systematically established in galaxy centers. The central 650\,pc of M82 are described as a `giant PDR' by \cite{Garcia-Burillo2002}, with a very high HCO abundance at the outer edge of the molecular torus. However, velocity-resolved observations with \herschel\ HIFI of the nuclear region of M82 reveal a highly inhomogeneous medium, with the presence of multiple components within a beam, which can be modelled with a low density component (70\%), a high density-low $G_0$ component (29\%), and high density-high $G_0$ component covering only 1\% of the beam area \citep{Loenen2010}.
Similarly, three main ISM phases are identified in the central region of the starburst galaxy NGC\,253 by \cite{Rosenberg2014a} and \citet{PerezBeaupuits2018}: a diffuse, warm component and a high density, low temperature component reproducing the bulk of the low- and mid-$J$ CO emission, and a third, high-density, high temperature component accounting for the higher-$J$ CO emission and HCN emission, potentially heated by shock or turbulence. The importance of cosmic-ray heating in this region remains debated.
In the center of 9 nearby active galaxies, \cite{Liu2017} also identified two ISM components, with a warm ($40-70$\,K), dense ($10^5-10^6$\,cm$^{-3}$) phase (dominating the CO intensity up to $J=8$, the FIR emission, and the emission of medium-excitation H$_2$O lines), and a more extended cold ($20-30$\,K), dense ($10^4-10^5$\,cm$^{-3}$) phase. In addition, the presence of a compact ($\lesssim$100\,pc), hotter component is identified for the two ULIRGs of the sample.  

In addition to presenting a mix of ISM components, the extreme physical conditions found in galactic centers are likely to affect line emissions in different ways. Most galaxy centers are CO bright relative to their low molecular gas content, leading to CO-to-\HH\ conversion factors approximately ten times lower than the standard Milky Way value \citep[][]{Israel2020}.
The higher gas-phase carbon abundances, elevated kinetic gas temperatures, and high molecular gas velocity dispersion in extragalactic molecular zones seem to contribute equally to this decrease of the CO-to-\HH\ conversion factor. 
Moreover, enhanced HCN/HCO$^+$ ratios in the central $\sim$100 pc region of AGN host galaxies \citep[e.g.,][]{Krips2008,Izumi2016, Imanishi2019} have been proposed as a signpost of the effect of X-rays on the gas heating, but recent analysis \citep[e.g.,][]{Privon2015, Privon2020} find no correlation between elevated HCN/HCO$^+$ and AGN activity traced by X-rays. Finally, bright HC$_3$N has been found in AGN nuclei such as Mrk\,231 \citep{Aalto2012} and NGC\,1063 \citep{Rico-Villas2021}. Given that the strong radiation from the super massive black hole is expected to destroy this molecule, its detection implies large column densities \citep[e.g., N$_{\rm HC_3N} \sim 10^{14}-10^{16}$ cm$^{-2}$ in the starburst ring of NGC\,1068; see][]{Rico-Villas2021}. 

In summary, the heating in galactic centers is caused by a mix of processes, which are difficult to model with a limited number of observational constraints, while keeping in mind that diagnostics can be ambiguous (see also Section~\ref{subsec:XDR_diagnostics}). In addition, the relative importance of these different heating mechanisms (FUV radiation, presence of an EUV/X-ray source, cosmic rays, turbulence, and shocks) likely vary between galaxies. 

\subsection{Low metallicity environments}
\label{subsec:LowMetallicityEnv}

Nearby, low-metallicity environments are often suggested as good laboratories to better understand the physical processes taking place in unresolved, high-redshift galaxies. The FIR fine-structure lines tracing the cooling and the physical conditions of the gas indeed indicate that the structure of the low-metallicity ISM is qualitatively and quantitatively different from the one in higher metallicity galaxies \citep[e.g.,][]{Kennicutt2003, Brauher2008}.
While it is becoming clear that a combination of XDR and PDR modelling is necessary to understand the observations at high redshift (see Section~\ref{sec:high-z}), the heating in most star-forming regions in moderately-low metallicity nearby galaxies is often dominated by UV radiation.

The LMC and SMC (0.5\,\Zsun\ and 0.3\Zsun),  being our closest neighbors, have been extensively studied to understand the interplay between gas, star formation and feedback at low-metallicity \citep[e.g.,][]{Leroy2009, Okada2015, Chevance2016, Jameson2018, Lee2016, Lee2019, Okada2019}. On the scale of individual star-forming regions, these studies reveal a highly porous ISM structure (with a low volume filling factor of the dense gas) and a change in the relative abundances of C$^+$, C and CO compared to higher metallicity regions: an extended PDR envelope exists around dense molecular clouds, where CO is photodissociated to deeper column densities but where \HH\ can survive through efficient self-shielding.

These findings are confirmed on larger scales, where several surveys have targeted the neutral gas in a large number of galaxies at intermediate-to-low metallicity, covering a wide range of galaxy type and star formation activity. A few examples are the sample of 22 Blue Compact Galaxies (BCGs) observed by \citet{Hunt2010}, the LITTLE THINGS survey \citep{Hunter2012} and the Dwarf Galaxy Survey \citep[DGS,][]{Madden2013}. These studies indicate that [\cii ] remains a good tracer of the neutral atomic and CO-dark gas
at low metallicity. Furthermore, there is a trend for increasing fraction of [\cii ] from the ionized gas with increasing metallicity \citep{Croxall2017}.

Several structural changes of the ISM at low metallicity are only indirectly linked to the reduced abundance of metals (including a reduced dust abundance), but more directly to a global increase of the strength of the radiation field. In the DGS, \citet{Cormier2015} find high ratios [\oiii] 88\,\mic /[\nii] 122\,\mic\ and [\niii] 57\,\mic /[\nii] 122\,\mic, indicating a harder radiation field at low metallicities. On average, [\oiii] 88\,\mic\ is the brightest line globally in these dwarf galaxies, indicating that the emission from the star forming regions dominates the emission even on galaxy scales. This fact is also reflected by the high [\oiii] 88\,\mic/[\oi] 63\,\mic\ ratio, a factor of $\sim 4$ higher than in higher metallicity galaxies, revealing a decreasing filling factor of the PDRs with decreasing metallicity. The same trend seems to be confirmed for high-$z$ galaxies \citep[][and Section~\ref{sec:high-z}]{Harikane2020}.
Using the Cloudy PDR model, higher radiation fields and densities in the PDRs are measured by \citet{Cormier2019}. On average, this hard radiation field is also found responsible for the destruction of small PAHs \citep[e.g.,][]{Hunt2010}.
We note however, that the ratio [\oiii]/[\cii] is not noticeably elevated in the 5 galaxies from the LITTLE THINGS survey analyzed by \citet{Cigan2016}, indicating harder radiation fields are not necessarily ubiquitous at low metallicity.
In addition to a potential change of the hardness of the radiation field, the photoelectric heating from FUV radiation and the X-ray heating from high-mass X-ray binaries seem to scale in opposite directions with metallicity (or the dust-to-gas ratio). The comparison of FIR fine-structure lines with photoionization models indeed suggests that X-rays from binaries could be an important heating process in extremely metal-poor environments (e.g., IZw18 at 1/30\,\Zsun ; \citealt{Lebouteiller2017}).

Finally, the deficit of CO emission at low metallicity also confirms a highly structured and porous ISM, with a very clumpy distribution of the dense molecular gas limited to small volume clumps surrounded by diffuse gas \citep[e.g.,][]{Indebetouw2013, Lebouteiller2017, Vallini2017, Jameson2018, Chevance2020b}. \HH\ might be able to survive outside of these dense clumps, due to efficient self-shielding, implying potentially large reservoirs of CO-dark gas \citep[e.g.,][]{Maloney1988, Pak1998, Glover2011, Shetty2011, Narayanan2012}. In the DGS, \citet{Madden2020} estimates that $>70$\% of the molecular gas is not traced by CO(1-0).
If high porosity is also characteristic of the high-redshift Universe, it could facilitate the escape of ionizing photons during cosmic reionization \citep{Stark2016}.

One extensively-studied region to probe a potential reservoir of CO-dark gas is the high-mass star forming region of 30 Doradus in the LMC, which is often considered as one of the best laboratories to study the impact of a super star cluster (SSC) on the sub-solar metallicity ISM. In its center is located the SSC R136, which contains a large population of massive stars \citep[see the review by][and references therein]{Crowther2019} and creates an extreme environment over tens of parsecs. The ionizing radiation propagates far from the cluster due to the lower dust abundance of the LMC and a porous environment, creating extended PDR regions \citep[e.g.,][]{Chevance2016}. ALMA observations reveal a clumpy structure of the molecular gas in this region, showing small $^{12}$CO filaments and clumps (0.1 pc) covering only about 15\% of the total area mapped by \citet{Indebetouw2013}. 
Direct high spatial resolution \HH~1-0~S(1) observations reveal that the \HH\ emission originates from the PDRs, with no evidence for shock excitation \citep{Yeh2015}.
However, the molecular gas mass inferred from these observations of the warm \HH\ is a strong lower limit on the total molecular gas mass (it does not include the mass associated with other levels of \HH\ nor does it trace the cold molecular gas associated with star formation).
Several [\cii] components not associated with CO emission were identified around R136 by \cite{Okada2019}, but the low spatial resolution of the \hi\ observations prevents their clear association with atomic gas or CO-dark molecular gas.
The comparison between the observed CO emission and the total molecular gas mass predicted by the Meudon PDR model based on SOFIA/FIFI-LS observations led \citet{Chevance2020b} to suggest the existence of a large reservoir of CO-dark molecular gas ($\gtrsim$ 75\%).
\citet{Melnick2021} caution however that this value is likely an upper limit on the fraction of CO-dark gas in 30 Doradus, if part of the dust emission originates from nebular dust not associated with the molecular gas. 
Even in that case, the combination of the strong radiation field with the half-solar metallicity of the surrounding gas does create a significant reservoir of CO-dark molecular gas around R136, as well as most likely in other high-mass star-forming regions at low metallicity. This has important implications for the inferred star formation efficiencies in these environments (biased towards higher values if a large fraction of the gas is undetected), the rate at which feedback from massive stars evaporate the reservoir of molecular gas, and the extent to which the associated shielding of molecular gas enables ongoing star formation in other parts of the cloud complex.

\section{PDRs and XDRs in the high-redshift Universe}
\label{sec:high-z}
The advent of ALMA opened a new window
on the characterization of PDR and XDR properties in the high-redshift
($z>3$) Universe
\citep[see][for a recent review]{Hodge2020}. Resolving  $<100$\,pc scales is possible only for
bright gravitationally lensed sources \citep[e.g.,][]{Rybak2020b}. Nevertheless, the
unprecedented sensitivity and resolution of ALMA enable observations
approaching sub-kpc scales, thus providing a first glimpse on the overall
ISM conditions of distant sources. The [\cii] 158\,\mic\ line is by far the
most widely targeted PDR tracer in normal (SFR$=1-100\, \rm M_{\odot}
yr^{-1}$) sources at high-$z$, while [\oi] 63 and 145\,\mic\
\citep[e.g.,][]{Rybak2020a, Lee2021}, [\ci] 609 and 370\,\mic\
\citep[e.g.,][]{Strandet2017, Valentino2020}, CO
\citep[e.g.,][]{Dodorico2018, Pavesi2019, Apostolovski2019}, and $\rm
H_2O$, $\rm HF$, and $\rm OH^+$ lines \citep[e.g.,][]{Casey2019,
Lehnert2020, Richers2021, Pensabene2021} are mostly detected in
rare/massive and highly star forming galaxies and quasars.\begin{marginnote}
 \entry{Hubble Deep Field}{{\it Hubble Space Telescope} observation 
 carried out with  long integration time that revealed
 many faint galaxies at high redshift.}
 \end{marginnote}
 
\begin{figure}[ht]
    \centering
    \includegraphics[scale=0.4]{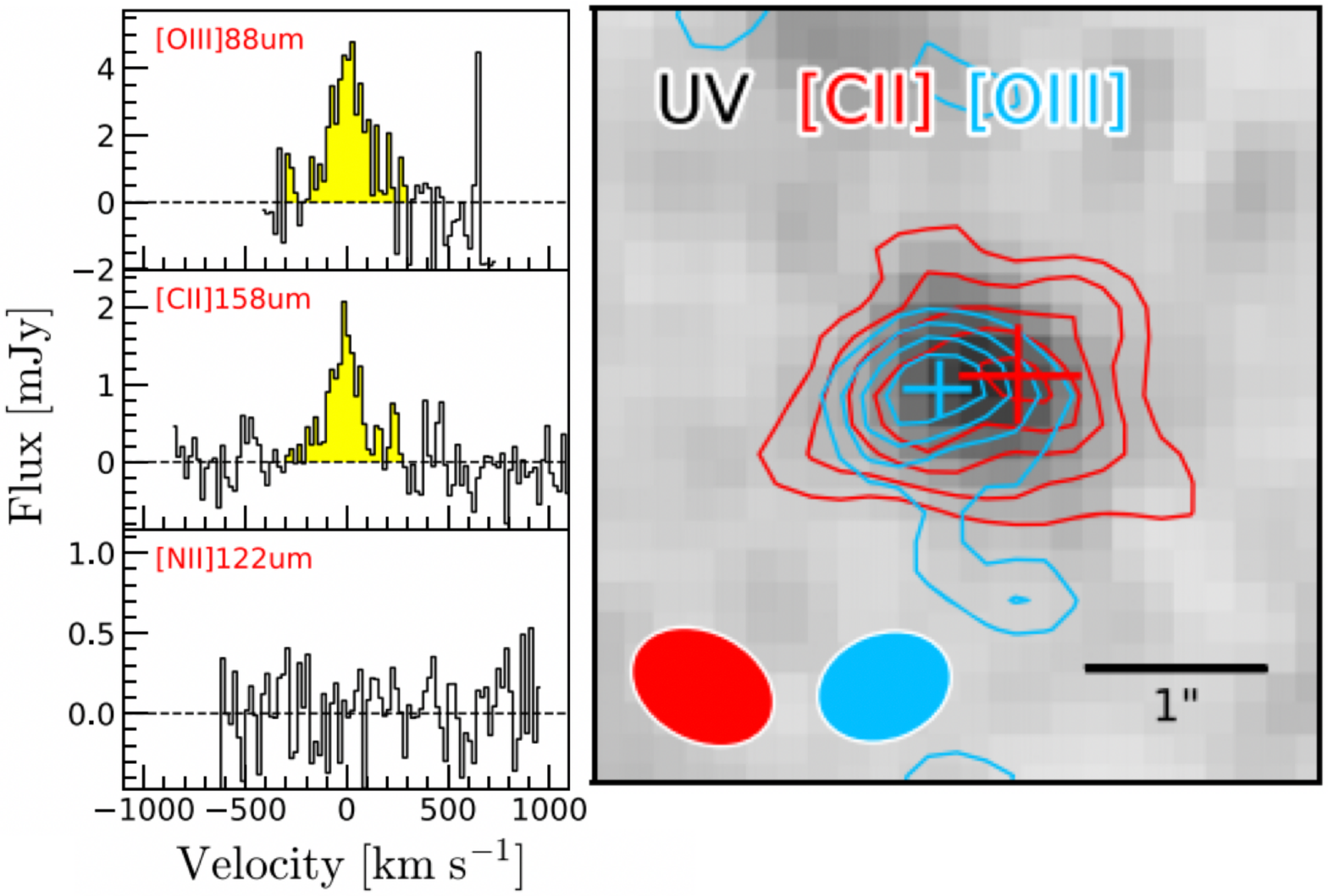}
    \caption{ALMA [CII] 158\,\mic\ and [OIII] 88\,\mic\ spectra (left panels) and emission contours (at 1$\sigma$ intervals from 2$\sigma$, right panel) of J1211-0118, a normal (SFR$=84\, \rm M_{\odot}\, yr^{-1}$) Lyman Break Galaxy at $z=6.02$ (1"$\approx5.8$ kpc). The contours are overplotted on the rest-UV Subaru image of the source. The [\nii] 122$\mu$m line was targeted but not detected. Figure adapted with permission from \citet{Harikane2020}, \copyright AAS.}
    \label{fig:ALMA_highz}
\end{figure}

 Large [\cii] surveys at $z\approx 3-5.5$, such as the ALMA-SPT \citep{Spilker2016}, ALPINE \citep{LeFevre2020}, and REBELS \citep{Bouwens2021} targeted samples of galaxies with homogeneous properties, while ASPECS \citep{Walter2016} blindly searched for CO \citep{Boogard2020} and [\cii] emitters \citep{Uzgil2021} in the Hubble Deep Field scanning the broad $1.0\lesssim z \lesssim 8$ range. Importantly, by exploiting [\cii] as a SFR indicator (see discussion in Section~\ref{sec:obsgalaxy_scale}), \citet{Loiacono2021} estimated the cosmic SFR density in the $z\approx4-5$ range from the ALPINE survey. Before ALMA, only the rest-frame UV was accessible at such high-redshifts with deep imaging campaigns conducted with the Hubble Space Telescope and ground-based observatories \citep[e.g.][and references therein]{Stark2016}. Unlike [\cii] though, the UV continuum needs to be corrected, as it is affected by dust extinction. Apart from large surveys, dozens of targeted [\cii] detections in the Epoch of Reionization (EoR) have been reported \citep[see][for a compilation]{Harikane2020}, with MACS1149-JD1 at $z=9.1$ being the farthest [\cii] emitter so far discovered \citep{Carniani2020}.\begin{marginnote}
\entry{EoR}{Epoch of Reionization is the time 
(translated to redshift $z\approx 6-15$) in 
which H between
galaxies became ionized by first stars, galaxies and quasars.}
\end{marginnote}Several peculiarities are emerging from EoR FIR line detections, including the presence of low surface brightness [\cii] halos arising from the circumgalactic medium \citep{Fujimoto2019, Ginolfi2020, Herrera-Camus2021}, large scatter with respect to the local [\cii]-SFR relation \citep{DeLooze2014}, and remarkably low [\cii]/[\oiii] 88$\mu$m ratios (see {\bf Figure \ref{fig:ALMA_highz}}).

To infer the overall conditions of PDRs (and possibly XDRs), PDR/XDR models that provide an accurate treatment of atomic and molecular microphysics on small scales 
are coupled with a rapidly increasing number of zoom-in cosmological simulations able to resolve $\approx 10$ pc scales in the ISM \citep{Vallini2013,Vallini2015, Olsen2017, Pallottini2017, Pallottini2019, Katz2017, Katz2019, Arata2020, Lupi2020} and with semi-analytic models, describing the cosmic evolution of galaxies with $z$ \citep{Lagache2018, Popping2019, Ferrara2019}.  Emission lines are computed in post-processing by interpolating pre-tabulated PDR/XDR calculations at the local FUV/X-ray flux, gas density, column density, and chemical abundances resulting from the hydro+radiative transfer simulations.
The cosmic microwave background (CMB) has non-negligible effects on the thermal equilibrium of cold gas in the EoR. Hence, any PDR/XDR model must include the appropriate CMB at the relevant redshift \citep{DaCunha2013}. The CMB sets a lower limit of $T\approx 20\, \rm K$ at $z\approx 6$ for the gas temperature, and provides a stronger background against which the lines are observed. The [\cii] emission from the diffuse neutral medium \citep{Vallini2015, Olsen2017}, and the low-$J$ CO lines \citep{DaCunha2013} are affected by these two competing effects. 
The comparison between simulations and the observed line luminosities, ratios, and morphology suggests early galaxies to be characterized by high FUV fluxes \citep[$G_0=10^2-10^4$ on kpc scales,][]{Katz2017, Rybak2020b}, high gas densities \citep[$n \approx 10^2-10^3\, \rm cm^{-3}$ on kpc scales,][]{Pallottini2017}, high turbulence {\citep[$\mathcal{M}=30$ on $\approx$30 pc scales,][]{Leung2020, Kohandel2020}}, and bursty star formation episodes during which PDRs are efficiently photoevaporated \citep{Vallini2017, Decataldo2019}. The low filling factor of PDRs in a predominantly ionized ISM on large scales, along with low C/O abundance ratios at low-metallicities, have been suggested as possible causes for the low [\cii]/[\oiii] ratios  observed at $z>7$ \citep{Harikane2020, Arata2020}.
The occurrence of extended [\cii] halos is, however, 
far from being fully understood. SN-driven cooling outflows in which the gas cools very rapidly to $T\approx 100$ K and recombines at the same time  is suggested as a viable mechanism \citep{Pizzati2020}. In this regime the formation and survival of C$^{+}$ is guaranteed, and the [\cii] emission comes from PDRs in the cold neutral outflowing gas.

The expected low metallicity and dust content of high-$z$ sources, and the corre\-spon\-dingly more widespread CO photodissociation \citep[comparable to that of low-$z$ dwarf galaxies,][and discussion in Section~\ref{subsec:LowMetallicityEnv}]{Madden2020} make the [\cii] likely a better tracer of the total \HH~mass than CO in the high-$z$ Universe \citep{Zanella2018, Dessauges2020}. \HH~mass derivations from low-$J$ CO lines are still scarce \citep{Pavesi2019} because the CMB \citep{DaCunha2013} makes low-$J$ CO line detections challenging. Instead, simulations suggest mid-$J$ CO lines to be enhanced in EoR galaxies
\citep{Vallini2018, Inoue2020}, boosted by the high turbulence, density, and molecular gas temperature characterizing these sources.
As discussed in Section \ref{subsec:XDR_diagnostics}, high-$J$ CO lines and the [\ci]/[\cii] ratios can be useful diagnostics to determine whether, on global scales, FUV photons from star formation or X-rays by AGN accretion influence the ISM heating. Given that the host galaxies of massive $z>6$ quasars are now routinely detected in [\cii] \citep[e.g.,][]{Decarli2018, Li2020_cii}, [\ci], and CO \citep[e.g.,][]{Gallerani2014, Venemans2017, Wang2019, Carniani2019, Li2020_co}, an overall estimate of the physical properties (gas density, gas temperature, dominant heating mechanism) in their ISM is now possible. In particular, the low [\ci]/[\cii] ratio observed by \citet{Venemans2017} and \citet{Pensabene2021} in $z>6$ quasars  suggest that, despite the powerful accretion rates onto the central black hole, the heating of the bulk of molecular gas in the host galaxy is likely provided by star formation producing PDRs. \citet{Gallerani2014} exploited instead the CO(17-16) line to infer the presence of a substantial XDR component contributing to the molecular emission from the host galaxy of a $z=6.8$ quasar. \citet{Pensabene2021} also concluded that high-$J$ CO detections in two quasars at $z>6$ show evidence of an XDR component, although the ${\rm H_2O}$ emission in the same sources points towards a significant contribution of IR pumping from star formation.

\begin{summary}[SUMMARY POINTS]
\begin{enumerate}

\item
Velocity resolved PDR lines provide the kinematics of the neutral 
atomic gas. These line observations 
point  to stellar winds as having a prominent role in stellar feedback, although theoretical simulations  suggest that stellar winds are much less important
due to turbulent mixing of cool material into
the wind shocked gas. In addition, 
it has long been suspected that
foreground absorption has reduced the emitted
PDR line intensities. The velocity resolved observations have
confirmed this for [\oi] and [\cii]
along diffuse lines of sight and towards dense PDRs.

\item 
1-D models set the ground work for understanding the chemistry,
thermal processes, radiation transfer, and line diagnostics. In general,
these are in  good agreement with observations of the dominant
cooling lines. 
Integrating PDR models with hydrodynamic codes add an additional tool to
understand the 
complex geometries, velocity fields,
and time dependence in FUV 
illuminated turbulent gas. 
These are 
especially important for simulations
of the diffuse ISM,  cloud evolution, galactic disks, and 
global simulations of ISM phases. 

\item A correlation between the thermal pressure and the strength of the radiation field is observed over several orders of magnitude. It is particularly interesting to note that Galactic and extragalactic star forming regions as well as full galaxies covering a wide range of environments seem to follow a similar trend. 

\item A significant fraction of molecular mass
resides in CO-dark gas
especially in low-metallicity/highly irradiated environments. The fraction is estimated to be $30-50$\% in the solar neighborhood but is $>70$\% 
for galaxies in the \herschel\ DGS.

\item
New observational facilities, reaching high spectral and spatial resolution, help to identify and characterize the mix of different components in the Galactic and extragalactic ISM. 

\item 
The CO ladder excitation 
 and [\ci]/[\cii] ratios are useful diagnostics to determine if FUV or X-rays dominate the ISM heating of extragalactic sources. 
High spatial resolution observations of HCN and HCO$^+$ lines can disentangle XDRs from PDRs 
in the galactic center of nearby sources, albeit 
 time-dependence, shock heating, and IR pumping must be included in the modelling of the line ratios.

\item 
PDR and XDR tracers are now routinely detected
on galactic scales over cosmic time using ALMA. The combination with multi-wavelength observations enables linking 
the star formation history of the Universe to the evolution of the 
physical and chemical properties of the gas.
Current PDR observations in high-$z$ galaxies suggest high $G_0$, and high density. The high ratios between ionized vs PDR gas tracers point towards low PDR filling factors. In luminous high-$z$ quasars the bulk of molecular gas heating seems to be provided by star formation producing PDRs but high-$J$ CO lines are consistent with an enhancement produced by an XDR component. High spatial resolution observations can assess if the high-$J$ lines are confined to the galactic center.

\end{enumerate}
\end{summary}


\begin{issues}[FUTURE ISSUES]
\begin{enumerate}

\item 
Wide-field, high spatial and spectral resolution IR observations are crucially needed to examine the kinematics of large scale PDR
structures, and to  bridge Galactic and extragalactic observations, and improve the interpretation of lower-resolution observations. Several planned surveys and new instruments will make the first steps in this direction in the coming years (e.g., SOFIA, JWST, GUSTO). In the nearby Milky Way, JWST will resolve PDR line emission on ${\sim 100}\,{\rm  AU}$ scales so that thermal and chemical processes  are well constrained. 

\item Several crucial chemical processes need further
refinement through laboratory, theoretical, or observational
work. These include \HH\ formation at high gas and grain
temperatures and rates for ion recombination on grains.

\item Another workshop comparing PDR models would be beneficial to
understand more recent developments and to
examine differences in temperature structure. This could also
include hydrodynamic simulations to compare directly 
steady-state and time dependent codes.
\item Further studies on time dependence in XDRs, with a specific focus on galactic centers are needed. Currently this aspect is mainly treated in the context of protoplanetary disks.

\item An XDR-focused workshop 
would be beneficial to compare and benchmark models developed for different purposes (AGN impact on  galactic centers, protoplanetary disk characterization) and to settle the capability of diagnostic line ratios to discriminate between X-ray and FUV induced heating.
\end{enumerate}
\end{issues}



\section*{DISCLOSURE STATEMENT}
The authors are not aware of any affiliations, memberships, funding, or financial holdings that
might be perceived as affecting the objectivity of this review. 

\section*{ACKNOWLEDGMENTS}
We would like to acknowledge helpful 
discussions with D. Hollenbach, A. Tielens, A. Sternberg,
S. Bialy,
G. Esplugues,
A. Ferrara,
S. Gresens,
B. Godard,
M. Kaufman,
D. Kruijssen,
D. Neufeld, 
 C. Pabst,
 A. Pallottini,
 M. Pound, 
 M. R\"ollig,
 D. Seifried, 
 E. Tarantino,
 M. Tiwari,
 help with figures from C. Pabst, M. Pound, and M. Tiwari, and a careful reading of the manuscript and insightful comments from E. van Dishoeck and E. Ostriker. 
MGW was supported in part by 
SOFIA Legacy Programs FEEDBACK and HyGal
provided by NASA through awards SOF070077 and SOF080038, 
and by the balloon project GUSTO through NASA award  NNG16FC08C.
LV gratefully acknowledges support from the European Research Council (ERC) under the European Union's Horizon 2020 research and innovation programme via the ERC Advanced Grant INTERSTELLAR (grant agreement number 740120, PI Ferrara). MC gratefully acknowledges funding from the Deutsche Forschungsgemeinschaft (German Research Foundation) through an Emmy Noether Research Group (grant number KR4801/1-1) and from the European Research Council (ERC) under the European Union's Horizon 2020 research and innovation programme via the ERC Starting Grant MUSTANG (grant agreement number 714907).


\end{document}